\documentclass[a4paper, 11pt]{article}
\usepackage[utf8]{inputenc}
\usepackage[T1]{fontenc}
\usepackage{amsfonts}
\usepackage{lmodern}
\RequirePackage{amsthm,amsmath,amssymb}
\usepackage{mathrsfs}
\RequirePackage[authoryear]{natbib}

\RequirePackage{graphicx}
\graphicspath{{figures/}}
\usepackage{mathtools}
\usepackage{booktabs}
\usepackage{caption}

\usepackage{setspace}
\usepackage{xcolor}
\usepackage{utopia}
\usepackage{geometry}
\numberwithin{equation}{section}

\theoremstyle{remark}

\newcommand{\rp}{\textsf{p}}

\newcommand{\mathdis}[1]{\mathsf{#1}}

\newcommand{\bigO}{\mathrm{O}}

\newcommand{\bs}[1]{\boldsymbol{#1}}

\renewcommand{\P}[2][]{{\mathsf P}_{#1}\left(#2\right)}
\newcommand{\code}[1]{\texttt{#1}}
\DeclareMathOperator{\sign}{sign}
\DeclareMathOperator{\se}{se}
\newcommand{\Rlang}{\ensuremath{\mathsf{R}}\ }

\newcommand{\SM}{Supplementary Material}
\RequirePackage[colorlinks,linkcolor=black,citecolor=black,urlcolor=black]{hyperref}
\usepackage{cleveref}

 \title{Improved inference on risk measures for univariate extremes}
 \author{Léo R. Belzile\footnote{HEC Montréal, 
 3000, chemin de la C\^{o}te-Sainte-Catherine, Montr\'eal (QC), Canada H3T 2A7, \texttt{leo.belzile@hec.ca}}, 
 Anthony C. Davison\footnote{Institute of Mathematics, Ecole Polytechnique F\'ed\'erale de Lausanne, Station 8, 1015 Lausanne, Switzerland, \texttt{anthony.davison@epfl.ch}}}
 
\begin{document}

\maketitle 

\begin{abstract}
We discuss the use of likelihood asymptotics for inference on risk measures in univariate extreme value problems, focusing on estimation of high quantiles and similar summaries of risk for uncertainty quantification. We study whether higher-order approximation based on the tangent exponential model can provide improved inferences, and conclude that inference based on maxima is generally robust to mild model misspecification and that profile likelihood-based confidence intervals will often be adequate, whereas inferences based on threshold exceedances can be badly biased but may be improved by higher-order methods, at least for moderate sample sizes.  We use the methods to shed light on catastrophic rainfall in Venezuela, flooding in Venice, and the lifetimes of Italian semi-supercentenarians.\footnote{Keywords: Extreme value distribution, Generalized Pareto distribution, Higher order asymptotic inference, Poisson process, Profile likelihood, Tangent exponential model.}

\end{abstract}

\section{Introduction}

\subsection{Risk measures} 
Estimating worst-case scenarios is important for risk management and policy making, but the hypothetical events that keep decision-makers awake at night can lie far outside the available data. Large-sample likelihood approximations are routinely used for inference based on extreme observations, but the numbers of rare events can be small, which raises the question of the adequacy of standard asymptotic approximations. Improved approximations are used in other domains \citep[e.g.,][]{Brazzale/Davison/Reid:2007}, but thus far they have had limited impact in extreme-value statistics. 

An important parameter in modelling univariate extremes, commonly known as the extremal index or shape parameter and denoted by $\xi$, determines how the tail probability declines at extreme levels. A negative value of $\xi$ yields a distribution with bounded support, whereas zero or positive values of $\xi$ yield unbounded extremes, with the distribution tail of form $x^{-1/\xi}$ for large $x$, so larger values of $\xi$ correspond to increasingly heavy tails. Authors who have studied likelihood inference for $\xi$ and other parameters of extreme value distributions include \citet{Pires:2018}, who focused on the shape parameter of the generalized Pareto distribution, and \citet{Giles:2016} and \citet{Roodman:2018}, who considered bias-corrected estimates of extreme-value parameters. Although inferences on $\xi$ give qualitative insights into rare event probabilities, the focus in applications is typically on measures of risk, such as exceedance probabilities for particular values of $x$, quantiles or related summaries of the tail distribution.  Accurate small-sample inference for these risk measures is the topic of this paper.

Risk measures are typically small probabilities or high quantiles, which may have very asymmetric sampling distributions, so classical `estimate$\pm c\times $standard error' confidence intervals may be appallingly bad: they may contain inadmissible parameter values, and the empirical probability that such an interval contains the true parameter value may be much less than the nominal probability.   A standard approach to dealing with this is to compute the confidence interval on a transformed scale, for example by considering the logit of a probability, but even this can perform very badly, as we shall see below.  In such settings it is natural to focus on confidence intervals that are invariant to so-called `interest-preserving transformations', which transform in a natural way: if $(L,U)$ is a $(1-\alpha)$ confidence interval for a scalar parameter $\psi$ in a model with other parameters $\lambda$, then for any monotone increasing transformation $g$, $(g(L),g(U))$ is the corresponding confidence interval for $g(\psi)$, even if the remaining parameters are transformed from $\lambda$ to $\zeta(\lambda, \psi)$.   Confidence intervals based on the profile likelihood have this property, and this is an important reason to use them in preference to other risk measures, but we shall see below that modified versions of these intervals may be preferred in some cases.  

\subsection{Vargas tragedy} \label{subsec_vargas1}

Cumulative rainfall of around 911mm over a three-day period in mid-December 1999 led to landslides and debris flow that caused an estimated 30,000 deaths in the Venezuelean coastal state of Vargas. Daily cumulated rainfall data recorded at the Maiquetía \textsl{Simón Bolívar International Airport} for the years 1961--1999 were analyzed in \cite{Coles:2003} and \cite{Coles:2003b}, whose  fit to annual maxima up to 1998 suggested that the return period for such an event would be approximately 18 million years, though more sophisticated models led to much more reasonable risk estimates. Yearly maxima for 1951--1960 and anecdotal records are also available: for example, during the floods of February 1951, a reported 282 mm of rain fell in Maiquetía over consecutive days, while the neighbouring station of El Infiernito in the Cordillera de la Costa, between Caracas and Maiquetía, recorded 529 mm for the same day \citep{Wieczorek:2001}. These events suggests potential for extremes well beyond the range seen in the daily records.

\begin{figure}[t!] 
\centering 
\includegraphics[width=\textwidth]{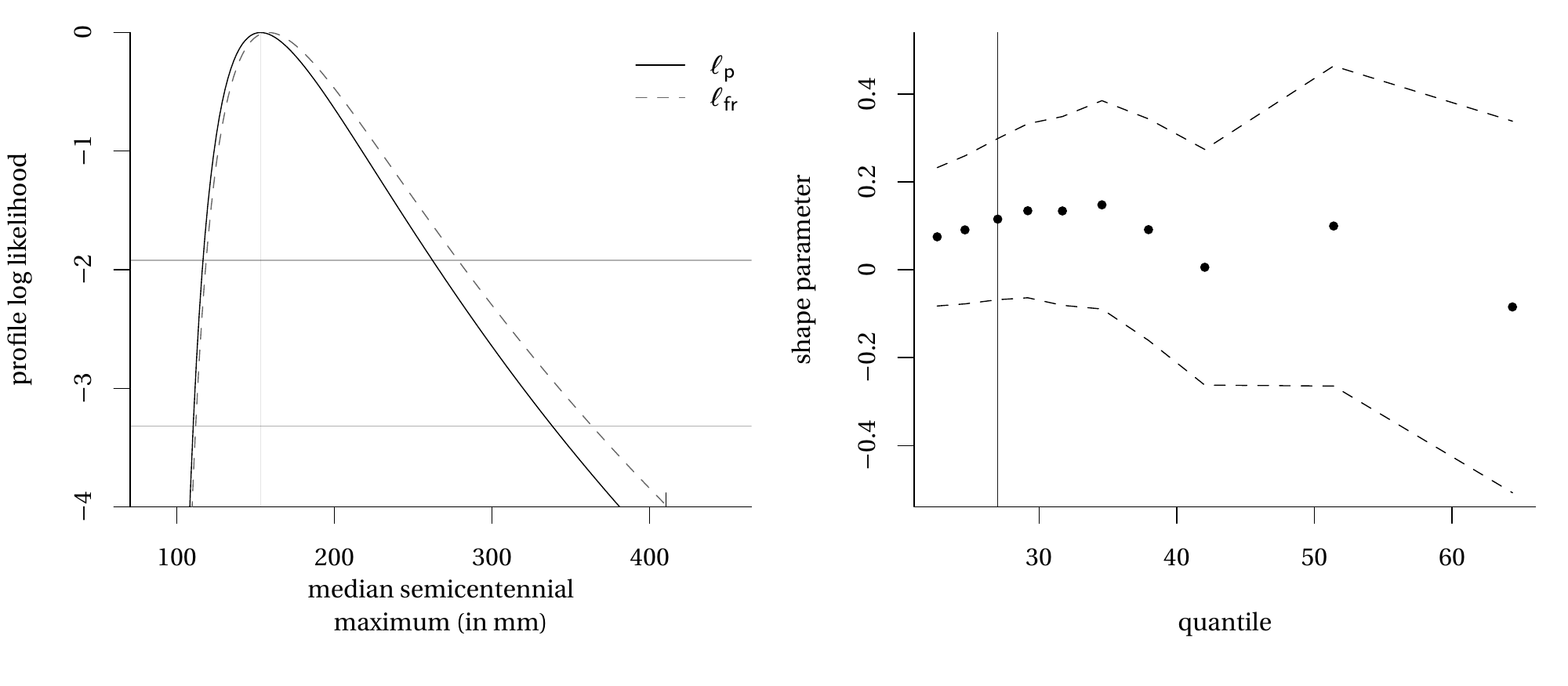}
\caption{Left panel: profile likelihood and higher-order version discussed in \Cref{subsec:tem} for the median semicentennial maximum daily rainfall at Maiquetía based on threshold exceedances above 27mm, using daily data from 1961--1998. The dashed grey horizontal line at $-1.92$ indicates cutoff values for 95\% confidence intervals based on the asymptotic $\chi^2_1$ distribution. Right panel: threshold stability plot of \cite{Wadsworth:2016} for the shape parameter $\xi$ of the generalized Pareto distribution, with 95\% simultaneous confidence intervals.}
\label{fig1}
\end{figure}

To motivate the practical need for modified likelihood approximations, we fit the generalized Pareto model described in~\S\ref{sect:EVT} to daily rainfall totals from 1961 to November 1999 that exceed 27mm, and estimate the median of the semicentennial maximum  distribution $\psi$. The threshold stability plot in the right-hand panel of \Cref{fig1} suggests that a threshold of $u=27$mm is appropriate, and the extremogram \citep{Davis/Mikosch:2009},  which estimates the conditional probabilities $\P{Y_t > u \mid Y_{t-h} > u}$ for a threshold $u$ and lags $h$, suggests that high rainfall   on successive days is only very weakly dependent. {This leaves $n_u=142$ exceedances for inference, on average 3.74 per year.} The profile likelihood for $\psi$ shown in Figure~\ref{fig1} is highly asymmetric; the maximum likelihood estimator is $\hat\psi=148$mm, and the 95\% profile and higher-order confidence intervals are respectively $(109, 247)$mm and $(120, 261)$mm.  In view of the shape of the log likelihood, any symmetric interval would be highly inappropriate and could lead to severe underestimation of the risk of rare events. 
Simulations summarised in Section~\ref{sect:sims} suggest that the modified profile likelihood interval has better properties than the usual profile likelihood interval, both in this case and for related measures of risk. 

In later sections we first outline the necessary elements of extreme-value statistics and its use for the estimation of risk.  Then we explain the construction of variants of the profile likelihood using the tangent exponential model approximation and show how they influence the conclusions to be drawn in three applied settings:  rainfall extremes in coastal Venezuela, probabilities of severe flooding in Venice and excess lifetimes of Italian semi-supercentenarians.  Finally we use simulation to assess when the higher-order methods provide improved inferences. We conclude that the coverage properties of profile likelihood intervals are adequate overall, though small-sample bias appears for extrapolation too far into the tail. The improved methods yield wider confidence intervals with more accurate error rates, though slightly larger samples are needed for them to be effective. 

\section{Basic notions}

\subsection{Extremal models} \label{sect:EVT}

Extreme value analysis is concerned with two main problems: estimating the probability of extremes of given sizes, and estimating a typical worst-case scenario over a given period. A standard approach is to fit specific distributions justified by asymptotic arguments to maxima (or minima) over specific time periods or to exceedances of a high (or low) threshold. The limiting distributions for high and low extremes are related by a simple change of sign, so we can consider only maxima and exceedances of a high threshold.  


The extremal types theorem \citep{Fisher:1928,Gnedenko:1943} characterizes the limiting distribution of maxima under very mild conditions, but we use slightly stronger assumptions for ease of exposition. Let $F(y)$ denote a thrice-differentiable distribution function with density $f(y)$ whose support has upper endpoint $y^*$, define $s(y)=-F(y)\log\{F(y)\}/f(y)$, let $b_m$ denote the solution of the equation $-\log F(b_m) = m^{-1}$ and let $a_m = s(b_m)>0$. If $M_m$ denotes the maximum of a block of $m$ independent observations from $F$, then the existence of $\xi^{\star} = \lim_{m \to \infty} s'(b_m)$ implies the existence of the limit 
\begin{align}
 \lim_{m \to \infty} \mathsf{P}{\left\{ (M_m-b_m)/a_m\leq y\right\}} = \lim_{m \to \infty} F^m(a_mx+b_m) = \exp\left\{-(1+\xi^{\star} y)_+^{-1/\xi^{\star}}\right\}, \label{prop:evattract}
\end{align}
where $a_+=\max(a,0)$ for real $a$, and also implies convergence of both the corresponding density function
and of its derivative uniformly in $y$ on all finite intervals \citep[][Theorem 5.2]{Pickands:1986}. Thus, if $a_m$ and $b_m$ were known, we might approximate the distribution of $(M_m-b_m)/a_m$ by the right-hand side of~\eqref{prop:evattract}. In practice they are unknown, so we fit the generalized extreme value distribution $\mathdis{GEV}(\mu, \sigma, \xi)$ with location parameter $\mu \in \mathbb{R}$, scale parameter $\sigma \in \mathbb{R}_{+}$ and shape parameter $\xi \in \mathbb{R}$, 
 \begin{align}
 G(y) = 
\begin{cases}
\exp\left\{-\left(1+\xi \frac{y-\mu}{\sigma}\right)^{-1/\xi}\right\}, & \xi \neq 0,\\
\exp \left\{ -\exp \left(-\frac{y-\mu}{\sigma}\right)\right\},& \xi = 0,
\end{cases}
\label{eq:gevdist}
 \end{align}
which is defined on $\{y \in \mathbb{R}: \xi(y-\mu)/\sigma > -1\}$. Setting $\xi=0$ yields the Gumbel distribution. 

The $\mathdis{GEV}(\mu, \sigma, \xi)$ is max-stable, and this allows extrapolation beyond the observed data into the tail of the distribution: if $Y_1,\ldots, Y_T \sim \mathdis{GEV}(\mu,\sigma, \xi)$ are independent, then $\max\{Y_1, \ldots, Y_T\} \sim \mathdis{GEV}(\mu_T, \sigma_T, \xi)$ with $\mu_T = \mu + \sigma(T^\xi-1)/\xi$ and $\sigma_T = \sigma T^\xi$ when $\xi \neq 0$, and with $\mu_T = \mu +\sigma \log(T)$ and $\sigma_T = \sigma$ when $\xi=0$. Thus if $\mu$, $\sigma$ and $\xi$ have been estimated from $n$ independent block maxima and the fit of~\eqref{eq:gevdist} appears adequate, the distribution of a maximum of $T$ further independent observations can also be estimated, even if $T \gg n$, though the uncertainty generally grows alarmingly as $T$ increases.

If \cref{prop:evattract} holds, then the linearly rescaled conditional distribution of an exceedance over a threshold $u < y^*$ also converges \citep[e.g.,][Theorem~3.4.5]{EKM:1997}. Let $r(y) = \{1-F(y)\}/f(y)$ denote the reciprocal hazard function; then
\begin{align}
\lim_{u \to y^*} &\frac{1-F\{u+r(u)y\}}{1-F(u)} = 1-H(y;1,\xi), \label{gpd.eq}
\shortintertext{where}
H(y; \tau, \xi) &= 
\begin{cases}
1-\left(1+\xi {y}/{\tau}\right)_{+}^{-1/\xi}, & \xi \neq 0,\\ 1-
\exp \left(-{y}/{\tau}\right)_{+},& \xi = 0, 
\end{cases} \label{eq:gpdist}
 \end{align}
is the generalized Pareto (GP) distribution function with scale $\tau \in \mathbb{R}_{+}$ and shape $\xi \in \mathbb{R}$, denoted $\mathdis{GP}(\tau, \xi)$. If $Y \sim \mathdis{GP}(\tau, \xi)$, straightforward calculations show that $Y-u \mid Y>u \sim \mathdis{GP}(\tau + \xi u, \xi)$ for any $u \in \mathbb{R}$ such that $\tau+\xi u>0$, so exceedances above a threshold $u$ also follow a $\mathdis{GP}$ distribution. This property is termed threshold-stability, and its consequences parallel those of max-stability. 

If the data consist of independent and identically distributed random variables that arrive regularly, for example on a daily basis, and that satisfy the conditions above, then the times of events that exceed the threshold $u$ can be approximated by a homogeneous Poisson process and the exceedances themselves are independent generalized Pareto variables, and this induces a Poisson process of independent event times and sizes $(t_1,y_1),\ldots, (t_n,y_n)$.

 \subsection{Risk measures}
 \label{subsec:retlev} 
Max- and threshold-stability allow the estimation of risks associated with rare events, by extrapolating the fits of the~$\mathdis{GEV}$ or~$\mathdis{GP}$ distributions.   The most common risk measure is the $T$-year return level,\ i.e., the quantile of $F$ corresponding to an event of probability $p=1-1/T$ for an annual maximum, often interpreted as ``the level exceeded by an annual maximum on average once every $T$ years''. The probability $p_l$ that a $T$-year return level is exceeded $l$ times in $T$ years of independent annual maxima may be computed using a binomial distribution with $T$ trials and success probability $1-1/T$. For large $T$, a Poisson approximation yields $p_0=p_1=0.368$, $p_2=0.184$, $p_3=0.061$ and $p_4=0.015$, so the probability of at least one exceedance over $T$ years is in fact roughly 0.63. {Perhaps more to the point, any return level is a parameter of a distribution.  Even if this was known perfectly, risk would be associated with the uncertain nature of future events.  In the Bayesian paradigm, one could measure risk using the posterior predictive distribution of the $T$-year maximum, which can be approximated by higher-order techniques \citep{Davison:1986}.  \citet[\S~3(b)]{Cox:2002} suggested using direct summaries of the distribution of the $T$-year maximum also in a frequentist setting; the $T$-year return level approximates the $0.368$ quantile of this distribution.  } 


\label{prop:extrapolation}
We mentioned above that the distribution $G_T(y)$ of the maximum of $T$ independent and identically distributed $\mathdis{GEV}(\mu,\sigma, \xi)$ variates is $\mathdis{GEV}(\mu_T,\sigma_T, \xi)$.
 Denote the expectation and $p$ quantile of the $T$-year maximum by $\mathfrak{e}_T$ and $\mathfrak{q}_p = G_T^{-1}(p)$ and the associated return level by $z_T = G^{-1}(1-1/T)$. These may all be expressed in the form
\begin{align*}
\begin{cases}
\mu+\sigma\left(\kappa_{\xi}-1\right)/\xi, & \xi <1, \xi \neq 0, \\
\mu+\sigma\kappa_0, & \xi =0,
\end{cases}
\end{align*}
where $\kappa_{\xi}$ equals $T^\xi\Gamma(1-\xi)$ for $\mathfrak{e}_T$, $\{-T/\log(p)\}^{\xi}$ for $\mathfrak{q}_p$ and $\left\{-\log 
\left(1-{1}/{T}\right)\right\}^{-\xi}$ for 
$z_T$, and 
$\kappa_0 $ equals $\log(T)+\gamma_{e}$ for $\mathfrak{e}_T$, $\log(T)-\log\{-\log(p)\}$ for $\mathfrak{q}_p$
and 
$-\log\{-\log(1-1/T)\}$ for $z_T$.

Threshold exceedances are related to maxima as follows: suppose we fit a $\mathdis{GP}(\tau, \xi)$ distribution  to exceedances above a threshold $u$, and let $\zeta_u$ denote the unknown proportion of points above $u$. If there are on average $N_y$ observations per year, then we take $H^{\zeta_uTN_y}$ as an approximation to the distribution of the $T$-year maximum above $u$.

\subsection{Penultimate approximation} \label{sec:penult}


When the  $\mathdis{GEV}$ or  $\mathdis{GP}$ distribution is fitted to maxima or threshold exceedances, the best approximating extremal distribution will generally depend on the block size $m$ or threshold $u$ and the shape parameters will differ from the limiting values arising when $m\to\infty$ or $u\to x^*$. \cite{Smith:1987} shows that, if $\xi^{\star} = \lim_{m \to \infty} s'(b_m)$ exists, then for any $x \in \{y:1+\xi^{\star} y >0\}$ there exists $z$ such that
\begin{align*}
\frac{-\log[F\{v+s(v)x\}]}{-\log\{F(v)\}} = \left\{1+s'(z)x\right\}^{-1/s'(z)}, \qquad v < z < v+s(v)x.
\end{align*}
For each $m \geq 1$, setting $v=b_m$ and $a_m=s(b_m)$ yields 
\begin{align*}
 F^m(a_mx+b_m)=\exp\left[-\left\{1+s'(z)x\right\}^{-1/s'(z)}\right] + \bigO(m^{-1}),
\end{align*}
which can be regarded as a finite-$m$, or penultimate, version of the approximation stemming from \cref{prop:evattract}. \citeauthor{Smith:1987} shows that the Hellinger distance between $F^m(a_mx+b_m)$ and the penultimate approximation $\mathdis{GEV}\{0,1, s'(b_m)\}$ approaches zero as $m\to \infty$ and that it is smaller than that between $F^m(a_mx+b_m)$ and $\mathdis{GEV}(0,1, \xi^{\star})$. Similar statements hold for the generalized Pareto distribution: unless $r'(x)$ is constant, there exists $y$ such that a finite-$u$ version of \cref{gpd.eq},
\begin{align*}
 \frac{1-F\{u+r(u)x\}}{1-F(u)} = \left\{1+r'(y)x\right\}_{+}^{-1/r'(y)}, \qquad u < y < u+r(u)x,
\end{align*}
holds. One can replace the limiting shape $\lim_{v \to x^*} r'(v)=\xi^{\star}$ by $r'(u)$, thereby reducing the Hellinger distance between the true conditional distribution of exceedances and the generalized Pareto approximation. Penultimate appproximations for specific parametric models are straightforward to obtain, as one only needs to compute the scale $a_m$, location $b_m$ and shape $s'(b_m)$ parameters for the $\mathdis{GEV}$ approximation or the scale $r(u)$ and shape $r'(u)$ parameters for the $\mathdis{GP}$ approximation.

When the limiting parametric models are fitted to finite samples, maximum likelihood estimates of the shape parameter will tend to be closer to their penultimate counterparts than to $\xi^{\star}$; moreover the estimator of $\xi$ will converge to a target that changes as the threshold or the block size increases, depending on the curvature of $s'$ or $r'$. Extrapolations far beyond the data will inevitably be biased due to the incorrect assumption that the max- or threshold-stability property that holds for infinite $m$ or the limiting threshold also applies at finite levels.  

\subsection{Finite-sample bias}
Although many approaches to estimation of the generalized Pareto and generalized extreme-value distribution have been proposed, we shall consider likelihood-based estimation, which can easily be extended to complex settings and sampling schemes. Consistency and asymptotic normality of maximum likelihood estimators for the extremal distributions have been established \citep[][and references therein]{Bucher/Segers:2017,Dombry/Ferreira:2019}, but such studies consider infinite sample size, while small-sample biases can arise even when the assumed model is correct. 

\begin{figure}
 \centering 
 \includegraphics[width = \textwidth]{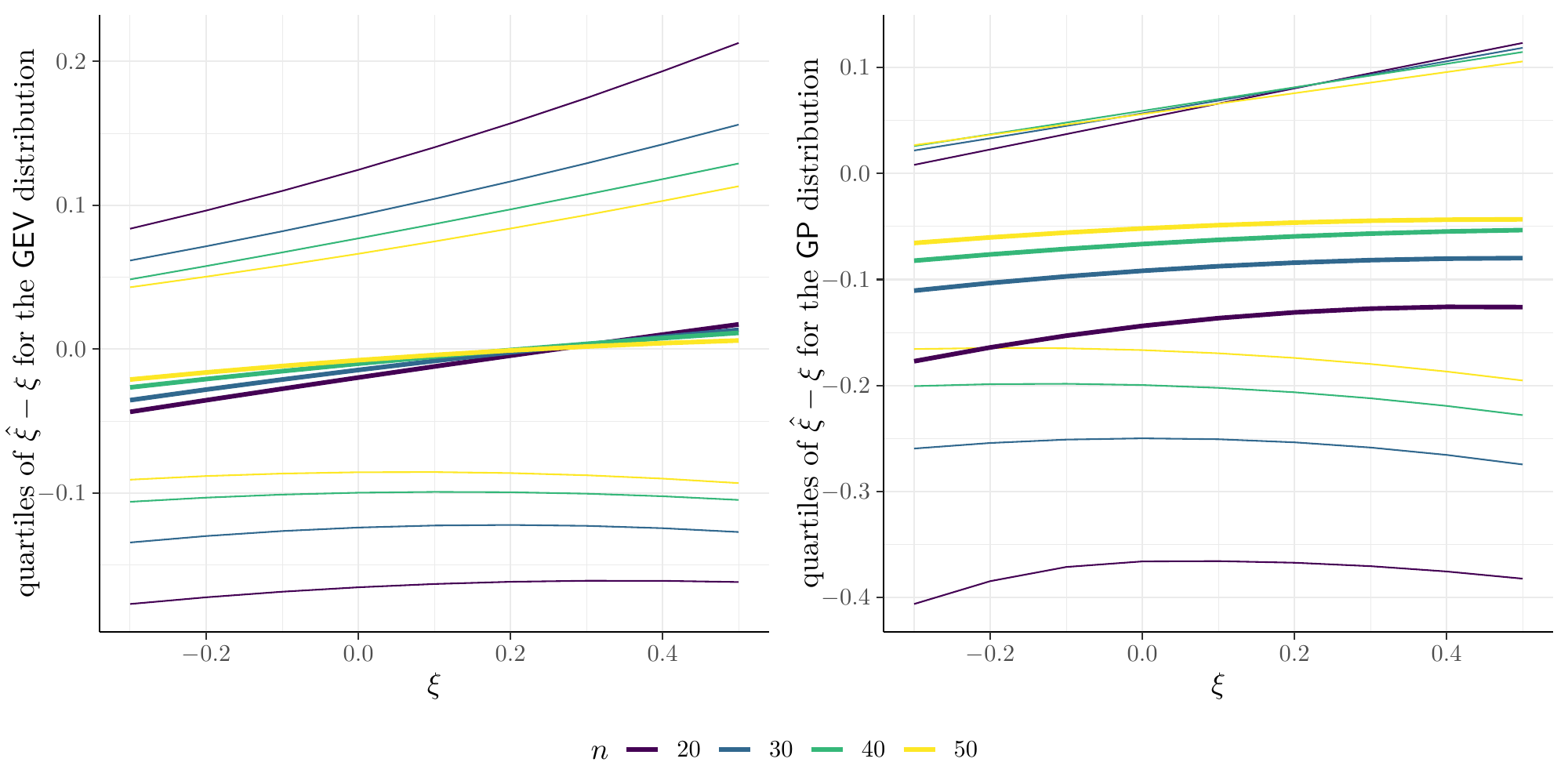}
 \caption{Smoothed quartiles of the distribution of differences between maximum likelihood estimates and true shape parameter, $\widehat{\xi}-\xi$, based on 13000 simulations from $\mathdis{GEV}(0,1,\xi)$ and $\mathdis{GP}(1,\xi)$ distributions for samples of sizes $n=20,30,40,50$.}
 \label{fig:biasEVD}
 \end{figure}

The finite-sample properties of maximum likelihood estimators for extreme value distributions can be poor \citep[e.g.,][Table~5]{Hosking:1987} due to their small-sample bias \citep{Giles:2016,Roodman:2018}. \Cref{fig:biasEVD} illustrates this for the shape parameter $\xi$. Apart from becoming more concentrated as the sample size $n$ increases, the distribution of $\widehat{\xi}-\xi$ when fitting the $\mathdis{GEV}$ distribution to maxima depends little on $n$, and in particular its median barely changes.  By contrast, the distribution of $\widehat{\xi}-\xi$ based on the $\mathdis{GP}$ distribution shows strong negative skewness, with its median and lower quartile systematically increasing as $n$ increases, while the upper quartile barely changes. It turns out that the scale estimator $\hat\sigma$ for the $\mathdis{GP}$ distribution is upwardly biased, and this partially compensates for the downward bias of $\widehat{\xi}$, but extrapolation too far into the tail based on a $\mathdis{GP}$ fit tends to underestimate the sizes of extreme events anyway. Bias-correction can mitigate this, but analytical first-order corrections of the type pioneered by \citet{Cox:1968} are applicable only when $\xi < -1/3$ and are unbounded near this. Analytical bias correction is quite different from bootstrap bias correction \citep{Belzile:thesis} and furthermore bias-correction formulae for risk measures are currently unavailable. We thus consider implicit bias corrections below.

\section{Inference}

\subsection{Likelihood}
\label{subsec:lik}

Consider a parametric model for observations $y_1,\ldots, y_n$ with log-likelihood function $\ell(\bs{\theta})$ whose $p$-dimensional 
parameter vector $\bs{\theta}=(\bs{\psi}, \bs{\lambda})$ can be decomposed into a $q$-dimensional parameter of interest $ \bs{\psi} $ and a $(p-q)$-dimensional nuisance vector $\bs{\lambda}$.
The score vector $U(\bs{\theta})$, the observed information matrix $j(\bs{\theta})$ and its inverse are partitioned accordingly and the maximum likelihood estimate is $\widehat{\bs{\theta}}=(\widehat{\bs{\psi}}, \widehat{\bs{\lambda}})$.
The profile log-likelihood for the parameter of interest is 
\begin{align*}
 \ell_\rp(\bs{\psi})=\max_{\bs{\lambda}}\ell(\bs{\psi}, \bs{\lambda})=\ell(\widehat{\bs{\theta}}_{\bs{\psi}})=\ell(\bs{\psi}, \widehat{\bs{\lambda}}_{\bs{\psi}}).
\end{align*}
The asymptotic properties of this and related statistics stem from those of the full likelihood and are standard under mild conditions \citep[cf.][p.~128]{Severini:1999}: for example, if $\psi$ is scalar with true value $\psi_0$, 
the likelihood root
\begin{align}
R(\psi_0) = \sign(\widehat{\psi}-\psi_0)[2\{\ell_{\rp}(\widehat{\psi}) - \ell_{\rp}(\psi_0)\}]^{1/2} \label{eq:likroot} \end{align}
 has an asymptotic standard normal distribution to order $\bigO(n^{-1/2})$. Confidence limits $\psi_\alpha$ for $\psi$ are obtained by solving the equations $R(\psi_\alpha) = \Phi^{-1}(\alpha)$ for $\alpha\in(0,1)$, with $\Phi$ denoting the standard normal cumulative distribution function.  {Such intervals are invariant to interest-preserving reparametrisations.}   The same two-sided equi-tailed confidence intervals are obtained via $\chi^2$ approximation to the distribution of the likelihood ratio statistic based on $ \ell_\rp(\psi)$, but these and other first-order methods may perform badly in small samples if the dimension of $\bs{\lambda}$ is large, and higher-order methods may then provide more accurate tests and confidence intervals. 
 
 The standard asymptotic approximations described above apply under regularity conditions whereby the score statistic satisfies the first two Bartlett equalities. The $r$th moment of the score for the $\mathdis{GEV}$  and $\mathdis{GP}$  distributions exists only for $\xi>-1/r$ \citep{Smith:1985}, so  the above discussion applies to them only for $\xi>-1/2$.  Estimates of $\xi$ are rarely less than $-0.3$ in applications, and very often are close to zero, so the failure of regularity conditions is rarely of practical importance.  

In the case of scalar $\psi$, one improvement is via normal approximation to a modified likelihood root \citep[][\S 6.6.1]{Barndorff-Nielsen/Cox:1994}
 \begin{align}
 R^\star(\psi) = R(\psi)+ \frac{1}{R(\psi)} \log\left\{\frac{Q(\psi)}{R(\psi)}\right\},
 \label{eqmodiflikroot}
 \end{align}
 where $Q(\psi)$ is discussed below. If the response distribution is continuous, then $R^\star(\psi_0)$ is asymptotically standard normal to order $\bigO(n^{-3/2})$; this is known as a third-order approximation. In many ways more important than the reduction of the error rate from $n^{-1/2}$ to $n^{-3/2}$ is the fact that the error when using $R^\star(\psi)$ is relative, leading to improved inferences even when $\widehat{\psi}$ is distant from $\psi_0$. Confidence limits are  obtained by solving the equations $R^\star(\psi_\alpha)=\Phi^{-1}(\alpha)$, {and these too are invariant to interest-preserving reparametrization to the given order}.
 
Estimators of $\psi$ can be obtained by solving the equations $R(\psi)=0$ and $R^\star(\psi)=0$. The first yields the maximum likelihood estimator $\widehat{\psi}$, while the second, $\widehat{\psi}^{\star}$, yields an implicitly debiased version of $\widehat{\psi}$; {both transform appropriately}. The maximum likelihood estimator could also be debiased directly by subtracting an estimated bias, or indirectly by modifying the corresponding score equation \citep{Firth:1993,KennePagui.Salvan.Sartori:2017,Belzile:thesis}.
 
 The use of~\cref{eqmodiflikroot} hinges on the ready computation of $Q(\psi)$. This involves sample space derivatives of the log-likelihood, which are awkward in general, and a variety of approaches to their computation have been proposed. In the next section we sketch a simple general approach developed by D.~A.~S.~Fraser, N.~Reid and colleagues.
 
%

\subsection{Tangent exponential model}
\label{subsec:tem}

The tangent exponential model  (\textsc{tem}) provides a general formula for the quantity $Q(\psi)$ that appears in \cref{eqmodiflikroot} \citep{Fraser/Reid/Wu:1999}. The idea is to approximate the probability density function of the data by that of an exponential family, for which highly accurate inference is possible. 
Following \citet[Chapter 8]{Brazzale/Davison/Reid:2007}, we outline its construction. The presence of sample space derivatives makes it necessary to distinguish a generic response vector $\bs{y}=(y_1,\ldots, y_n)^\top$ from the responses actually observed, $\bs{y}^{\textrm{o}} = (y^{\textrm{o}}_1,\ldots, y^{\textrm{o}}_n)^\top$. 
 
The tangent exponential model depends on an $n \times p$ matrix $\mathbf{V}$, whose $i$th row 
equals the derivative of $y_i$ with respect to $\bs{\theta}^\top$, evaluated at $\widehat{\bs{\theta}}$ and $\bs{y}^{\textrm{o}}$; the $p$ columns of $\mathbf{V}$ correspond to vectors in $\mathbb{R}^n$ that are informative about the variation of $\bs{y}$ with $\bs{\theta}$. The \textsc{tem} implicitly conditions on an $(n-p)$-dimensional approximate ancillary statistic whose value lies in the space orthogonal to the columns of $\mathbf{V}$, and constructs a local exponential family approximation at $\widehat{\bs{\theta}}$ and $\bs{y}^{\textrm{o}}$ with canonical parameter
\begin{align*}
 \bs{\varphi}(\bs{\theta}) 
 = \mathbf{V}^\top\left.\frac{\partial \ell(\bs{\theta}; 
\bs{y})}{\partial \bs{y}} 
\right|_{\bs{y}=\bs{y}^{\textrm{o}}}.
\end{align*}
The components of  $\bs{\varphi}(\bs{\theta})$ can be interpreted as the directional derivatives of $\ell(\bs{\theta}; \bs{y}^{\textrm{o}}+\mathbf{V}\bs{t})$ with respect to the columns of $\mathbf{V}$, obtained by differentiating with respect to the components of the $p\times 1$ vector $\bs{t}$ and setting $\bs{t}=\bs{0}$.

In models for continuous scalar responses the $\mathbf{V}_i$ can be obtained by using the probability integral transform to write $y_i=F^{-1}(u_i;\bs{\theta})$ in terms of a uniform variable $u_i$, yielding 
\begin{align}
\mathbf{V}_i=\left.\frac{\partial y_i}{\partial\bs{\theta}^\top}\right|_{\bs{y}=\bs{y}^{\textrm{o}},\bs{\theta} = \widehat{\bs{\theta}}} = \left.-\frac{\partial F\big(y_i^{\textrm{o}}; \bs{\theta}\big)}{\partial \bs{\theta}^\top}\frac{1}{f\big(y_i^{\textrm{o}}; \bs{\theta}\big)}\right|_{\bs{\theta}=\widehat{\bs{\theta}}}, \quad i=1,\ldots, n;
 \label{eq:Vmat}
\end{align}
equivalently we may take the total derivative of the pivotal quantity $F(y_i;\bs{\theta})$.   Discrete responses cannot be differentiated and are replaced by their means, leading to an error of order $n^{-1}$ for inferences based on~\cref{eqmodiflikroot} \citep{Davison.Fraser.Reid:2006}. 

The approximate pivot in \cref{eqmodiflikroot} stemming from the \textsc{tem} is 
\begin{align}
 Q({\psi})=\frac{\left|\begin{matrix} \bs{\varphi}(\widehat{\bs{\theta}})-\bs{\varphi}(\widehat{\bs{\theta}}_{\psi}) &\quad \partial \bs{\varphi}/ 
\partial \bs{\lambda}^\top(\widehat{\bs{\theta}}_{\psi})\end{matrix}\right|}{\left|\partial \bs{\varphi}/ \partial 
\bs{\theta}^\top(\widehat{\bs{\theta}})\right|} 
\frac{\left|j(\widehat{\bs{\theta}})\right|^{1/2}}{\left|j_{\bs{\lambda}\bs{\lambda}}(\widehat{\bs{\theta}}_{{\psi}})\right|^{1/2}
}, \label{eq:qtem}
\end{align}
where the first matrix in the numerator is formed by binding the $p \times 1$ vector $\bs{\varphi}(\widehat{\bs{\theta}})-\bs{\varphi}(\widehat{\bs{\theta}}_{\psi})$ to the $p \times (p-1)$ matrix $\partial \bs{\varphi}/ 
\partial \bs{\lambda}^\top$.
%
A modified profile likelihood $\ell_{\mathsf{fr}}({\psi}) \propto -\{R^\star(\psi)\}^2/2$ may be constructed by using \cref{eq:qtem} and treating $R^\star(\psi)$ as standard normal.

\subsection{Modified profile likelihoods}

In the previous section, the likelihood root was derived from the profile log-likelihood function via the likelihood ratio statistic, then modified and used to construct the modified profile likelihood $\ell_{\mathsf{fr}}({\psi})$. An alternative is direct modification of the profile log-likelihood, two approaches to which are listed by \citet[][\S~9.5.3--9.5.4]{Severini:2000}.
The first approach uses elements of the tangent exponential model and is of the form 
\begin{align}
 \ell_{\mathsf{m}}^{\mathrm{tem}}(\bs{\psi}) = \ell_{\rp}(\bs{\psi}) + \frac{1}{2}\log 
\left\{\left|j_{\bs{\lambda\lambda}}(\widehat{\bs{\theta}}_{\bs{\psi}})\right|\right\} - 
\log\left\{\left|\ell_{\bs{\lambda};\bs{y}}(\widehat{\bs{\theta}}_{
\bs { \psi } } )\mathbf{V}_{\bs{\lambda}}(\widehat{\bs{\theta}})\right|\right\}, \label{eq:severtem}
\end{align}
where $\ell_{\bs{\lambda};\bs{y}} = \partial^2 \ell/ \partial \bs{\lambda} \partial \bs{y}^\top$ is the derivative of the log-likelihood with respect to the nuisance parameter and observations and $\mathbf{V}_{\bs{\lambda}}$ denotes the columns of $\mathbf{V}$ in \cref{eq:Vmat} corresponding to derivatives with respect to $\bs{\lambda}$. 
The second approach, due to Severini and similar in spirit to ideas in \cite{Skovgaard:1996}, uses empirical covariances, yielding 
\begin{align}
 \ell_{\mathsf{m}}^{\mathrm{cov}}(\bs{\psi}) &= \ell_{\rp}(\bs{\psi}) + \frac{1}{2} \log \left\{\left| 
j_{\bs{\lambda\lambda}}(\widehat{\bs{\theta}}_{\bs{\psi}})\right|\right\} - \log \left\{\left|\;\widehat{j}_{\bs{\lambda}; 
\bs{\lambda}}(\widehat{\bs{\theta}}_{\bs{\psi}}; \widehat{\bs{\theta}})\;\right|\right\}, \label{eq:severcov}
\end{align}
where
\begin{align*}
\widehat{j}_{\bs{\lambda}; 
\bs{\lambda}}(\widehat{\bs{\theta}}_{\bs{\psi}}; \widehat{\bs{\theta}}) &= \sum_{i=1}^n \ell_{\bs{\lambda}}^{(i)}(\widehat{\bs{\theta}}_{\bs{\psi}})
\ell_{\bs{\lambda}}^{(i)}(\widehat{\bs{\theta}})^\top;
\end{align*}
here $\ell_{\bs{\lambda}}^{(i)}$ denotes the component of the score statistic due to the $i$th observation.  It is straightforward to check that these are invariant to interest-preserving reparametrisation. 
\subsection{Example}
 We illustrate the derivation of the computations underlying Figure~\ref{fig1}, in which we fit a generalized Pareto distribution \eqref{eq:gpdist} to $n_u$ threshold exceedances above $u$ to make inference for the median of the semicentennial maximum. Approximating the distribution of the latter by $H^{\zeta_u TN_y}$, we obtain 
 \begin{align*}
\mathfrak{q}_p \approx \kappa_p = u + \frac{\tau}{\xi} \left[\left\{1 - p^{1/(\zeta_uTN_y)}\right\}^{-\xi} - 1\right], \qquad p = 0.5.
\end{align*}
For simplicity, we treat $\zeta_u$ as fixed and reparametrize the model in terms of $(\kappa_p, \xi)$. The sufficient directions are $\mathrm{V}_{i,\kappa_p} = y_i^{\textrm{o}}/(\widehat{\kappa}_p-u)$ and 
\begin{align*}
\mathrm{V}_{i,\xi} = \tau(\widehat{\kappa}_p, \widehat{\xi}) \widehat{r}_i {\left(\frac{\log \widehat{r}_i}{\widehat{\xi}^{2}} + \frac{y_i \log q}{\tau(\widehat{\kappa}_p, \widehat{\xi}) {\widehat{r}_i} {\left(1 - {q}^{\widehat{\xi}}\right)}}\right)}, 
\intertext{where we used the short-hand notation}
r_i =1+\frac{\xi}{\tau(\kappa_p, \xi)}y_i^{\mathrm{o}}, \qquad q = 1-p^{1/(TN_y \zeta_u)};
\end{align*}
with $\widehat{r}_i$ denoting $r_i$ evaluated at the maximum likelihood estimate ($\widehat{\kappa}_p, \widehat{\xi}$). 
The canonical parameter is
\begin{align*}
\bs{\varphi}_i(\kappa_p, \xi) &=  -\sum_{i=1}^{n_u}\mathbf{V}_i \times 
\frac{{\left(1+{\xi}\right)}}{\tau(\kappa_p, \xi) {r_i}},
\intertext{while the mixed partial derivatives appearing in \cref{eq:qtem} are}
\bs{\varphi}_{i, \kappa_p} &= \mathbf{V}_i \times 
\frac{\xi \left(1+\xi\right)}{\tau(\kappa_p, \xi)^{2} {r_i} \left(q^{-\xi} - 1\right)}\left(\frac{1-\xi y_i^{\textrm{o}}}{\tau(\kappa_p, \xi) r_i}\right), \\
\bs{\varphi}_{i, \xi} &= \mathbf{V}_i \times \left\{
\frac{(1+\xi) \log q}{ \tau(\kappa_p, \xi) {r_i} \left(1-{q}^{\xi}\right)}\left(1 - \frac{\xi y_i^{\textrm{o}}}{\tau(\kappa_p, \xi) r_i}\right) + \frac{1}{\tau(\kappa_p, \xi) \xi{r_i} }\right\}.
\end{align*}
The components of the score vectors and information matrices are readily obtained, if necessary using computer algebra, though care is needed to interpolate them for the $\mathsf{GEV}$ at $\xi=0$, since they can be numerically unstable.

\section{Data analyses}

\subsection{Vargas tragedy} \label{subsec_vargas}
To deepen our analysis of the daily rainfall totals discussed in Section~\ref{subsec_vargas1}, we consider the impact of stopping the data collection after the Maiquetía disaster and include the yearly maxima  
for 1951--1960. Failing to account for the fact that the last observation corresponds to the largest event ever observed leads to upwardly biased risk estimates \citep{Barlow:2020}, so to formalise the stopping rule we consider that sampling would have ended at the first time a daily value exceeded $s=282$mm, the largest two-day sum previously reported.

We consider a Poisson process $\mathcal{P}$ with measure $\nu$ on $\mathcal{X}=(0,\infty)\times(u,\infty)$, where the first axis represents the times $t$ of extreme events and the second axis represents their sizes $y$, which are presumed to exceed some threshold $u$.  If the events $x=(t,y)$ arrive at a constant unit rate in time, then 
\begin{equation}
\label{mu.eq}
\nu\{[t_1,t_2]\times (y,\infty)\} = (t_2-t_1)\Lambda(y), 
\end{equation}
where $\Lambda(y) = \{1+\xi(y-\mu)/\tau\}_+^{-1/\xi}$, and the intensity of the process is $\dot{\nu}(x) = -\dot\Lambda(y)=-\partial\Lambda(y)/\partial y$, say. 
Our stopping rule presupposes that for some given $s>u$, $\mathcal{X}$ is partitioned into a stopping set $\mathcal{S}=(0,\infty)\times (s,\infty) $ and its complement $\mathcal{S}^{\mathsf{c}}$, and the data are analysed just after the random time $T$ at which $\mathcal{P}$ first falls into $\mathcal{S}$.  The probability of no event in $\mathcal{S}$ before time $t$ is $\exp\{-t\Lambda(s)\}$, so  $T$ is an exponential random variable with mean $\Lambda(s)^{-1}$.  Let $N_t$ denote the number of events in $\mathcal{S}^{\mathsf{c}}$ before time $t$.  Since $\mathcal{S}$ and $\mathcal{S}^{\mathsf{c}}$ are disjoint, events in them are independent, so $T$ is independent of events in $\mathcal{S}^{\mathsf{c}}$, and the probability element corresponding to successive events $(t_1,y_1),\ldots, (t_n,y_n)$ in $\mathcal{S}^{\mathsf{c}}$ followed by $(t,y_t)$ in $\mathcal{S}$ is
\begin{equation}\label{pp-lik.eq}
\dot{\nu}(t, y_t)\prod_{j=1}^n \dot{\nu}(t_j,y_j) \times \exp\left[ -\nu\{[0,t]\times [u,\infty)\}\right],\quad u<y_1,\ldots,y_n<s< y_t. 
\end{equation}
As the stopping rule enters this expression only through the constraints on the data values, it affects the repeated sampling properties of estimators but not inferences based directly on the likelihood.  
It is helpful to rewrite~\eqref{pp-lik.eq} as
\begin{equation}
\label{Lstd.eq}
\exp\left[-t\{\Lambda(u)-\Lambda(s)\}\right]\prod_{j=1}^n \left\{ - \dot\Lambda(y_j)\right\} \times \exp\{-t\Lambda(s)\}\left\{ - \dot\Lambda(y_t)\right\},
\end{equation}
where the first term corresponds to the $n$ events in $\mathcal{S}^{\mathsf{c}}$ before time $t$, and the second term to the event in $\mathcal{S}$ that terminates the sampling. In the application ot the Vargas data we take $t=39$ years, so an annual maximum has distribution $\exp\{-\Lambda(y)\}$. The probability that sampling stops at time $t$ and there are then $n$ events in $\mathcal{S}^{\mathsf{c}}$ is
\begin{align*}
\mathsf{P}(N_{t} = n, T=t) = \frac{[t\{\Lambda(u)-\Lambda(s)\}]^n}{n!} \exp\left[-t\{\Lambda(u)-\Lambda(s)\}\right]\times \exp\left\{-t\Lambda(s)\right\}\times \Lambda(s).
\end{align*}

Expression~\eqref{Lstd.eq} corresponds to what \cite{Barlow:2020} call $\mathscr{L}_{\rm std}$, and which they find gives biased inferences, 
and the equivalent of their `full conditional' likelihood, which uses the joint density of the event sizes $y_1,\ldots, y_n,y_t$ conditional on $T=t$ and $N_{t}=n$, i.e., 
\begin{equation}
\label{Lfc.eq}
\mathscr{L}_{\rm fc} = \left\{ \frac{- \dot\Lambda(y_t)}{\Lambda(s)}\right\}\prod_{j=1}^n \left[ \frac{- \dot\Lambda(y_j)}{t\{\Lambda(u)-\Lambda(s)\}}\right], \quad u<y_1,\ldots,y_n<s< y_t,
\end{equation}
should be preferable.   If $y>u$ then $-\dot\Lambda(y)/\Lambda(u)$ equals a generalised Pareto density with parameters $\xi$ and $\tau_u= \sigma+\xi(u-\mu)$ evaluated at $y-u$, so~\eqref{Lfc.eq} reduces to a product of truncated generalised Pareto densities.   \citet{Barlow:2020}'s `partial conditional' likelihood $\mathscr{L}_{\rm pc}$ corresponds to replacing the terms $\Lambda(u)-\Lambda(s)$ in~\eqref{Lfc.eq} by $\Lambda(u)$, i.e., ignoring the right-truncation of  $y_1,\ldots, y_n$.  


Our functional of interest, the median of the semicentennial daily maximum, is given by taking $T=50$, $p=0.5$ and
\begin{align*}
\mathfrak{q}_{p} = \mu- \frac{\sigma}{\xi} \left\{1-\left(\frac{-T}{\log p}\right)^\xi\right\}. 
\end{align*}
The log likelihood has three components: the yearly maxima for 1951--1960, the right-truncated exceedances of $u=27$mm for 1961--1999 and finally a left-truncated largest record that exceeds the stopping rule threshold $s=282$mm.  As these are independent, they make additive contributions to $ \bs{\varphi}(\bs{\theta}) $.  The component for the maxima is readily obtained using their assumed generalized extreme-value distribution, and so are those for the exceedances in the full and partial conditional likelihoods,  which correspond to independent generalized Pareto variables, possibly truncated.  The form of $ \bs{\varphi}(\bs{\theta}) $ for the tangent exponential model approximation for a Poisson process, which is needed for the higher-order version of~\eqref{Lstd.eq},  seems not to have been derived previously and may be found in \Cref{sec:temPP}. 

\begin{figure}[t!] 
\centering 
\includegraphics[width=\textwidth]{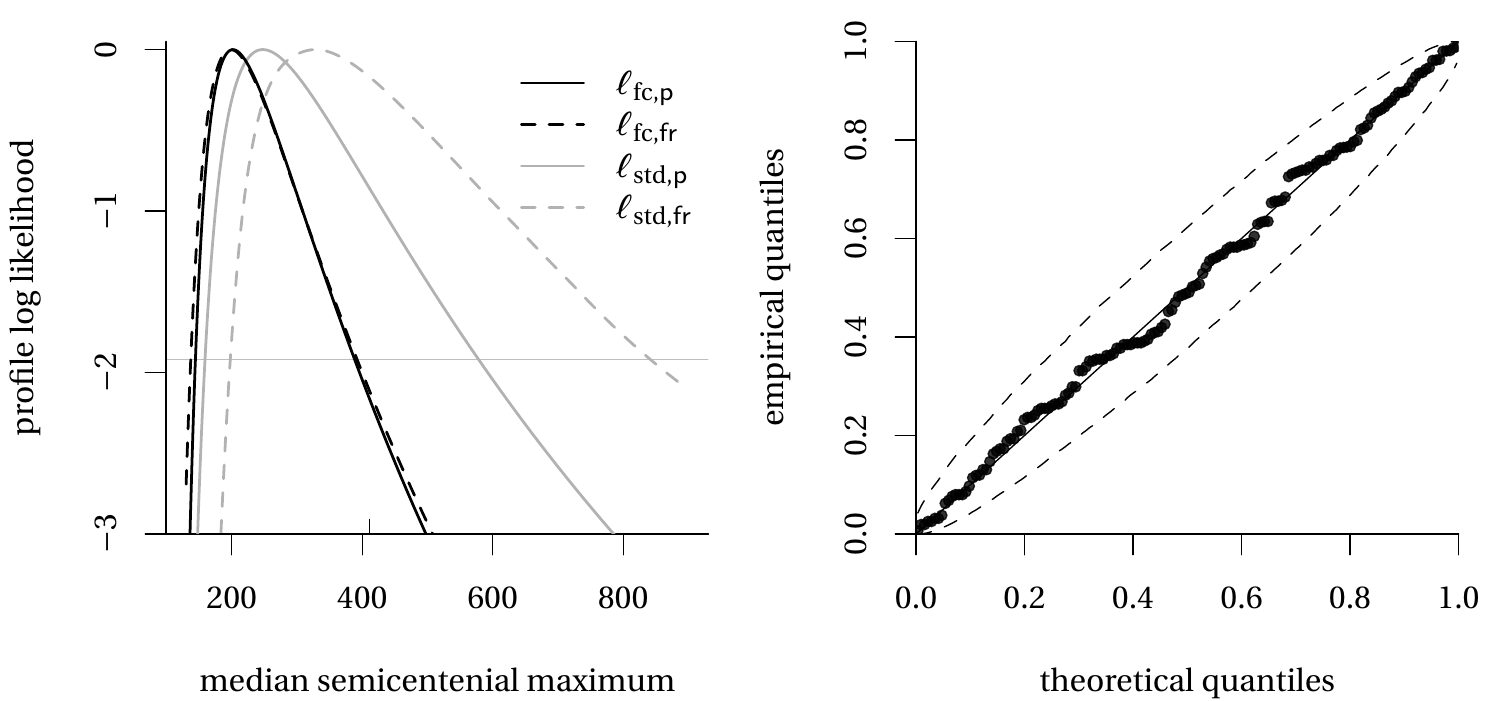}
\caption{Left panel: profile log likelihood based on $\mathscr{L}_{\text{std}}$ (gray) and $\mathscr{L}_{\text{fc}}$ (black) for the Maiquetía data using threshold exceedances up to and including data for December 15th, 1999. The curves show the shifted regular profile likelihood (full) and the \textsc{tem} approximation $\ell_{\mathsf{fr}} = -R^{\star2}/2$ (dashed).  The dashed grey horizontal line at $-1.92$ indicates cutoff values for 95\% confidence intervals based on the asymptotic $\chi^2_1$ distribution. The mark at 410.4mm indicates the record of December 15th, 1999. Right panel: probability-probability plot for the full conditional likelihood fit, with approximate simultaneous 95\% confidence intervals.}
\label{maiquetia_pot}
\end{figure}

The ordinary and \textsc{tem}-based profile likelihoods for $\mathfrak{q}_p$ with full conditioning and the standard likelihood, shown in the left panel of \Cref{maiquetia_pot}, are strikingly different: the conditional likelihood is much more concentrated and the higher-order \textsc{tem} version is just slightly less precise, whereas the \textsc{tem} version of the standard likelihood~(\ref{Lstd.eq}) gives much larger point estimates and upper confidence interval limit than does the standard profile likelihood.  Thus not allowing for the implicit stopping rule can have a dramatic effect not only on standard but also on higher-order inferences. If we consider the one-sided likelihood ratio test for $\mathscr{H}_0: \mathfrak{q}_p > 410.4$mm, the respective $p$-values obtained from  $R/R^\star$ are $0.095/0.28$ for the standard likelihood and $0.019/0.02$ for the full conditional likelihood, suggesting that the magnitude of the 1999 event was indeed significantly larger than the median 50-year maximum.

\subsection{Venice sea level}

\begin{figure}[t] 
\centering 
\includegraphics[width=\textwidth]{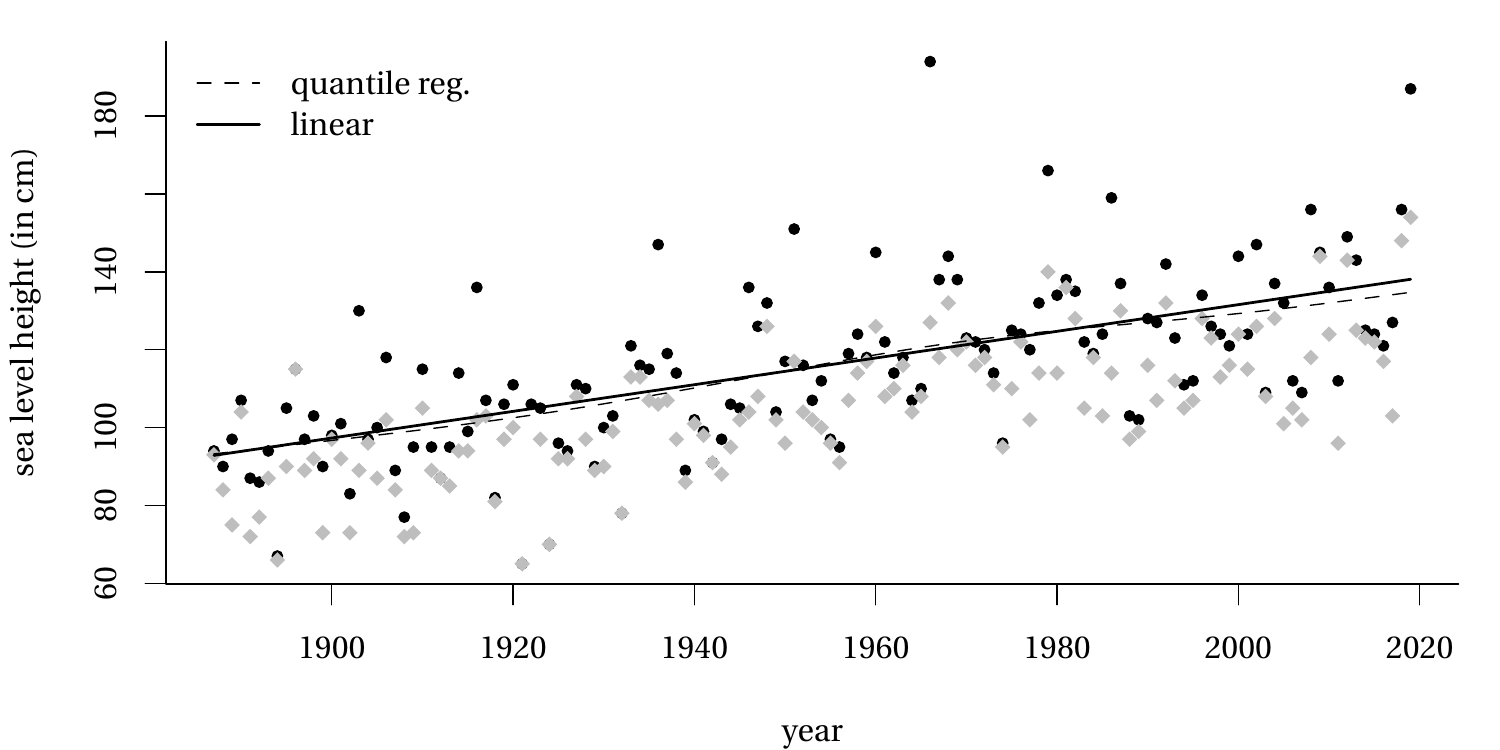}
\caption{First (black) and second (grey) largest yearly observations for the Venice sea level data (in cm),  a smooth additive quantile regression model for the median (with a smooth term for time and a different intercept for each order statistic). The predicted values for the largest order statistic and those for the corresponding linear regression are superimposed.}
\label{fig:venicedata}
\end{figure}

The Italian city of Venice is threatened by sea-level rise and subsidence, and is increasingly at risk from flooding in so-called \emph{acqua alta}\ events. To quantify this risk we consider data analyzed by \cite{Smith:1986} and \cite{Pirazzoli:1982} containing large annual sea level measurements from 1887 until 1981, complemented with series for 1982--2019 extracted from the \href{https://www.comune.venezia.it/it/content/archivio-storico-livello-marea-venezia-1}{City of Venice website} (accessed June 2020 and available under the CC BY-NC-SA 3.0 license). Only the yearly maximum is available for 1922 and only the six largest observations for 1936. \Cref{fig:venicedata} shows the two largest annual order statistics; while there is a clear trend, we detected no change when the measurement gauge was relocated in 1983. In addition to the simple straight-line model suggested by the plot we fitted a smooth additive nonparametric quantile regression \citep{Fasilio:2020} with 50 knots, and a smooth term for years and different intercepts for the two largest order statistics: the resulting fits, shown in \Cref{fig:venicedata},  suggest that a straight line is adequate.  

If the extremal types theorem holds, then the log-likelihood corresponding to the joint limiting distribution of the $r$ largest observations of a sample, $Y_{1} \geq \cdots \geq Y_{r}$, is 
\begin{multline}
\ell(\mu,\sigma,\xi; \bs{y}) = -r\log(\sigma) - \left(1+\frac{1}{\xi}\right)\sum_{j=1}^r \log\left(1 + \xi\frac{y_{j}-\mu}{\sigma}\right)_{+} \\ - \left(1 + \xi\frac{y_{r}-\mu}{\sigma}\right)^{-1/\xi}_+, \quad \mu,\xi\in\mathbb{R}, \sigma>0. \label{eq:rlarglik}
\end{multline}
The ten largest sea levels are available for almost each year, but one might ask whether they should be used.  The model presupposes that they arise from independent underlying variables, but in practice many are due to combinations of high tides and bad weather during the winter months.  The data source for recent years allows apparently independent events to be identified, but this is harder for the earlier data.  

\begin{figure}[t!] 
\centering 
\includegraphics[width=\textwidth]{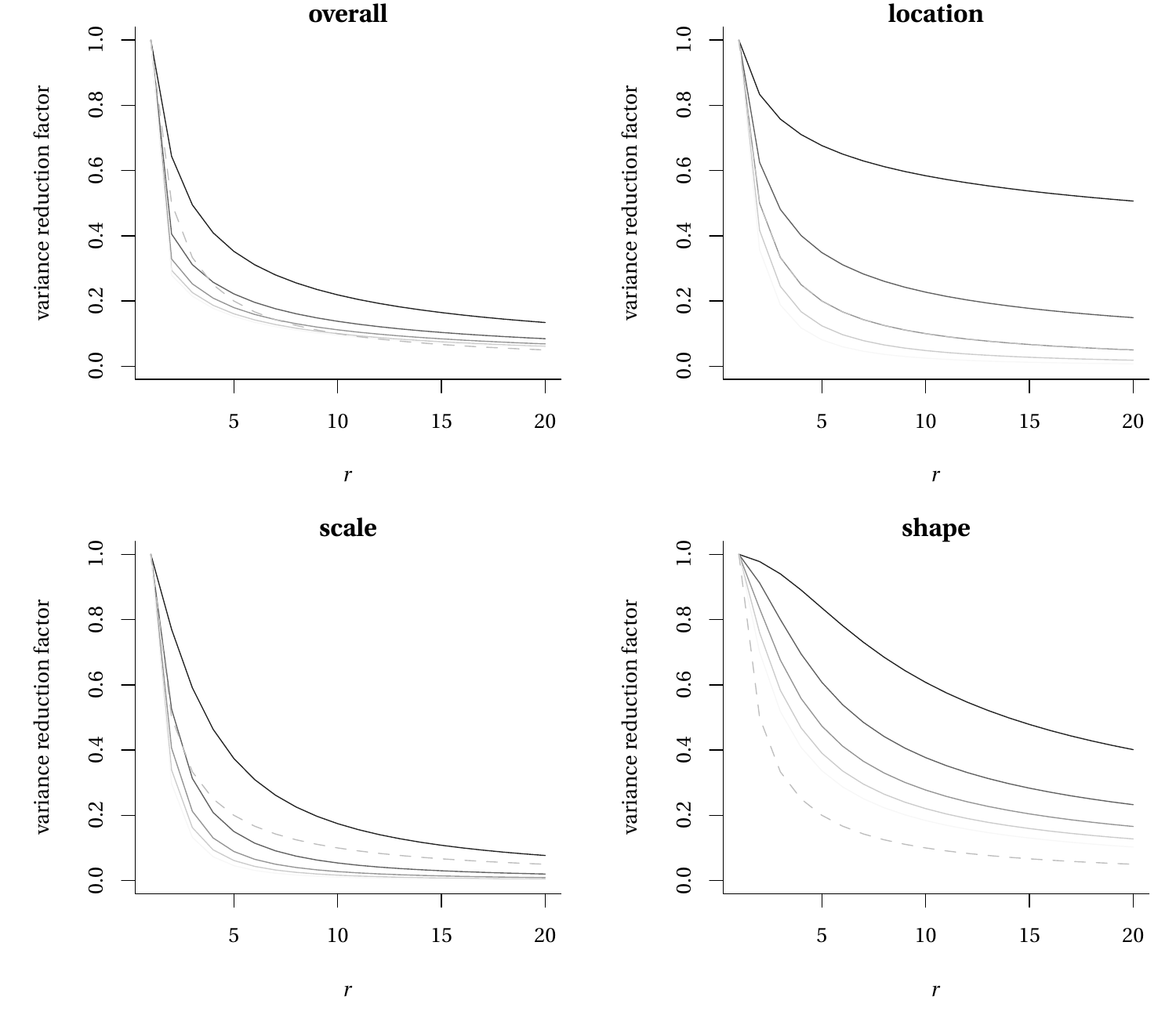}
\caption{Variance reduction factors for inference based on the $r$-largest order statistic for the location, scale, shape parameters and overall efficiency (clockwise from top right).  The dashed grey line shows the ideal efficiency gain for independent observations. The value of the shape parameter ranges from $\xi=-0.4$ (full black) to $\xi=0.4$ (full pale grey) in increments of $0.2$.}
\label{rlarginfo}
\end{figure}

One purely statistical basis for choosing $r$ is by balancing the information added as $r$ increases against the potential for bias when $r$ is too large.  Calculations in \Cref{section:infomatrlargest} establish that the $3\times 3$ Fisher information matrix based on~\eqref{eq:rlarglik} is of the form $I_r(\mu,\sigma,\xi)+(r-1)I(\mu,\sigma,\xi)$, where $I_r(\mu,\sigma,\xi)$ stems from $Y_{r}$, and $(r-1)I(\mu,\sigma,\xi)$ is the contribution for the other observations.  These matrices can be used to compute the information gain due to basing inference on $Y_{1},\ldots, Y_{r}$ rather than only on the sample maximum, $Y_{1}$.  To do so, we calculate the ratios of the diagonal elements of $I_1^{-1}(\mu,\sigma,\xi)$ to those of $\{I_r(\mu,\sigma,\xi)+(r-1)I(\mu,\sigma,\xi)\}^{-1}$; an overall variance reduction for a given $r$ is 
\[
\left\{\frac{|I_1(\mu,\sigma,\xi)|}{|I_r(\mu,\sigma,\xi)+(r-1)I(\mu,\sigma,\xi)|}\right\}^{1/3}.
\]
\Cref{rlarginfo} shows the variance reduction factors for $\mu$, $\sigma$, $\xi$ and the overall efficiency. There seems to be little gain from taking $r>5$ for estimation of $\mu$ and $\sigma$, while for $\xi$ the decline is closer to that of independent generalized extreme value data. This is because the parameters $\mu$ and $\sigma$ cannot be estimated based only on $I(\mu,\sigma,\xi)$, which has rank two, whereas both $I(\mu,\sigma,\xi)$ and $I_r(\mu,\sigma,\xi)$ contain information on $\xi$.  Hence as $r$ increases the information gain for the location and scale parameters becomes more limited. 

\begin{figure}
 \centering
 \includegraphics[width = 0.99\textwidth]{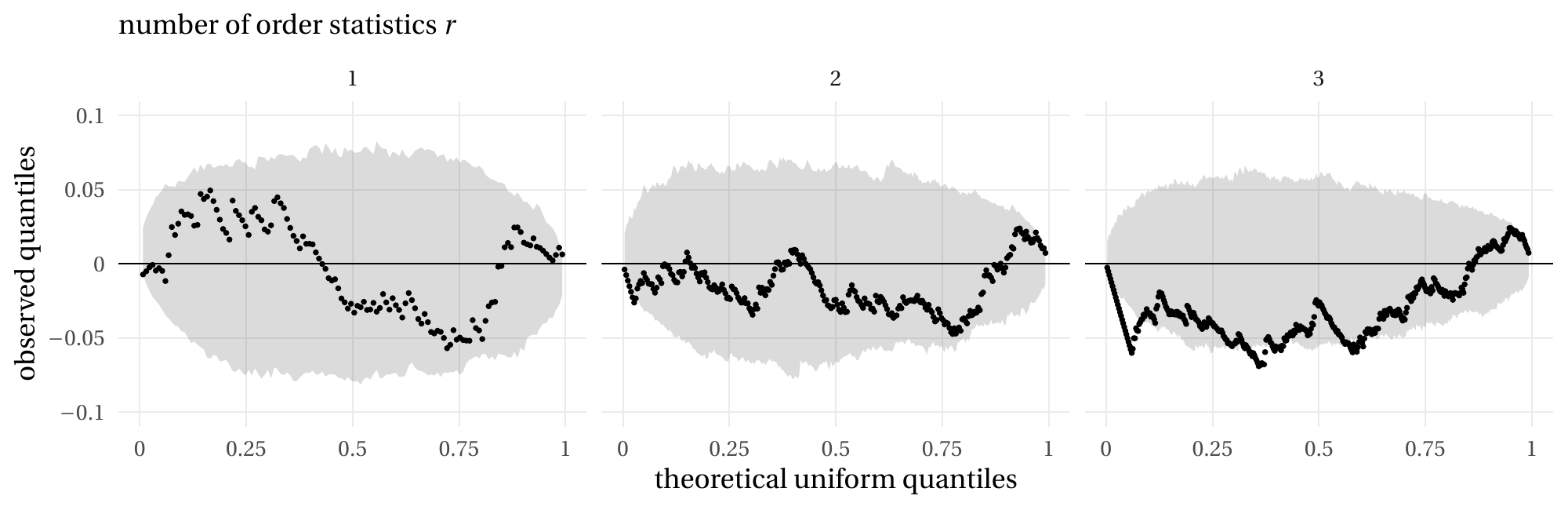}
 \caption{Tukey's detrended probability-probability plot for the spacings of the $r$-largest order statistics of the Venice data, with $r=1,2,3$, with approximate simultaneous 95\% confidence intervals obtain by using a parametric bootstrap and the envelope method \citep[][\S~4.2.4]{Davison/Hinkley:1997}.}\label{venice_ppplot}
\end{figure}

The fit can be checked by noting that if the model is correct, then 
\begin{equation}\label{Poisson.eqn}
0 < \Lambda_{\bs{\theta}}(y_{1})<\Lambda_{\bs{\theta}}(y_{2})<\cdots <\Lambda_{\bs{\theta}}(y_{r}), \quad \Lambda_{\bs{\theta}}(y) = \left\{ 1 + \xi(y-\mu)/\sigma\right\}^{-1/\xi}_+,
\end{equation}
are a realisation of the first $r$ points of a unit rate Poisson process on the positive half-line.  This implies that the spacings $\Lambda_{\bs{\theta}}(y_{(1}),\Lambda_{\bs{\theta}}(y_{2})-\Lambda_{\bs{\theta}}(y_{1}),\ldots$ have standard exponential distributions, and systematic departures from this will indicate model failure.  The $r$ largest observations from the asymptotic model can be generated by simulating a unit rate Poisson process $0<U_1<U_2<\cdots$, where $U_j=E_1+\cdots+E_j$ and $E_j\sim \mathsf{Exp}(1)$, and setting $Y_{j} = \mu + \sigma\big(U_j^{-1/\xi}-1\big)/\xi$. The estimated inverse transformation $\Lambda_{\widehat{\bs{\theta}}}$
can be used to obtain empirical spacings.  These should be approximately independent and can be used to construct probability-probability plots such as 
\Cref{venice_ppplot}.  The spacings for $r=3$ are suggestive of model misspecification for the Venice data, so it seems that just two extrema each year should be used. The linear decay in \Cref{venice_ppplot} is seemingly due to ties or near values for the lower records, as the observations are rounded to the nearest cm. 

Below we use the $r=2$ largest observations for each year and treat data for different years as independent. Our chosen risk measure is the probability that in year $t$ the annual maximum sea level exceeds the level $z=194$cm reached in the catastrophic flooding of 1966, based on a non-stationary extremal model with location parameter $\mu_0 +\mu_1 \texttt{year}$, $\sigma$ and $\xi$. 

In order to compute the terms necessary for the \textsc{tem} approximation, suppose that we have data $(y_1,\ldots, y_r)$ and pivots 
\begin{align*}
u_1(y_1;\bs{\theta}), \quad u_2(y_1,y_2;\bs{\theta}), \quad \ldots, \quad u_r(y_1,\ldots, y_r;\bs{\theta}).                                                                                                           \end{align*}
Total differentiation of $u_1(y_1;\bs{\theta})$ yields
\begin{align*}
0 = \frac{\partial u_1(y_1;\bs{\theta})}{\partial\bs{\theta}} + \frac{\partial y_1}{\partial\bs{\theta}} \frac{\partial u_1(y_1;\bs{\theta})}{\partial y_1},
\end{align*}
and therefore
\begin{align*}
\frac{\partial y_1}{\partial\bs{\theta}} = -\left\{\frac{\partial u_1(y_1;\bs{\theta})}{ \partial y_1}\right\}^{-1} \frac{\partial u_1(y_1;\bs{\theta})}{ \partial\bs{\theta}}.
\end{align*}
 Total differentiation of $u_j(y_1,\ldots, y_j;\bs{\theta})$  likewise yields
\begin{align*}
\frac{\partial y_j}{\partial\bs{\theta}} = -\left\{\frac{\partial u_j(y_1,\ldots, y_j;\bs{\theta})}{ \partial y_j}\right\}^{-1}\left\{
\frac{\partial u_j(y_1,\ldots, y_j;\bs{\theta})}{ \partial \bs{\theta}} + \sum_{i=1}^{j-1} \frac{\partial y_i}{ \partial \bs{\theta}} \frac{\partial u_i(y_1,\ldots, y_i;\bs{\theta})}{ \partial y_i}\right\},
\end{align*}
with all these expressions evaluated at $y_1^{\rm{o}},\ldots, y_j^{\rm{o}}$ and $\widehat{\bs{\theta}}$.
In the present case the differences in~\eqref{Poisson.eqn} are pivots, $u_j(y_1,\ldots, y_j;\bs{\theta})=\Lambda_{\bs{\theta}}(y_j)-\Lambda_{\bs{\theta}}(y_{j-1})$, and the resulting expressions for $\partial y_j/\partial \bs{\theta}$ involve at most two of the $y_i$.

\begin{figure}[t] 
\centering 
\includegraphics[width=\textwidth]{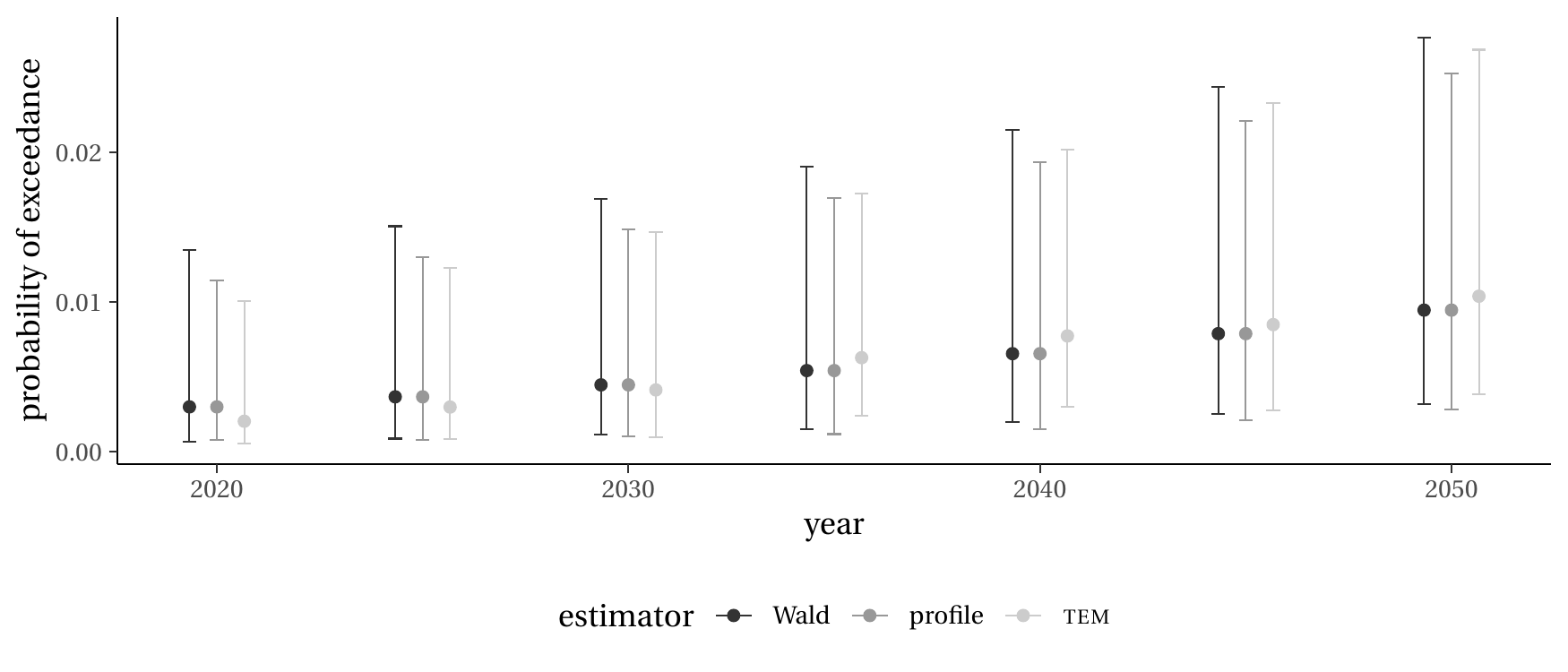}
\caption{Probability of exceedance of 194cm with 95\% pointwise confidence based on Wald (logit-scale), likelihood root $R$ and modified likelihood root $R^\star$ statistics.}
\label{fig:veniceprobexc}
\end{figure}

\Cref{fig:veniceprobexc} shows that the profile- and \textsc{tem}-based point estimates and 95\% confidence intervals for the probability of a flood exceeding the 1966 level for various years are quite similar, though the higher-order estimates vary slightly more over time.  The Wald-based confidence intervals, computed on the logit scale and back-transformed, are somewhat wider.   Despite the increase in sea level, it appears that even without interventions, an event as rare as that in 1966 will remain unlikely for at least the next two decades.  The recent inauguration of the \href{https://www.mosevenezia.eu}{Mose} system of flood barriers, which can be raised in order to prevent Venice from flooding when there are adverse tides in the Adriatic sea, should reduce this probability yet further, at least in the medium term. 


\subsection{Old age in Italy}
The existence or not of a finite upper limit for human lifetimes has recently sparked interest in the extreme value community \citep{Hanayama/Sibuya:2016,Rootzen/Zholud:2017,Einmahl:2019}. The Italian centenarian data set, kindly provided by Holger Rootz\'en, contains the birth dates and ages of 3836 individuals from a study of semi-supercentenarians conducted by the Istituto Nazionale di Statistica (Istat); see \cite{Barbi:2018}. Individuals are included if they were aged 105 years or more at some point between January 1st, 2009 ($c_1$) and January 1st, 2016 ($c_2$); the survival time is censored for individuals alive at $c_2$. The cohort comprises persons born between 1896 and 1910 with excess lifetimes above 105 years measured in days above $u=38351$ days. It is natural to fit the generalized Pareto model to these excess lifetimes, but it is important to account for the potential left-truncation and right-censoring. Failure to account for the censoring would lead to negative bias for the shape parameter $\xi$, for example,  since individuals born after 1910 could not attain 116 years. A negative shape parameter corresponds to a finite upper limit $\iota=-\tau/\xi$, whereas $\xi\geq 0$ means there is no upper limit.

We consider excess lifetime of individuals whose age exceeded $u$ between calendar times $c_1$ and $c_2$: let $S$ and $f$ denote the survival and the density functions of lifetimes, let $x_i$ denote the calendar date at which individual $i$ reached $u$ years, let $t_i$ denote the excess lifetime above $u$ at calendar time $c_2$, and let $a_i$ be an indicator variable taking value $1$ if individual $i$ was alive at calendar time $c_2$ and zero otherwise. 
Then the likelihood is 
\begin{align*}
 L(\bs{\theta}; \bs{t}, \bs{s}) = \prod_{i=1}^n \left[\frac{f(t_i)}{S\{(c_1-x_i)_{+}\}}\right]^{1-a_i} \left[\frac{S(t_i)}{S\{(c_1-x_i)_{+}\}}\right]^{a_i},
\end{align*}
with the first and second terms in the product corresponding to those individuals seen to die and to those whose lifetimes are censored at $c_2$. We fit a generalized Pareto distribution to excess lifetimes over a range of thresholds starting from $105$ years and give the maximum likelihood estimates in \Cref{tab_italcent_gpdparam}. The largest excess lifetime, for Emma Morano, who died aged 117 years in 2017, after $c_2$, is censored, and the estimated shape $\widehat{\xi}_u$ for threshold $u$ is typically close to zero, although its variability is large for high $u$.

\begin{table}[t!]
\centering
\begin{tabular}{lrrrr}
  \toprule
\multicolumn{1}{c}{$u$} & \multicolumn{1}{c}{$n_u$} & \multicolumn{1}{c}{$\widehat{\sigma}$} & \multicolumn{1}{c}{$\widehat{\xi}$} & \multicolumn{1}{c}{$\ell(\widehat{\boldsymbol{\theta}})$} \\ 
  \midrule
105 & 3836 & 1.67 (0.04) & $-0.04\; (0.02)$ & $-4253.7$ \\ 
  106 & 1874 & 1.70 (0.06) & $-0.07\; (0.03)$ & $-2064.3$ \\ 
  107 & 946 & 1.47 (0.08) & $-0.02\; (0.04)$ & $ -999.3$ \\ 
  108 & 415 & 1.47 (0.11) & $-0.01\; (0.06)$ & $ -440.6$ \\ 
  109 & 198 & 1.33 (0.15) & $ 0.03\; (0.09)$ & $ -202.9$ \\ 
  110 & 88 & 1.22 (0.23) & $ 0.12\; (0.17)$ & $  -85.4$ \\ 
  111 & 34 & 1.50 (0.47) & $ 0.06\; (0.30)$ & $  -34.9$ \\ 
   \bottomrule
\end{tabular}
\caption[Parameter estimates for the generalized Pareto distribution for the Italian super-centenarian data.]{Maximum likelihood estimates of the generalized Pareto for the Italian super-centenarian data. From left to right, threshold $u$ (in years), number of threshold exceedances $n_u$, estimates (standard errors) of the scale $\sigma$, shape $\xi$ parameters, log-likelihood at \textsc{mle} $\ell(\widehat{\theta})$.} 
\label{tab_italcent_gpdparam}
\end{table}

\Cref{tab_italcent_confint} gives the point estimates and 95\% confidence interval; the numbers of exceedances at these thresholds are appreciable, but nevertheless higher-order correction  substantially increases the upper confidence limit for $\iota$.

\begin{figure}[t] 
\centering 
\includegraphics[width=\textwidth]{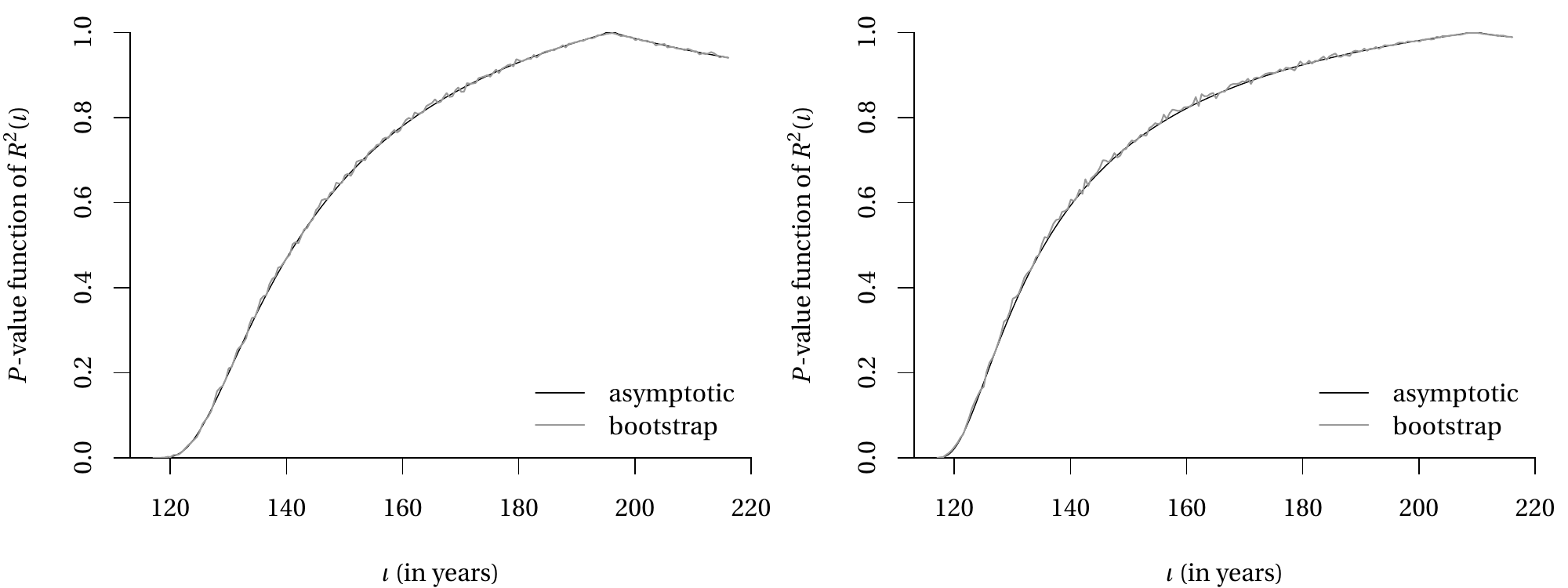}
\caption{Italian semi-supercentenarian data: $p$-value function for the profile likelihood ratio statistic $R^2(\iota) =2\{\ell(\widehat{\bs{\theta}})-\ell(\widehat{\xi}_{\iota})\}$ using excess lifetime {above $u=107$ (left) and $u=108$ (right) years} based on the asymptotic $\chi^2_1$ distribution (black) and the bootstrap distribution $\#\{b: R^{2(b)}(\iota) > R^2(\iota)\}/B$ for bootstrap replications $b=1, \ldots, B$ (grey).}
\label{pvalfun}
\end{figure}

To confirm our findings we can estimate the distribution of the likelihood root for $\iota$ using the bootstrap \citep[cf.][]{Lee/Young:2005}. We did not consider this approach in the simulation study, as its good properties have been checked in other contexts and its calibration entails a costly double parametric bootstrap. The $b$th bootstrap likelihood root $R^{(b)}(\iota)$ is computed at each value of $\iota$ based on a sample simulated from a generalized Pareto distribution with parameters $(\widehat{\xi}_\iota, \iota)$. \Cref{pvalfun} shows that the bootstrap $p$-value and the $p$-value obtained from the asymptotic $\chi^2_1$ distribution of the profile likelihood ratio test agree up to Monte Carlo variability, and suggests that this approach may be useful more widely in the context of extremal inference. 
 
%

\begin{table}[t!]
\centering
\begin{tabular}{lll}
  \toprule
\multicolumn{1}{c}{$u$} & \multicolumn{1}{c}{$R$} & \multicolumn{1}{c}{$R^\star$} \\ 
  \midrule
105 & $142.2$ ($128.5$, $213.7$) & $143.6$ ($129.3$, $235.3$) \\ 
  105.5 & $127.5$ ($122.1$, $138.5$) & $127.7$ ($122.3$, $140.1$) \\ 
  106 & $131.5$ ($123.8$, $159.3$) & $132.9$ ($124.2$, $166.0$) \\ 
  106.5 & $138.4$ ($125.3$, $300.4$) & $143.1$ ($126.5$, $596.6$) \\ 
   \bottomrule
\end{tabular}
\caption[Estimates and 95\% confidence intervals for the upper limit to lifetime]{Point estimates (95\% confidence intervals) for the upper limit to lifetime $\iota$ (in years) based on the profile likelihood ratio statistic $R(\iota)$ (middle) and the modified likelihood ratio statistic $R^{\star}(\iota)$ for the tangent model approximation (right) using threshold exceedances of $u$ for the Italian semi-super centenarian data set. } 
\label{tab_italcent_confint}
\end{table}

The apparent stability of the estimates and the large standard errors for $\widehat{\xi}_u$ seen in  \Cref{tab_italcent_gpdparam} do not allow us to rule out the exponential tail for thresholds $u>107$ years, though the $p$-values for the four lowest thresholds 105--106.5 years, of 4.1\%, 0.5\%, 1.4\% and 6.0\%, are smaller.
 Under the exponential model, the probability of surviving one additional year conditional on survival up to $u$ years is $\exp(-1/\tau)$. Based on exceedances of $u=110$ years, the model would yield an estimated probability of surviving an additional year of 0.476 with 95\% confidence interval (0.416, 0.537):  fewer than four in a thousand supercentenarians would be expected to live older than Emma Moreno. 
These results are quite coherent with those of \citet{Rootzen/Zholud:2017}, who analysed a smaller dataset on individuals who lived over 110 years.

\section{Simulation study}\label{sect:sims}
The higher order methods highlighted in the data illustrations typically lead to much wider confidence intervals for high quantiles. Despite their theoretically appealing properties, one may inquire whether the additional effort is worth it, and if in particular whether profile likelihood intervals have adequate coverage properties. Sample sizes for extremes, whether block maxima or threshold exceedances, are often small. We used Monte Carlo simulation to investigate small-sample inference for risk measures based on the profile log-likelihood, the tangent exponential model approximation and Severini's corrections.

The \citet{Weissman:1978} estimator can be used to estimate extreme quantiles based on the largest observations of a sample and is popular among practitioners who use nonparametric estimators of $\xi$. In this case, uncertainty statements that are attached seem to be almost exclusively Wald-based confidence intervals \citep[e.g., ][\S~4.3]{deHaan/Ferreira:2006}; see \cite{Buitendag:2020} for a recent alternative.

\subsection{General setup}
In a typical data analysis, one may attempt to predict the 100-year maximum temperature based on 20 years of daily records, where restricting attention to summer months yields around 90 observations per year.
To mimic this scenario, we generated $1800$ independent observations from a parametric model and targeted the expectation and median of the distribution of 9000-observation maximum from that same distribution, with  benchmarks computed using penultimate approximations.

The choice of block size or threshold compromises between closeness of approximation (and thus reduced asymptotic bias) and small-sample effects. For larger block size/thresholds, the extreme-value approximation is in principle better, but estimation uncertainty is larger because of the smaller sample size.  
We divided the 1800 simulated values into blocks of sizes $m=30, 45, 90$, and fitted the  $\mathdis{GEV}$ distribution to the block maxima. We also fitted the  $\mathdis{GP}$  distribution to the largest $n_u=20, 40, 60$ order statistics of a sample of size 1800 from the  $\mathdis{GP}$  distribution. We likewise generated data from six other distributions mentioned in \Cref{sec:penult} and applied both block maximum and threshold methods to these data; see the \SM.

For each sample, we obtained four estimates and five sets of confidence limits for $\psi$, based on the Wald statistic; the likelihood root $R(\psi)$ and the modified likelihood root  $R^\star(\psi)$ defined in \cref{eq:likroot} and \cref{eqmodiflikroot}; and the modified profile likelihoods \eqref{eq:severtem} and \eqref{eq:severcov}. The Wald statistic was  computed on the log scale and back-transformed, i.e., with limits $\exp\{ \log(\widehat{\psi}) \pm \Phi^{-1}(1-\alpha/2)\se(\widehat{\psi})/\widehat{\psi}\}$; the log transformation is intended to mitigate the poor properties of this statistic in highly asymmetric situations. For each target (return level, median and expectation of the $T$-year maximum), distribution and threshold or block size, we also calculated the relative bias of the point estimators, and the overall coverage and the average widths of two-sided confidence intervals. The full results are in the \SM, and we summarise the main findings below, focusing on properties of one-sided confidence limits. 

The maximum likelihood estimator of the shape parameters can occasionally be very large, leading to very wide confidence intervals. To avoid this unduly affecting the results, we use trimmed mean estimates for the relative width and the relative bias, with 10\% trimmed proportion in each tail.

\subsection{Summary of findings}

\subsubsection*{Relative error of one-sided confidence intervals}
\begin{figure}[t]
\centering
 \includegraphics[width =\textwidth]{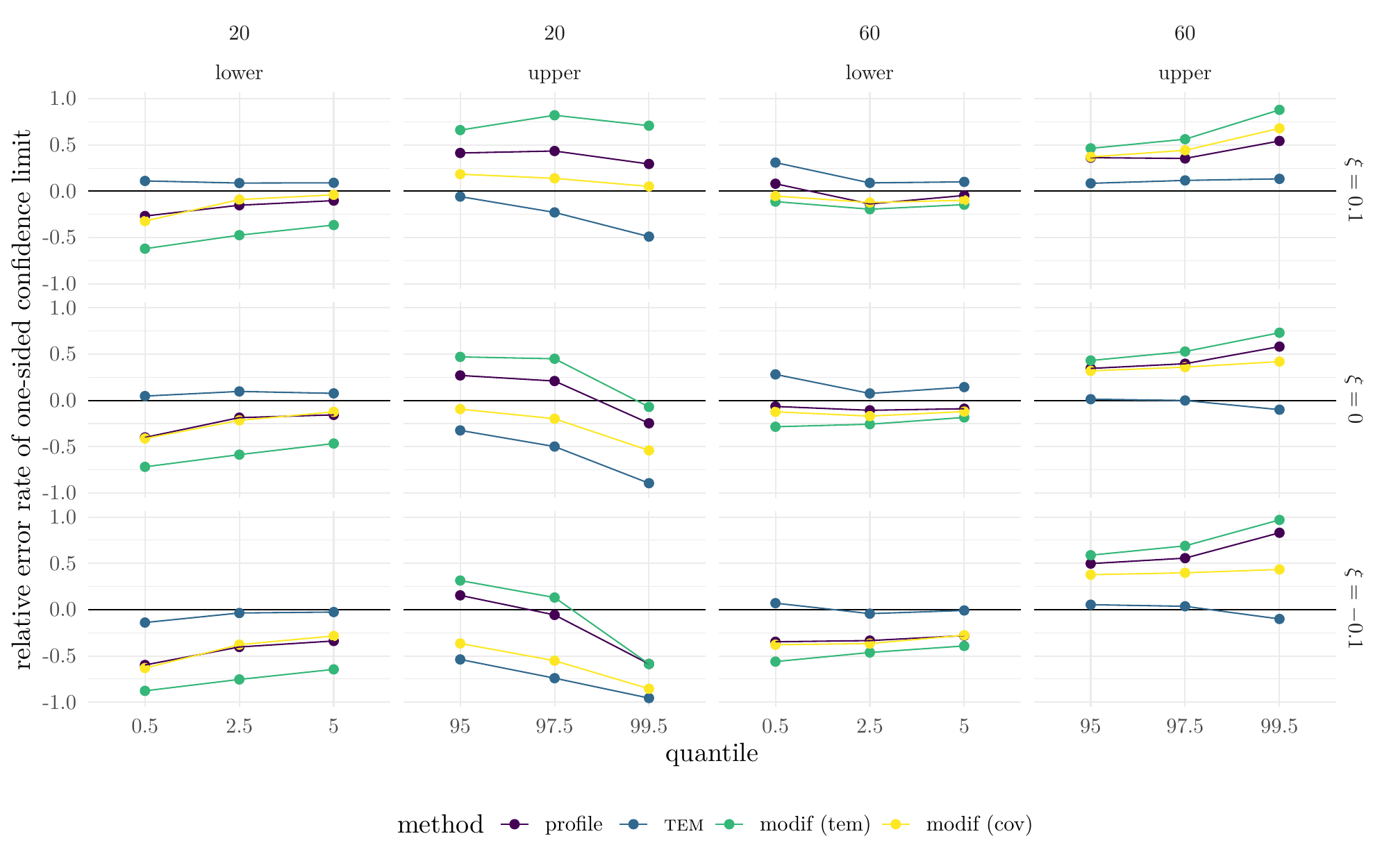}
 \caption{Relative coverage errors for one-sided lower and upper confidence limits with nominal error rates $0.5$\%, $2.5$\% and $5$\% for the expected $9000$-observation maximum estimated using maxima of generalized extreme value  samples of sizes $n=20$ (left) and $n=60$ (right) with shape parameter $\xi=-0.1, 0, 0.1$ (bottom to top). An ideal method would have zero relative error in both tails, whereas methods with relative error $\pm 0.5$ have empirical error rates 1.5 ($+$) or 0.5 ($-$) times the nominal rate.  The upper and lower tail errors for intervals whose relative errors have opposite signs will cancel to some extent when a two-sided interval is computed. If both upper and lower tail errors are positive, the corresponding two-sided intervals have empirical coverage that is too low, whereas  negative upper and lower tail errors correspond to conservative two-sided confidence intervals.}
 \label{fig:errorratebm}
\end{figure}

\begin{figure}[!t]
\centering
 \includegraphics[width = \textwidth]{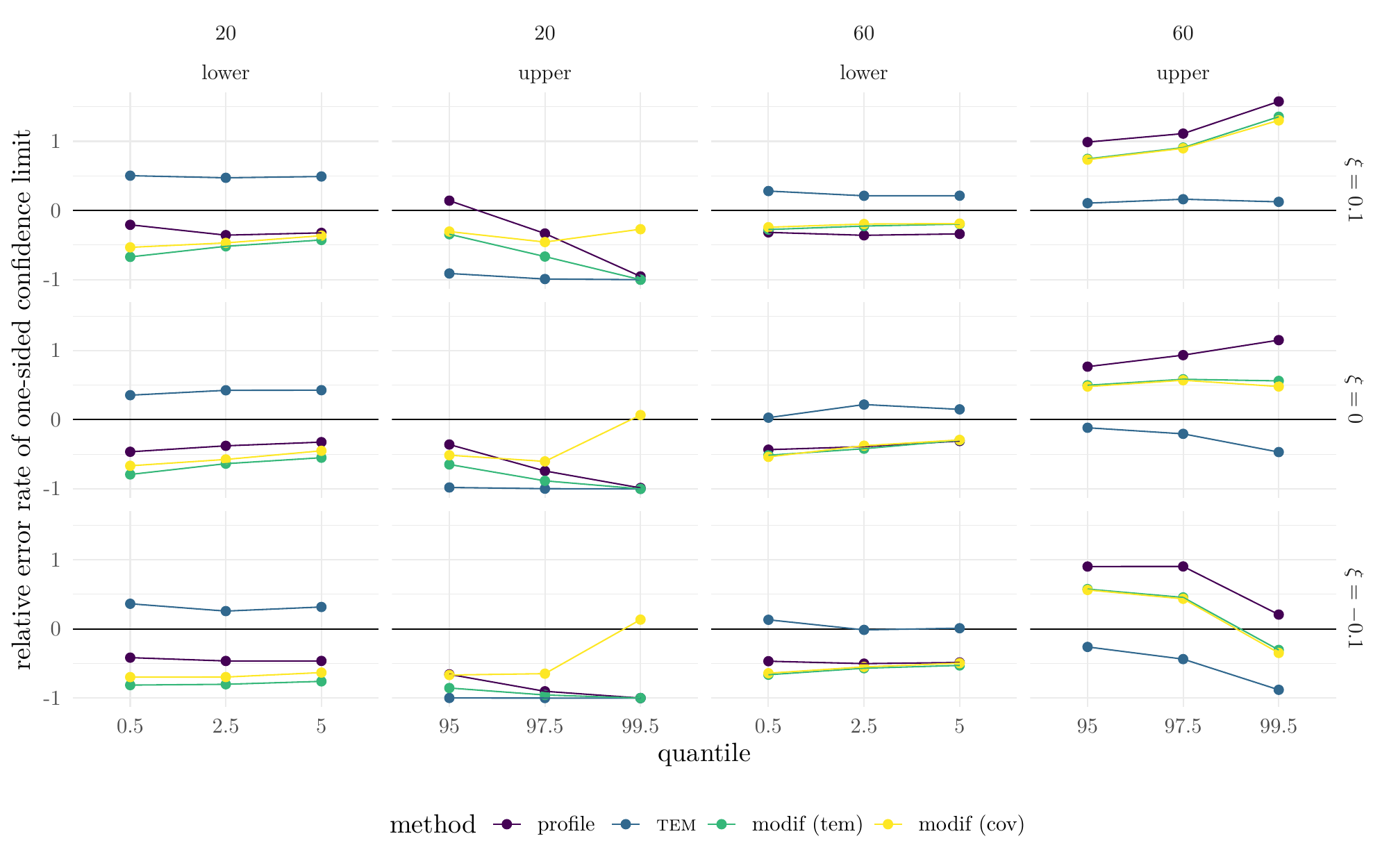}
 \caption{Relative coverage errors for one-sided lower and upper confidence limits for the expected $9000$-observation maximum estimated using  generalized Pareto  samples of sizes $n=20$ (left) and $n=60$ (right) with $\xi =-0.1,0,0.1$ (bottom to top). See the caption to \Cref{fig:errorratebm} for explanation.}
 \label{fig:errorratepot}
\end{figure}

\Cref{fig:errorratebm,fig:errorratepot} display one-sided relative coverage errors for the expected $N$-observation maximum; similar results hold for $N$-observation median and $N$-observation return level. 
Despite the log-transformation, the Wald intervals fail to capture the positive skewness of the estimators of $z_{N}$, $\mathfrak{q}_{1/2}$ (not shown) and $\mathfrak{e}_N$ defined in \Cref{subsec:retlev}. The one-sided relative coverage errors for the Wald statistics are so large that they fall outside the limits of \Cref{fig:errorratebm,fig:errorratepot}: for example, applying the block maximum method with 20 observations to samples from a $\mathdis{GEV}$ distribution (\Cref{tab_nomerr_l3_bm}), the Wald-based 99\% confidence intervals contain the true value roughly $85$\% and $81$\% of the time when $\xi=0.1$ and $\xi=-0.1$, respectively, but the $5$\% empirical error rate for the lower limit is 0\%, indicating that the interval is too wide on the left and too short on the right. 

If the data are generated from the generalized extreme value distribution, the empirical error rates for the \textsc{tem} are closer to nominal, but no method is universally best. Perhaps unexpectedly, the penultimate effects are not really visible for the other distributions (\Cref{tab_nomerr_l3_bm}). The profile and higher-order methods for block maxima seem impervious to the effects of extrapolation and their coverage is excellent overall.

\Cref{fig:errorratepot} shows that the results based on threshold exceedances are more variable. The performance of Wald-based intervals remains calamitous: the empirical upper error rate for the nominal 5\% limit is around 30\% in all scenarios for the untransformed Wald statistic and improves only to 20--30\% after transformation. 
With $k=20$ observations (\Cref{fig:errorratepot,tab_nomerr_l1_pot}), most higher-order methods overcover even when the model is correctly specified. The \textsc{tem} interval is shifted to the right, whereas Severini's corrections display higher empirical error in the lower tail. This breakdown of the \textsc{tem} could be due to penultimate effects and small-sample bias, as it vanishes as the sample size grows; the \textsc{tem} performs very well when $k=60$ (\Cref{tab_nomerr_l3_pot}). 
Two-sided profile likelihood intervals typically have good coverage, but their upper empirical error rates can be more than double the nominal values, as the intervals tend to lie too far to the left. Thus the price paid for intervals with better coverage is increased uncertainty stemming from their greater width. 

\subsubsection*{Width of confidence intervals}

When $\xi>0$ the expected $N$-observation maximum is larger than both the median of the $N$-observation maximum and the $N$-year return level, and its confidence intervals are the widest of those for all three risk measures due to the extrapolation in the upper tail.
The higher-order intervals, especially those based on $R^\star$, overcover slightly when the sample size is smaller and the blocks are larger, e.g., for $m=90$ with $k=20$ (\Cref{tab_nomerr_l3_bm}). The average widths of two-sided confidence intervals for the block maximum method with $m=30$, $k=60$ are comparable (not shown). For this setting, the intervals based on $\ell_{\mathsf{m}}^{\mathrm{tem}}$ are the shortest among those implemented. 

Higher-order methods for threshold exceedances give wider confidence intervals, often because they have better coverage in the upper tail: for example, the \textsc{tem} confidence intervals are between $1.75$ and $2$ times wider than those based on the profile likelihood when $k=20$ and about $1.25$ times wider when $k=60$.

\subsubsection*{Bias of point estimators of risk measures}

When using threshold exceedances, maximum likelihood estimators of $\xi$ are negatively biased for any sample size $k\leq 60$ (\Cref{fig:biasEVD}) and risk estimators are likewise downwardly biased. For $k=20$, the \textsc{tem} point estimators obtained by solving the equation $R^\star(\psi)=0$ are positively biased, but they have the lowest bias of all point estimators considered  when $k\geq 40$. The point estimators derived using Severini's modified profile log-likelihoods have lower bias than the maximum likelihood estimator. 

\subsection{Practical guidelines} \label{practicalguidelines}
 
Wald-based confidence intervals for the risk measures considered here should never be used; their coverage is appallingly low, even after transformation.  For block maxima, profile likelihood-based confidence intervals have good two-sided coverage overall for the risk measures we considered, and there seems to be little gain in using higher-order methods: the discrepancy between the empirical error rates in the lower and upper tails seems to be due to the bias of the risk estimators themselves.  For both types of data, the \textsc{tem}-based estimator is systematically larger than the maximum likelihood estimator. For threshold exceedances, \textsc{tem}-based confidence intervals have very good coverage and the corresponding point estimators have smaller bias  when the sample size is larger than around $50$; while no higher-order method is always better, \textsc{tem}-based intervals usually improve on the others. 

\subsubsection*{Acknowledgements}

The work was financially supported by the Natural Sciences and Engineering Research Council of Canada and the Swiss National Science Foundation. 


\clearpage
\setlength{\bibsep}{0.0pt}
 {\small 

 }
 \clearpage 
\appendix 
\section*{Supplementary material}
\section{\texorpdfstring{Derivation of the information matrix of the $r$-largest order statistics}{Derivation of the information matrix of the r-largest order statistics}}\label{section:infomatrlargest}
{{
To derive the information matrix for the $r$-largest likelihood given in \Cref{eq:rlarglik}, we note that the marginal density of $Y_{r}$ is
\begin{align*}
f_{Y_{r}}(y_{r};\mu, \sigma, \xi) = \frac{1}{(r-1)!\sigma} \left(1 + \xi\frac{y_{r}-\mu}{\sigma}\right)_{+}^{-r/\xi-1}\exp\left\{-\left(1 + \xi\frac{y_{r}-\mu}{\sigma}\right)_+^{-1/\xi}\right\},
\end{align*}
so the joint density of $Y_{1},\ldots, Y_{r}$  may be written as
\begin{multline*}
\frac{1}{(r-1)!\sigma} \left(1 + \xi\frac{y_{r}-\mu}{\sigma}\right)_+^{-r/\xi-1}\exp\left\{-\left(1 + \xi\frac{y_{r}-\mu}{\sigma}\right)_+^{-1/\xi}\right\} \\
\times (r-1)!\prod_{j=1}^{r-1} \frac{1}{ \sigma} \frac{\left(1 + \xi\frac{y_{j}-\mu}{ \sigma}\right)_+^{-1/\xi-1} }{ \left(1 + \xi\frac{y_{r}-\mu}{ \sigma}\right)_+^{-1/\xi\hphantom{-1}}};
\end{multline*}
that is, we write the joint density as the product of the density of $Y_{r}$ and the joint conditional density of $Y_{1}, \ldots, Y_{r-1}$ conditional on $Y_{r}=y_{r}$.  This conditional density equals that of the order statistics of $r-1$ independent variables with generalized Pareto density
\[
h(\bs{y}_{-r}-y_{r}; \tau,\xi) = \prod_{j=1}^{r-1}\frac{1}{\tau} \left(1 + \xi\frac{y_{j}-y_{r}}{\tau}\right)^{-1/\xi-1}_+, 
\]
where $\tau = \sigma + \xi(y_{r}-\mu)$.  Thus the overall log likelihood is
\begin{align*}
\ell(\mu,\sigma,\xi; y_{j}, y_{r}) \equiv  \log \big\{f_{Y_{r}}(y_{r};\mu, \sigma,\xi)\big\} + \sum_{j=1}^{r-1} \log \big\{h(y_{j}-y_{r}; \tau,\xi)\big\},
\end{align*}
where $y_{1},\ldots, y_{r-1}$ represent the observed values of a random sample of generalized Pareto variables. 
We may thus write the joint density of the $r$-largest order statistics as the product of the density of $Y_{r}$ and the joint conditional density of $Y_{1}, \ldots, Y_{r-1}$ conditional on $Y_{r}=y_{r}$.
To obtain the observed information we first calculate the Hessian matrix of $-\ell$, then condition on $Y_{r}=y_{r}$ and take expectations over $X_j=Y_{j}-y_{r}$. It remains to write $\tau = \sigma + \xi(y_{r}-\mu)$ and integrate over $Y_{r}$.  The matrices themselves can be found in \Cref{sec:rlarginfomat} and a numerical implementation is available via the function \texttt{rlarg.infomat} in the \Rlang{} package \texttt{mev}.

\section{\texorpdfstring{Derivation of the \textsc{tem} for the Poisson process}{Derivation of the TEM for the Poisson process}} \label{sec:temPP}
Suppose we observe events of an inhomogeneous Poisson process ${\mathcal P}$ with intensity $\dot\nu(x;\boldsymbol{\theta})$ for $x\in{\mathcal X}$, where ${\mathcal X}$ is partitioned into subsets ${\mathcal X}_1,\ldots, {\mathcal X}_K$.  Let $N({\mathcal A})$ denote the number of events of ${\mathcal P}$ in a measurable set ${\mathcal A}\subset{\mathcal X}$, let $N_k=N({\mathcal X}_k)$, and suppose that $N({\mathcal X})$ has finite expectation
\begin{align*}
\nu({\mathcal X};\boldsymbol{\theta})=\int_{\mathcal X} \dot\nu(x;\boldsymbol{\theta})\, \mathrm{d} x.                                                                         
 \end{align*}

If $\dot\nu$ is constant on each of the ${\mathcal X}_k$, then the log likelihood is that of the independent Poisson variables $N_1,\ldots, N_K$, 
\begin{align*}
\sum_{k=1}^K n_k\log\left\{ |{\mathcal X}_k|\dot\nu(x_k;\boldsymbol{\theta})\right\} -  |{\mathcal X}_k|\dot\nu(x_k;\boldsymbol{\theta}) 
&\equiv \sum_{k=1}^K \left\{n_k\log \dot\nu(x_k;\boldsymbol{\theta}) -  |{\mathcal X}_k|\dot\nu(x_k;\boldsymbol{\theta})\right\} = \sum_{k=1}^K \ell_k(\boldsymbol{\theta}), 
\end{align*}
say, where $n_k$ is the realised value of $N_k$ and $x_k\in {\mathcal X}_k$. The terms $n_k\log|{\mathcal X}_k|$ dropped at the $\equiv$ sign do not depend on $\boldsymbol{\theta}$; retaining them leads to an affine transformation of $\varphi(\boldsymbol{\theta})$ and makes no difference to inferences. The arguments in \cite{Davison.Fraser.Reid:2006} imply that second-order inference is obtained on using
\begin{align*}
\varphi(\boldsymbol{\theta}) &= \sum_{k=1}^K \mathbf{V}_k \frac{\partial \ell_k(\boldsymbol{\theta})}{\partial n_k}, 
\shortintertext{where}
\mathbf{V}_k&= \left.\frac{\partial \mathsf{E}(N_k;\boldsymbol{\theta})}{\partial\boldsymbol{\theta}}\right|_{\boldsymbol{\theta}=\widehat{\boldsymbol{\theta}}} = \left.\frac{\partial  |{\mathcal X}_k|\dot\nu(x_k;\boldsymbol{\theta})}{\partial\boldsymbol{\theta}}\right|_{\boldsymbol{\theta}=\widehat{\boldsymbol{\theta}}}, 
\end{align*}
and this yields
\begin{equation}\label{eq2}
\varphi(\boldsymbol{\theta}) = \sum_{k=1}^K |{\mathcal X}_k| \left.\frac{\partial\dot\nu(x_k;\boldsymbol{\theta})}{\partial\boldsymbol{\theta}}\right|_{\boldsymbol{\theta}=\widehat{\boldsymbol{\theta}}}\log\dot\nu(x_k;\boldsymbol{\theta})
=
\int_{\mathcal X}  \left. \frac{\partial\dot\nu(x;\boldsymbol{\theta})}{\partial\boldsymbol{\theta}}\right|_{\boldsymbol{\theta}=\widehat{\boldsymbol{\theta}}}\log\dot\nu(x;\boldsymbol{\theta})\mathrm{d} x .
\end{equation}
This integral does not depend on the partition of ${\mathcal X}$, so must be the limit as $K\to\infty$.  

The integral in (\ref{eq2}) is intractable in general, but the numerical approximation
\begin{equation}\label{eq1}
 \sum_{j=1}^n \left. \frac{\partial\dot\nu(x_j;\boldsymbol{\theta})}{\partial\boldsymbol{\theta}}\right|_{\boldsymbol{\theta}=\widehat{\boldsymbol{\theta}}} \times  \frac{1}{\dot\nu(x_j;\boldsymbol{\theta})}\log \dot\nu(x_j;\boldsymbol{\theta})
\end{equation}
based on events $x_1, \ldots, x_n\in {\mathcal X}$ differs from $\varphi(\boldsymbol{\theta})$ by a term of order $\nu({\mathcal X};\boldsymbol{\theta})^{1/2}$ and therefore gives the same order of error.  To check this, note that, conditional on $N({\mathcal X})=n$, the $x_j$ are independent and identically distributed on ${\mathcal X}$ with density $\dot\nu(x;\boldsymbol{\theta})/\nu({\mathcal X};\boldsymbol{\theta})$. 
 Thus the expectation of~(\ref{eq1}), conditional on $N({\mathcal X})=n$, is 
\begin{align*}
n \int_{\mathcal X} \left. \frac{\partial\dot\nu(x;\boldsymbol{\theta})}{\partial\boldsymbol{\theta}}\right|_{\boldsymbol{\theta}=\widehat{\boldsymbol{\theta}}}\times \frac{1}{\dot\nu(x;\boldsymbol{\theta})} \log\dot\nu(x;\boldsymbol{\theta})\times \frac{\dot\nu(x;\boldsymbol{\theta})}{\nu({\mathcal X};\boldsymbol{\theta})}\mathrm{d} x  = \frac{n}{\nu({\mathcal X};\boldsymbol{\theta})}\varphi(\boldsymbol{\theta}),
\end{align*}
and as $N({\mathcal X})$ has expectation $ \nu({\mathcal X};\boldsymbol{\theta})$, the expectation of (\ref{eq1}) is $\varphi(\boldsymbol{\theta})$, as required.  One can check  that~(\ref{eq1}) has variance of order $\nu({\mathcal X};\boldsymbol{\theta})$ under mild conditions on the integrand.

\section{Details of the simulation study}
We used the infrastructure provided by the \textsf{R} package \code{simsalapar} \citep{Hofert/Maechler:2016} for the simulation study; the routines and higher-order methods described in the simulation study are implemented in the \textsf{R} package \texttt{mev} and the code used for the simulation study and the applications is available for download at \url{https://github.com/lbelzile/hoa-extremes}. 

We obtained the maximum likelihood estimates for the generalized extreme value using an augmented Lagrange optimization routine with boundary constraints to ensure that $\widehat{\xi}\geq -1$ and that $\widehat{\sigma} + \widehat{\xi}(x_i -\widehat{\mu}) > 0$ $(i=1, \ldots, n)$, so that the log likelihood was finite. The generalized Pareto distribution was fitted using the algorithm of \cite{Grimshaw:1993}. 
 
 The profile log-likelihood estimates were obtained using constrained optimization methods at selected values of $\psi$, using dedicated algorithms such as sequential quadratic programming to obtain profile log-likelihood values for each of the values of $\psi$ on a grid. The tangent space derivatives, canonical parameters, score and information matrices were derived analytically for every parameter and model of interest.
 The correction term $Q$ for the tangent exponential model was calculated via \cref{eq:qtem} and used to obtain $R^\star=R+R^{-1}\log(Q/R)$.
For the penalized likelihood of Severini, we computed penalty terms that were added to the profile log-likelihood values. 
Although the correction can be shown to be continuous, values of $R^\star(\psi)$ can be numerically unstable when $\psi\approx \widehat\psi$ \citep[cf.][p.149]{Brazzale/Davison/Reid:2007}:  the values of $R^\star$ can vary uncontrollably even though in principle $R^\star \approx 0$ in a neighborhood of $\widehat{\psi}$. The fact that $R^\star \sim \mathdis{No}(0,1)$ suggests fitting constrained quantile regression $B$ splines for the median of $R^\star-R$ as a function of $R$, downweighting any observations near $R=0$ with abnormal values. We then compare the predicted values for $\widehat{R}^*$ based on the spline fit with the calculated values, standardize the latter and exclude all points exceeding the 0.95 $\chi^2_1$ quantile. A second spline regression is fitted to the remaining values in order to interpolate the missing values of $R^\star$. This ad-hoc scheme usually affects estimates by less than $\bigO(10^{-3})$, but is effective at removing the outliers and works well in practice. 
For the confidence intervals, we fitted constrained $B$-splines with response $\psi$ and values $R$ (respectively $R^\star$) and predicted the quantile $\psi$ corresponding to $2\ell_{\rp}(\psi) = \pm q$, where ${q}$ is the ($1-\alpha$) $\chi^2_1$ quantile. For the penalized methods, we computed the equivalent of the signed likelihood root statistic and proceeded analogously. 

\subsection{Parametric models}
In addition to simulating data from a generalized extreme value distribution and from a generalized Pareto distribution with shape parameter $\xi \in \{-0.1,0, 0.1\}$, we considered parametric families of distribution satisfying the extremal types theorem. These are 
\begin{enumerate}
 \item the standard normal distribution,
 \item the standard log-normal distribution,
 \item the Student-$t$ distribution with $10$ degrees of freedom,
 \item the Burr distribution with survival function  $S(x)=(1+x^a)^{-b}$ with $a=5$, $b=2$,
 \item the Weibull distribution with survival function $S(x) = \exp(-x^a)$ with $a=2/3$,
 \item the generalized Gamma distribution with survival function \[S(x)=\frac{\Gamma\left\{\frac{\gamma_1}{\gamma_2}, \left(\frac{x}{\beta}\right)^{\gamma_2}\right\}}{\Gamma\left( \frac{\gamma_1}{\gamma_2}\right)}, \qquad x>0, \gamma_1, \gamma_2, \beta >0,\] where $\Gamma(a, z) \coloneqq \int_z^\infty x^{a-1}\exp(-x)\mathrm{d} x$. The values for these parameters ($\beta = 1.83, \gamma_1=1.16$ and $\gamma_2=0.54$) were taken from \cite{Papalexiou:2013}, so as to reflect values found by hydrologists in estimating the tail index for global rainfall.
\end{enumerate}
\Cref{table:penult_shape} gives the limiting and penultimate shape parameters for various thresholds and block sizes. The Burr and Student distributions are heavy-tailed and have positive penultimate shape parameters, whereas the other distributions have different penultimate behaviours even though $\xi^{\star}=0$. 

\begin{table}[t!]   \centering\small   \begin{tabular}{*{1}{l}*{7}{r}*{1}{l}}     \toprule     Distribution \textbar\ & \multicolumn{1}{c}{\small{$q=0.967$}} & \multicolumn{1}{c}{\small{$q=0.978$}} & \multicolumn{1}{c}{\small{$q=0.989$}} & \multicolumn{1}{c}{\small{$m = 30$}} & \multicolumn{1}{c}{\small{$m = 45$}} & \multicolumn{1}{c}{\small{$m = 90$}} & \multicolumn{1}{c}{\small{$m = 9000$}} & \multicolumn{1}{c}{$\xi_{\infty}$} \\     \midrule     Burr & $0.01$ & $0.03$ & $0.05$ & $0.03$ & $0.04$ & $0.06$ & $0.10$ & $0.1$ \\     Weibull & $0.15$ & $0.13$ & $0.11$ & $0.16$ & $0.14$ & $0.12$ & $0.05$ & $0$ \\     Gen. gamma & $0.15$ & $0.14$ & $0.12$ & $0.17$ & $0.15$ & $0.13$ & $0.07$ & $0$ \\     Gaussian & $-0.18$ & $-0.16$ & $-0.13$ & $-0.16$ & $-0.14$ & $-0.12$ & $-0.06$ & $0$ \\     Lognormal & $0.27$ & $0.26$ & $0.25$ & $0.28$ & $0.27$ & $0.25$ & $0.19$ & $0$ \\     Student & $-0.05$ & $-0.03$ & $0.00$ & $-0.03$ & $-0.02$ & $0.01$ & $0.07$ & $0.1$ \\     \bottomrule   \end{tabular}   \caption{Penultimate shape parameters for six distributions, based on threshold exceedances with threshold at $q$ percentile (first three columns), block maxima with maximum of $m$ observations (fourth to sixth column).  The penultimate shape parameter for the maximum of $9000$ observations is the reference, still far from the tail index $\xi_{\infty}$ in the last column.}   \label{table:penult_shape} \end{table}

\section{Simulation results}

The discussed below tables may also be found in Appendix~A  of \cite{Belzile:thesis}. \Cref{tab_nomerr_l2_bm,tab_nomerr_l3_bm,tab_nomerr_l1_pot,tab_nomerr_l3_pot} contain the estimated one-tailed error rates for the confidence intervals studied above, for data from a variety of underlying distributions, including the limiting generalized extreme-value and generalized Pareto distributions for maxima and threshold exceedances. \Cref{tab_RB_l2_bm,tab_RB_l3_bm,tab_RB_l1_pot,tab_RB_l3_pot} contain the trimmed means of the relative bias of the point estimators for those same scenarios, whereas \Cref{tab_CIW_l2_bm,tab_CIW_l3_bm,tab_CIW_l1_pot,tab_CIW_l3_pot} provide the  estimated relative width of the confidence intervals relative to those of the profile likelihood method. We use trimmed means and discard the smallest and largest 10\% of the simulations to remove confidence intervals that are extrapolated far beyond the range of the $\psi$ values at which the profile likelihood is calculated, as these implausibly large intervals would be discarded by practitioners.

For the risk measures we consider, the point estimators provided by the \textsc{tem} for the block maximum method are systematically larger than the maximum likelihood estimator, typically by 2\%. They are positively biased when the (penultimate) shape is positive. The bias is more pronounced for the threshold method: maximum likelihood estimators are negatively biased and the \textsc{tem} corrects for this in the case of correctly specified models $F_7$--$F_9$ unless the sample size is too small ($k=20$).

The \textsc{tem} confidence intervals are at most 6\% wider than profile likelihood intervals when $k=60$ for the block maximum method, but up to 15\% wider with $k=20$ exceedances. The intervals provided by Severini's modified profile likelihood based on the \textsc{tem} approximation are narrower, and those based on empirical covariances are wider, than their profile likelihood counterparts. For threshold exceedances with $k=20$, the confidence intervals based on the modified likelihood root are sometimes twice as wide as those of the profile, whereas all the modified profile likelihoods have 30\% longer intervals. The abysmal coverages for the Wald-based confidence intervals are due to the fact they are too short and to the asymmetry of the distributions of the corresponding estimators.

{\footnotesize 
\captionsetup{font=scriptsize}
\begin{table}[t!]   \centering\footnotesize   \begin{tabular}{*{2}{l}*{12}{r}}     \toprule      & Parameter & \multicolumn{6}{c}{Quantile} & \multicolumn{6}{c}{$N$-obs. mean} \\     \cmidrule(lr){3-8} \cmidrule(lr){9-14}     $F$ & Method \textbar\ Error rate & \multicolumn{1}{c}{0.5} & \multicolumn{1}{c}{2.5} & \multicolumn{1}{c}{5} & \multicolumn{1}{c}{5} & \multicolumn{1}{c}{2.5} & \multicolumn{1}{c}{0.5} & \multicolumn{1}{c}{0.5} & \multicolumn{1}{c}{2.5} & \multicolumn{1}{c}{5} & \multicolumn{1}{c}{5} & \multicolumn{1}{c}{2.5} & \multicolumn{1}{c}{0.5} \\     \midrule     $F_1$ & Wald & $0.0$ & $0.0$ & $0.0$ & $19.5$ & $15.5$ & $10.5$ & $0.0$ & $0.0$ & $0.0$ & $20.5$ & $17.0$ & $12.0$ \\     & profile & $0.5$ & $2.0$ & $4.0$ & $7.0$ & $3.0$ & $0.5$ & $0.5$ & $2.0$ & $4.0$ & $7.5$ & $3.5$ & $0.5$ \\     & \textsc{tem} & $0.5$ & $2.5$ & $5.0$ & $5.0$ & $2.0$ & $0.5$ & $0.5$ & $2.5$ & $4.5$ & $5.5$ & $2.5$ & $0.5$ \\     & Severini (\textsc{tem}) & $0.5$ & $1.5$ & $3.5$ & $8.0$ & $3.5$ & $0.5$ & $0.5$ & $1.5$ & $3.5$ & $8.5$ & $4.0$ & $0.5$ \\     & Severini (cov.) & $0.5$ & $2.0$ & $4.5$ & $5.0$ & $2.0$ & $0.5$ & $0.5$ & $2.5$ & $4.5$ & $5.5$ & $2.0$ & $0.5$ \\ \addlinespace[3pt]     $F_2$ & Wald & $0.0$ & $0.0$ & $1.5$ & $13.0$ & $10.5$ & $7.0$ & $0.0$ & $0.0$ & $1.0$ & $13.0$ & $10.5$ & $7.0$ \\     & profile & $0.5$ & $2.5$ & $5.0$ & $4.5$ & $2.5$ & $0.5$ & $0.5$ & $3.0$ & $6.0$ & $4.0$ & $2.0$ & $0.5$ \\     & \textsc{tem} & $0.5$ & $3.0$ & $6.0$ & $3.5$ & $1.5$ & $0.5$ & $1.0$ & $3.5$ & $7.0$ & $3.0$ & $1.5$ & $0.5$ \\     & Severini (\textsc{tem}) & $0.5$ & $2.0$ & $4.5$ & $5.0$ & $2.5$ & $0.5$ & $0.5$ & $2.5$ & $5.0$ & $4.5$ & $2.5$ & $0.5$ \\     & Severini (cov.) & $0.5$ & $2.5$ & $5.0$ & $4.5$ & $2.5$ & $0.5$ & $0.5$ & $3.0$ & $6.0$ & $4.0$ & $2.0$ & $0.5$ \\ \addlinespace[3pt]     $F_3$ & Wald & $0.0$ & $0.0$ & $1.0$ & $15.0$ & $12.0$ & $8.0$ & $0.0$ & $0.0$ & $0.5$ & $14.5$ & $12.0$ & $8.0$ \\     & profile & $0.0$ & $1.5$ & $3.5$ & $6.0$ & $3.5$ & $0.5$ & $0.5$ & $2.0$ & $4.0$ & $5.5$ & $3.0$ & $0.5$ \\     & \textsc{tem} & $0.0$ & $1.5$ & $4.0$ & $5.0$ & $2.5$ & $0.5$ & $0.5$ & $2.0$ & $5.0$ & $4.5$ & $2.0$ & $0.5$ \\     & Severini (\textsc{tem}) & $0.0$ & $1.0$ & $3.0$ & $6.5$ & $3.5$ & $1.0$ & $0.0$ & $1.5$ & $3.5$ & $6.0$ & $3.0$ & $0.5$ \\     & Severini (cov.) & $0.0$ & $1.5$ & $3.0$ & $6.0$ & $3.0$ & $0.5$ & $0.5$ & $2.0$ & $4.0$ & $5.5$ & $3.0$ & $0.5$ \\ \addlinespace[3pt]     $F_4$ & Wald & $0.0$ & $0.0$ & $0.0$ & $26.0$ & $22.5$ & $17.0$ & $0.0$ & $0.0$ & $0.0$ & $27.5$ & $24.0$ & $18.0$ \\     & profile & $0.5$ & $1.5$ & $3.5$ & $10.0$ & $5.5$ & $1.0$ & $0.5$ & $1.5$ & $3.0$ & $10.5$ & $5.5$ & $1.0$ \\     & \textsc{tem} & $0.5$ & $2.5$ & $4.5$ & $6.5$ & $3.0$ & $0.5$ & $0.5$ & $2.5$ & $4.5$ & $7.0$ & $3.5$ & $0.5$ \\     & Severini (\textsc{tem}) & $0.5$ & $1.5$ & $2.5$ & $10.5$ & $6.0$ & $1.5$ & $0.0$ & $1.0$ & $2.5$ & $11.0$ & $6.0$ & $1.5$ \\     & Severini (cov.) & $0.5$ & $1.5$ & $3.0$ & $7.5$ & $4.0$ & $0.5$ & $0.5$ & $1.5$ & $3.0$ & $8.0$ & $4.0$ & $0.5$ \\ \addlinespace[3pt]     $F_5$ & Wald & $0.0$ & $0.5$ & $2.0$ & $11.5$ & $8.5$ & $5.5$ & $0.0$ & $0.0$ & $1.0$ & $11.0$ & $8.5$ & $5.5$ \\     & profile & $0.5$ & $2.5$ & $5.5$ & $4.5$ & $2.5$ & $0.5$ & $1.0$ & $3.0$ & $7.0$ & $4.0$ & $2.0$ & $0.5$ \\     & \textsc{tem} & $0.5$ & $2.5$ & $5.5$ & $3.5$ & $2.0$ & $0.5$ & $1.0$ & $3.5$ & $7.0$ & $3.5$ & $1.5$ & $0.5$ \\     & Severini (\textsc{tem}) & $0.5$ & $2.0$ & $4.5$ & $5.0$ & $2.5$ & $0.5$ & $0.5$ & $2.5$ & $5.5$ & $4.5$ & $2.5$ & $0.5$ \\     & Severini (cov.) & $0.5$ & $2.5$ & $5.0$ & $4.5$ & $2.5$ & $0.5$ & $0.5$ & $3.0$ & $6.5$ & $4.0$ & $2.0$ & $0.5$ \\ \addlinespace[3pt]     $F_6$ & Wald & $0.0$ & $0.0$ & $0.5$ & $22.5$ & $19.0$ & $13.5$ & $0.0$ & $0.0$ & $0.0$ & $24.5$ & $20.5$ & $15.0$ \\     & profile & $0.5$ & $2.0$ & $4.0$ & $10.0$ & $5.5$ & $1.5$ & $0.5$ & $2.0$ & $3.5$ & $11.0$ & $6.0$ & $1.5$ \\     & \textsc{tem} & $0.5$ & $2.5$ & $5.0$ & $7.5$ & $4.0$ & $1.0$ & $0.5$ & $2.5$ & $4.5$ & $8.0$ & $4.5$ & $1.0$ \\     & Severini (\textsc{tem}) & $0.5$ & $1.5$ & $3.0$ & $11.0$ & $6.5$ & $1.5$ & $0.5$ & $1.5$ & $3.0$ & $11.5$ & $7.0$ & $2.0$ \\     & Severini (cov.) & $0.5$ & $2.0$ & $3.5$ & $9.5$ & $5.0$ & $1.0$ & $0.5$ & $2.0$ & $3.5$ & $10.0$ & $5.5$ & $1.5$ \\ \addlinespace[3pt]     $F_7$ & Wald & $0.0$ & $0.0$ & $1.0$ & $17.0$ & $13.5$ & $9.5$ & $0.0$ & $0.0$ & $0.5$ & $17.5$ & $14.5$ & $10.0$ \\     & profile & $0.5$ & $2.0$ & $4.5$ & $7.0$ & $3.5$ & $1.0$ & $0.5$ & $2.5$ & $4.5$ & $7.0$ & $3.5$ & $1.0$ \\     & \textsc{tem} & $0.5$ & $2.5$ & $5.5$ & $5.5$ & $2.5$ & $0.5$ & $0.5$ & $3.0$ & $5.5$ & $5.5$ & $3.0$ & $0.5$ \\     & Severini (\textsc{tem}) & $0.5$ & $2.0$ & $4.0$ & $8.0$ & $4.0$ & $1.0$ & $0.5$ & $2.0$ & $4.0$ & $8.0$ & $4.5$ & $1.0$ \\     & Severini (cov.) & $0.5$ & $2.0$ & $4.0$ & $7.0$ & $3.5$ & $1.0$ & $0.5$ & $2.5$ & $4.5$ & $7.0$ & $3.5$ & $1.0$ \\ \addlinespace[3pt]     $F_8$ & Wald & $0.0$ & $0.0$ & $0.5$ & $19.0$ & $15.5$ & $11.0$ & $0.0$ & $0.0$ & $0.5$ & $19.5$ & $16.5$ & $12.0$ \\     & profile & $0.5$ & $2.0$ & $4.5$ & $7.0$ & $3.5$ & $1.0$ & $0.5$ & $2.0$ & $4.5$ & $7.0$ & $3.5$ & $1.0$ \\     & \textsc{tem} & $0.5$ & $3.0$ & $5.5$ & $5.0$ & $2.5$ & $0.5$ & $0.5$ & $3.0$ & $5.5$ & $5.0$ & $2.5$ & $0.5$ \\     & Severini (\textsc{tem}) & $0.5$ & $1.5$ & $3.5$ & $7.5$ & $4.0$ & $1.0$ & $0.5$ & $1.5$ & $3.5$ & $7.5$ & $4.0$ & $1.0$ \\     & Severini (cov.) & $0.5$ & $2.0$ & $4.0$ & $6.5$ & $3.5$ & $0.5$ & $0.5$ & $2.0$ & $4.5$ & $6.5$ & $3.0$ & $0.5$ \\ \addlinespace[3pt]     $F_9$ & Wald & $0.0$ & $0.0$ & $0.0$ & $22.5$ & $19.0$ & $14.5$ & $0.0$ & $0.0$ & $0.0$ & $23.0$ & $19.5$ & $15.0$ \\     & profile & $0.5$ & $1.5$ & $3.5$ & $8.0$ & $4.0$ & $1.0$ & $0.5$ & $1.5$ & $3.5$ & $7.5$ & $4.0$ & $0.5$ \\     & \textsc{tem} & $0.5$ & $2.5$ & $5.0$ & $5.0$ & $2.5$ & $0.5$ & $0.5$ & $2.5$ & $5.0$ & $5.0$ & $2.0$ & $0.0$ \\     & Severini (\textsc{tem}) & $0.0$ & $1.0$ & $2.5$ & $8.5$ & $4.5$ & $1.0$ & $0.0$ & $1.0$ & $2.5$ & $8.0$ & $4.0$ & $0.5$ \\     & Severini (cov.) & $0.0$ & $1.5$ & $3.0$ & $6.5$ & $3.0$ & $0.5$ & $0.5$ & $1.5$ & $3.5$ & $6.5$ & $3.0$ & $0.5$ \\     \bottomrule   \end{tabular}   \caption{One-sided empirical error rates (\%) for lower (first to third columns) and upper (fourth to sixth columns) confidence limits, block maximum method with $m=45, k=40$.  
The distributions (from top to bottom) are Burr ($F_1$), Weibull ($F_2$), generalized gamma ($F_3$), normal ($F_4$), lognormal ($F_5$), Student $t$ ($F_6$), $\mathdis{GEV}(\xi =0.1)$ ($F_7$), Gumbel ($F_8$) and $\mathdis{GEV}(\xi=-0.1)$ ($F_9$).}   \label{tab_nomerr_l2_bm} \end{table}
\begin{table}[t!]   \centering\footnotesize   \begin{tabular}{*{2}{l}*{12}{r}}     \toprule      & Parameter & \multicolumn{6}{c}{Quantile} & \multicolumn{6}{c}{$N$-obs. mean} \\     \cmidrule(lr){3-8} \cmidrule(lr){9-14}     $F$ & Method \textbar\ Error rate & \multicolumn{1}{c}{0.5} & \multicolumn{1}{c}{2.5} & \multicolumn{1}{c}{5} & \multicolumn{1}{c}{5} & \multicolumn{1}{c}{2.5} & \multicolumn{1}{c}{0.5} & \multicolumn{1}{c}{0.5} & \multicolumn{1}{c}{2.5} & \multicolumn{1}{c}{5} & \multicolumn{1}{c}{5} & \multicolumn{1}{c}{2.5} & \multicolumn{1}{c}{0.5} \\     \midrule     $F_1$ & Wald & $0.0$ & $0.0$ & $0.0$ & $23.0$ & $20.0$ & $15.5$ & $0.0$ & $0.0$ & $0.0$ & $24.0$ & $21.0$ & $17.0$ \\     & profile & $0.5$ & $2.0$ & $4.0$ & $6.5$ & $3.0$ & $0.5$ & $0.5$ & $2.0$ & $4.0$ & $6.5$ & $3.0$ & $0.5$ \\     & \textsc{tem} & $0.5$ & $2.5$ & $5.0$ & $3.5$ & $1.5$ & $0.0$ & $0.5$ & $2.5$ & $5.0$ & $3.5$ & $1.5$ & $0.0$ \\     & Severini (\textsc{tem}) & $0.0$ & $1.0$ & $3.0$ & $7.5$ & $3.5$ & $0.5$ & $0.0$ & $1.0$ & $3.0$ & $7.5$ & $3.5$ & $0.5$ \\     & Severini (cov.) & $0.5$ & $2.0$ & $4.5$ & $3.0$ & $1.0$ & $0.0$ & $0.5$ & $3.0$ & $6.0$ & $3.0$ & $1.0$ & $0.0$ \\ \addlinespace[3pt]     $F_2$ & Wald & $0.0$ & $0.0$ & $0.5$ & $20.0$ & $17.5$ & $13.5$ & $0.0$ & $0.0$ & $0.0$ & $20.0$ & $17.5$ & $13.5$ \\     & profile & $0.5$ & $2.0$ & $4.0$ & $6.0$ & $3.0$ & $0.5$ & $0.5$ & $2.5$ & $4.5$ & $5.5$ & $2.5$ & $0.5$ \\     & \textsc{tem} & $0.5$ & $2.5$ & $5.0$ & $3.5$ & $1.5$ & $0.0$ & $0.5$ & $2.5$ & $5.5$ & $3.0$ & $1.5$ & $0.0$ \\     & Severini (\textsc{tem}) & $0.0$ & $1.0$ & $2.5$ & $7.0$ & $3.5$ & $0.5$ & $0.0$ & $1.5$ & $3.0$ & $6.5$ & $3.0$ & $0.5$ \\     & Severini (cov.) & $0.0$ & $1.5$ & $3.5$ & $5.0$ & $2.5$ & $0.5$ & $0.5$ & $2.0$ & $5.0$ & $4.5$ & $2.0$ & $0.5$ \\ \addlinespace[3pt]     $F_3$ & Wald & $0.0$ & $0.0$ & $0.5$ & $21.5$ & $19.0$ & $15.0$ & $0.0$ & $0.0$ & $0.0$ & $21.5$ & $19.0$ & $15.5$ \\     & profile & $0.0$ & $1.0$ & $3.0$ & $7.5$ & $4.0$ & $1.0$ & $0.0$ & $1.0$ & $3.5$ & $7.0$ & $3.5$ & $0.5$ \\     & \textsc{tem} & $0.0$ & $1.5$ & $3.5$ & $5.0$ & $2.5$ & $0.0$ & $0.5$ & $1.5$ & $4.0$ & $4.5$ & $2.0$ & $0.0$ \\     & Severini (\textsc{tem}) & $0.0$ & $0.5$ & $1.5$ & $9.0$ & $5.0$ & $1.0$ & $0.0$ & $0.5$ & $2.0$ & $8.5$ & $4.5$ & $1.0$ \\     & Severini (cov.) & $0.0$ & $1.0$ & $2.5$ & $6.5$ & $3.5$ & $0.5$ & $0.0$ & $1.5$ & $3.5$ & $6.0$ & $3.0$ & $0.5$ \\ \addlinespace[3pt]     $F_4$ & Wald & $0.0$ & $0.0$ & $0.0$ & $28.5$ & $25.5$ & $20.5$ & $0.0$ & $0.0$ & $0.0$ & $29.0$ & $26.0$ & $21.5$ \\     & profile & $0.5$ & $1.5$ & $3.5$ & $7.0$ & $3.5$ & $0.5$ & $0.5$ & $1.5$ & $3.5$ & $6.5$ & $3.0$ & $0.5$ \\     & \textsc{tem} & $0.5$ & $2.5$ & $5.0$ & $3.0$ & $1.0$ & $0.0$ & $0.5$ & $2.5$ & $5.0$ & $3.0$ & $1.0$ & $0.0$ \\     & Severini (\textsc{tem}) & $0.0$ & $1.0$ & $2.0$ & $8.0$ & $3.5$ & $0.5$ & $0.0$ & $1.0$ & $2.0$ & $8.0$ & $3.5$ & $0.5$ \\     & Severini (cov.) & $0.0$ & $1.5$ & $3.5$ & $3.5$ & $1.5$ & $0.0$ & $0.5$ & $1.5$ & $4.0$ & $3.0$ & $1.0$ & $0.0$ \\ \addlinespace[3pt]     $F_5$ & Wald & $0.0$ & $0.0$ & $1.0$ & $18.0$ & $15.0$ & $11.5$ & $0.0$ & $0.0$ & $0.0$ & $18.5$ & $15.5$ & $12.0$ \\     & profile & $0.5$ & $2.5$ & $4.5$ & $6.0$ & $3.0$ & $0.5$ & $0.5$ & $2.5$ & $5.5$ & $5.5$ & $3.0$ & $0.5$ \\     & \textsc{tem} & $0.5$ & $2.5$ & $5.0$ & $4.5$ & $2.0$ & $0.5$ & $1.0$ & $3.0$ & $6.0$ & $4.0$ & $2.0$ & $0.5$ \\     & Severini (\textsc{tem}) & $0.0$ & $1.5$ & $3.0$ & $7.0$ & $4.0$ & $1.0$ & $0.5$ & $1.5$ & $4.0$ & $7.0$ & $3.5$ & $1.0$ \\     & Severini (cov.) & $0.5$ & $2.0$ & $4.0$ & $5.5$ & $3.0$ & $0.5$ & $0.5$ & $3.0$ & $6.0$ & $5.0$ & $2.5$ & $0.5$ \\ \addlinespace[3pt]     $F_6$ & Wald & $0.0$ & $0.0$ & $0.0$ & $25.5$ & $22.0$ & $17.5$ & $0.0$ & $0.0$ & $0.0$ & $27.0$ & $23.5$ & $19.0$ \\     & profile & $0.5$ & $2.0$ & $4.0$ & $8.5$ & $4.5$ & $1.0$ & $0.5$ & $2.0$ & $4.0$ & $8.0$ & $4.0$ & $1.0$ \\     & \textsc{tem} & $0.5$ & $3.0$ & $5.5$ & $5.0$ & $2.5$ & $0.5$ & $0.5$ & $3.0$ & $5.5$ & $5.0$ & $2.0$ & $0.0$ \\     & Severini (\textsc{tem}) & $0.0$ & $1.5$ & $3.0$ & $9.5$ & $5.0$ & $1.0$ & $0.0$ & $1.5$ & $3.0$ & $9.5$ & $5.0$ & $1.0$ \\     & Severini (cov.) & $0.5$ & $2.0$ & $4.0$ & $6.0$ & $3.0$ & $0.5$ & $0.5$ & $2.5$ & $4.5$ & $5.5$ & $2.5$ & $0.5$ \\ \addlinespace[3pt]     $F_7$ & Wald & $0.0$ & $0.0$ & $0.5$ & $22.0$ & $19.0$ & $15.0$ & $0.0$ & $0.0$ & $0.0$ & $22.5$ & $20.0$ & $15.5$ \\     & profile & $0.5$ & $2.0$ & $4.5$ & $7.5$ & $4.0$ & $0.5$ & $0.5$ & $2.0$ & $4.5$ & $7.0$ & $3.5$ & $0.5$ \\     & \textsc{tem} & $0.5$ & $2.5$ & $5.5$ & $5.0$ & $2.0$ & $0.5$ & $0.5$ & $2.5$ & $5.5$ & $4.5$ & $2.0$ & $0.5$ \\     & Severini (\textsc{tem}) & $0.0$ & $1.5$ & $3.0$ & $8.5$ & $4.5$ & $1.0$ & $0.0$ & $1.5$ & $3.0$ & $8.5$ & $4.5$ & $1.0$ \\     & Severini (cov.) & $0.5$ & $1.5$ & $4.0$ & $6.5$ & $3.0$ & $0.5$ & $0.5$ & $2.5$ & $5.0$ & $6.0$ & $3.0$ & $0.5$ \\ \addlinespace[3pt]     $F_8$ & Wald & $0.0$ & $0.0$ & $0.0$ & $23.5$ & $20.5$ & $16.5$ & $0.0$ & $0.0$ & $0.0$ & $24.0$ & $21.0$ & $17.0$ \\     & profile & $0.5$ & $2.0$ & $4.0$ & $7.0$ & $3.5$ & $0.5$ & $0.5$ & $2.0$ & $4.0$ & $6.5$ & $3.0$ & $0.5$ \\     & \textsc{tem} & $0.5$ & $2.5$ & $5.5$ & $4.0$ & $1.5$ & $0.0$ & $0.5$ & $2.5$ & $5.5$ & $3.5$ & $1.5$ & $0.0$ \\     & Severini (\textsc{tem}) & $0.0$ & $1.0$ & $2.5$ & $8.0$ & $4.0$ & $0.5$ & $0.0$ & $1.0$ & $2.5$ & $7.5$ & $3.5$ & $0.5$ \\     & Severini (cov.) & $0.0$ & $1.5$ & $3.5$ & $5.0$ & $2.5$ & $0.5$ & $0.5$ & $2.0$ & $4.5$ & $4.5$ & $2.0$ & $0.0$ \\ \addlinespace[3pt]     $F_9$ & Wald & $0.0$ & $0.0$ & $0.0$ & $27.0$ & $24.0$ & $19.0$ & $0.0$ & $0.0$ & $0.0$ & $27.0$ & $24.5$ & $20.0$ \\     & profile & $0.0$ & $1.5$ & $3.0$ & $6.5$ & $3.0$ & $0.5$ & $0.0$ & $1.5$ & $3.5$ & $6.0$ & $2.5$ & $0.0$ \\     & \textsc{tem} & $0.5$ & $2.5$ & $5.0$ & $2.5$ & $1.0$ & $0.0$ & $0.5$ & $2.5$ & $5.0$ & $2.5$ & $0.5$ & $0.0$ \\     & Severini (\textsc{tem}) & $0.0$ & $0.5$ & $1.5$ & $7.5$ & $3.5$ & $0.5$ & $0.0$ & $0.5$ & $2.0$ & $6.5$ & $3.0$ & $0.0$ \\     & Severini (cov.) & $0.0$ & $1.0$ & $3.0$ & $4.0$ & $1.5$ & $0.0$ & $0.0$ & $1.5$ & $3.5$ & $3.0$ & $1.0$ & $0.0$ \\     \bottomrule   \end{tabular}   \caption{One-sided empirical error rates (\%)  for lower (first to third columns) and upper (fourth to sixth columns) confidence intervals, block maximum method with $m=90, k=20$.  
The distributions (from top to bottom) are Burr ($F_1$), Weibull ($F_2$), generalized gamma ($F_3$), normal ($F_4$), lognormal ($F_5$), Student $t$ ($F_6$), $\mathdis{GEV}(\xi =0.1)$ ($F_7$), Gumbel ($F_8$) and $\mathdis{GEV}(\xi=-0.1)$ ($F_9$).}   \label{tab_nomerr_l3_bm} \end{table}
\begin{table}[t!]   \centering\footnotesize   \begin{tabular}{*{2}{l}*{12}{r}}     \toprule      & Parameter & \multicolumn{6}{c}{Quantile} & \multicolumn{6}{c}{$N$-obs. mean} \\     \cmidrule(lr){3-8} \cmidrule(lr){9-14}     $F$ & Method \textbar\ Error rate & \multicolumn{1}{c}{0.5} & \multicolumn{1}{c}{2.5} & \multicolumn{1}{c}{5} & \multicolumn{1}{c}{5} & \multicolumn{1}{c}{2.5} & \multicolumn{1}{c}{0.5} & \multicolumn{1}{c}{0.5} & \multicolumn{1}{c}{2.5} & \multicolumn{1}{c}{5} & \multicolumn{1}{c}{5} & \multicolumn{1}{c}{2.5} & \multicolumn{1}{c}{0.5} \\     \midrule     $F_1$ & Wald & $0.0$ & $0.0$ & $0.0$ & $29.0$ & $26.0$ & $20.5$ & $0.0$ & $0.0$ & $0.0$ & $31.5$ & $28.5$ & $23.5$ \\     & profile & $0.5$ & $1.5$ & $3.0$ & $5.5$ & $1.5$ & $0.0$ & $0.5$ & $1.5$ & $3.5$ & $5.0$ & $1.0$ & $0.0$ \\     & \textsc{tem} & $1.0$ & $3.5$ & $6.5$ & $0.5$ & $0.0$ & $0.0$ & $0.5$ & $3.5$ & $7.5$ & $0.5$ & $0.0$ & $0.0$ \\     & Severini (\textsc{tem}) & $0.0$ & $1.0$ & $2.5$ & $3.0$ & $0.5$ & $0.0$ & $0.0$ & $1.0$ & $2.5$ & $3.0$ & $0.5$ & $0.0$ \\     & Severini (cov.) & $0.0$ & $1.0$ & $2.5$ & $4.0$ & $1.0$ & $0.0$ & $0.0$ & $1.5$ & $3.0$ & $3.0$ & $1.0$ & $0.5$ \\ \addlinespace[3pt]     $F_2$ & Wald & $0.0$ & $0.0$ & $0.0$ & $28.0$ & $24.5$ & $19.5$ & $0.0$ & $0.0$ & $0.0$ & $29.5$ & $26.0$ & $21.5$ \\     & profile & $0.5$ & $1.5$ & $3.0$ & $6.0$ & $2.0$ & $0.0$ & $0.5$ & $1.5$ & $3.0$ & $4.5$ & $1.0$ & $0.0$ \\     & \textsc{tem} & $0.5$ & $3.0$ & $6.0$ & $0.5$ & $0.0$ & $0.0$ & $0.5$ & $3.5$ & $7.0$ & $0.0$ & $0.0$ & $0.0$ \\     & Severini (\textsc{tem}) & $0.0$ & $1.0$ & $2.0$ & $3.0$ & $1.0$ & $0.0$ & $0.0$ & $1.0$ & $2.5$ & $2.5$ & $0.5$ & $0.0$ \\     & Severini (cov.) & $0.0$ & $1.0$ & $2.0$ & $3.5$ & $1.0$ & $0.0$ & $0.0$ & $1.0$ & $3.0$ & $2.5$ & $1.0$ & $0.5$ \\ \addlinespace[3pt]     $F_3$ & Wald & $0.0$ & $0.0$ & $0.0$ & $28.0$ & $25.5$ & $21.0$ & $0.0$ & $0.0$ & $0.0$ & $29.5$ & $26.5$ & $22.5$ \\     & profile & $0.0$ & $0.5$ & $2.0$ & $7.0$ & $2.5$ & $0.0$ & $0.0$ & $1.0$ & $2.5$ & $5.5$ & $2.0$ & $0.0$ \\     & \textsc{tem} & $0.5$ & $2.0$ & $4.5$ & $1.0$ & $0.0$ & $0.0$ & $0.5$ & $2.5$ & $5.5$ & $0.5$ & $0.0$ & $0.0$ \\     & Severini (\textsc{tem}) & $0.0$ & $0.5$ & $1.5$ & $4.0$ & $1.0$ & $0.0$ & $0.0$ & $0.5$ & $2.0$ & $3.5$ & $1.0$ & $0.0$ \\     & Severini (cov.) & $0.0$ & $0.5$ & $1.5$ & $4.5$ & $1.5$ & $0.0$ & $0.0$ & $1.0$ & $2.5$ & $3.5$ & $1.5$ & $0.5$ \\ \addlinespace[3pt]     $F_4$ & Wald & $0.0$ & $0.0$ & $0.0$ & $29.0$ & $25.5$ & $20.0$ & $0.0$ & $0.0$ & $0.0$ & $31.0$ & $28.0$ & $23.0$ \\     & profile & $0.5$ & $1.5$ & $3.0$ & $3.0$ & $0.5$ & $0.0$ & $0.5$ & $1.5$ & $3.0$ & $2.0$ & $0.5$ & $0.0$ \\     & \textsc{tem} & $1.0$ & $3.5$ & $7.0$ & $0.0$ & $0.0$ & $0.0$ & $1.0$ & $3.5$ & $6.5$ & $0.0$ & $0.0$ & $0.0$ \\     & Severini (\textsc{tem}) & $0.0$ & $1.0$ & $2.0$ & $1.5$ & $0.0$ & $0.0$ & $0.0$ & $1.0$ & $2.0$ & $1.0$ & $0.0$ & $0.0$ \\     & Severini (cov.) & $0.0$ & $0.5$ & $2.0$ & $2.5$ & $1.0$ & $0.0$ & $0.0$ & $1.0$ & $2.0$ & $2.0$ & $1.0$ & $0.5$ \\ \addlinespace[3pt]     $F_5$ & Wald & $0.0$ & $0.0$ & $0.0$ & $28.5$ & $25.0$ & $20.5$ & $0.0$ & $0.0$ & $0.0$ & $29.5$ & $26.5$ & $22.0$ \\     & profile & $0.0$ & $1.5$ & $3.0$ & $8.0$ & $3.5$ & $0.0$ & $0.5$ & $2.0$ & $4.0$ & $7.0$ & $3.0$ & $0.0$ \\     & \textsc{tem} & $0.5$ & $3.0$ & $6.0$ & $1.5$ & $0.0$ & $0.0$ & $1.0$ & $4.0$ & $7.5$ & $1.0$ & $0.0$ & $0.0$ \\     & Severini (\textsc{tem}) & $0.0$ & $1.5$ & $3.0$ & $5.5$ & $2.0$ & $0.0$ & $0.0$ & $1.5$ & $3.5$ & $4.5$ & $1.5$ & $0.0$ \\     & Severini (cov.) & $0.0$ & $1.5$ & $3.0$ & $5.5$ & $2.0$ & $0.0$ & $0.5$ & $2.0$ & $4.0$ & $4.5$ & $1.5$ & $0.5$ \\ \addlinespace[3pt]     $F_6$ & Wald & $0.0$ & $0.0$ & $0.0$ & $29.5$ & $26.5$ & $21.0$ & $0.0$ & $0.0$ & $0.0$ & $32.0$ & $28.5$ & $24.0$ \\     & profile & $0.5$ & $1.5$ & $3.5$ & $5.5$ & $1.5$ & $0.0$ & $0.5$ & $2.0$ & $3.5$ & $5.0$ & $1.0$ & $0.0$ \\     & \textsc{tem} & $1.0$ & $4.0$ & $7.5$ & $0.5$ & $0.0$ & $0.0$ & $1.0$ & $4.0$ & $7.5$ & $0.5$ & $0.0$ & $0.0$ \\     & Severini (\textsc{tem}) & $0.0$ & $1.5$ & $3.0$ & $3.0$ & $0.5$ & $0.0$ & $0.0$ & $1.0$ & $3.0$ & $3.0$ & $0.5$ & $0.0$ \\     & Severini (cov.) & $0.0$ & $1.0$ & $2.5$ & $4.0$ & $1.5$ & $0.0$ & $0.5$ & $1.5$ & $3.0$ & $3.5$ & $1.5$ & $0.5$ \\ \addlinespace[3pt]     $F_7$ & Wald & $0.0$ & $0.0$ & $0.0$ & $29.0$ & $25.5$ & $20.5$ & $0.0$ & $0.0$ & $0.0$ & $30.5$ & $27.5$ & $22.5$ \\     & profile & $0.5$ & $1.5$ & $3.0$ & $6.0$ & $2.0$ & $0.0$ & $0.5$ & $1.5$ & $3.5$ & $5.5$ & $1.5$ & $0.0$ \\     & \textsc{tem} & $1.0$ & $3.5$ & $6.5$ & $0.5$ & $0.0$ & $0.0$ & $1.0$ & $3.5$ & $7.5$ & $0.5$ & $0.0$ & $0.0$ \\     & Severini (\textsc{tem}) & $0.0$ & $1.0$ & $2.5$ & $4.0$ & $1.0$ & $0.0$ & $0.0$ & $1.0$ & $3.0$ & $3.5$ & $1.0$ & $0.0$ \\     & Severini (cov.) & $0.0$ & $1.0$ & $2.5$ & $4.5$ & $1.5$ & $0.0$ & $0.0$ & $1.5$ & $3.0$ & $3.5$ & $1.5$ & $0.5$ \\ \addlinespace[3pt]     $F_8$ & Wald & $0.0$ & $0.0$ & $0.0$ & $27.5$ & $24.0$ & $19.5$ & $0.0$ & $0.0$ & $0.0$ & $29.0$ & $26.0$ & $21.5$ \\     & profile & $0.5$ & $1.5$ & $3.5$ & $4.0$ & $1.0$ & $0.0$ & $0.5$ & $1.5$ & $3.5$ & $3.0$ & $0.5$ & $0.0$ \\     & \textsc{tem} & $0.5$ & $3.5$ & $6.5$ & $0.0$ & $0.0$ & $0.0$ & $0.5$ & $3.5$ & $7.0$ & $0.0$ & $0.0$ & $0.0$ \\     & Severini (\textsc{tem}) & $0.0$ & $1.0$ & $2.0$ & $2.0$ & $0.5$ & $0.0$ & $0.0$ & $1.0$ & $2.5$ & $2.0$ & $0.5$ & $0.0$ \\     & Severini (cov.) & $0.0$ & $0.5$ & $2.0$ & $3.0$ & $1.0$ & $0.0$ & $0.0$ & $1.0$ & $3.0$ & $2.5$ & $1.0$ & $0.5$ \\ \addlinespace[3pt]     $F_9$ & Wald & $0.0$ & $0.0$ & $0.0$ & $28.0$ & $24.0$ & $18.5$ & $0.0$ & $0.0$ & $0.0$ & $29.5$ & $26.5$ & $21.0$ \\     & profile & $0.5$ & $1.0$ & $2.5$ & $2.5$ & $0.5$ & $0.0$ & $0.5$ & $1.5$ & $2.5$ & $1.5$ & $0.0$ & $0.0$ \\     & \textsc{tem} & $1.0$ & $3.5$ & $6.5$ & $0.0$ & $0.0$ & $0.0$ & $0.5$ & $3.0$ & $6.5$ & $0.0$ & $0.0$ & $0.0$ \\     & Severini (\textsc{tem}) & $0.0$ & $0.5$ & $1.5$ & $1.0$ & $0.0$ & $0.0$ & $0.0$ & $0.5$ & $1.0$ & $0.5$ & $0.0$ & $0.0$ \\     & Severini (cov.) & $0.0$ & $0.5$ & $1.5$ & $2.0$ & $0.5$ & $0.0$ & $0.0$ & $1.0$ & $2.0$ & $1.5$ & $1.0$ & $0.5$ \\     \bottomrule   \end{tabular}   \caption{One-sided empirical error rates (\%) for lower (first to third columns) and upper (fourth to sixth columns) confidence limits, peaks-over-threshold method with $k=20$.  
The distributions (from top to bottom) are Burr ($F_1$), Weibull ($F_2$), generalized gamma ($F_3$), normal ($F_4$), lognormal ($F_5$), Student $t$ ($F_6$), $\mathdis{GP}(\xi =0.1)$ ($F_7$), exponential ($F_8$) and $\mathdis{GP}(\xi=-0.1)$ ($F_9$).}   \label{tab_nomerr_l3_pot} \end{table}
\begin{table}[t!]   \centering\footnotesize   \begin{tabular}{*{2}{l}*{12}{r}}     \toprule      & Parameter & \multicolumn{6}{c}{Quantile} & \multicolumn{6}{c}{$N$-obs. mean} \\     \cmidrule(lr){3-8} \cmidrule(lr){9-14}     $F$ & Method \textbar\ Error rate & \multicolumn{1}{c}{0.5} & \multicolumn{1}{c}{2.5} & \multicolumn{1}{c}{5} & \multicolumn{1}{c}{5} & \multicolumn{1}{c}{2.5} & \multicolumn{1}{c}{0.5} & \multicolumn{1}{c}{0.5} & \multicolumn{1}{c}{2.5} & \multicolumn{1}{c}{5} & \multicolumn{1}{c}{5} & \multicolumn{1}{c}{2.5} & \multicolumn{1}{c}{0.5} \\     \midrule     $F_1$ & Wald & $0.0$ & $0.0$ & $0.0$ & $24.0$ & $20.0$ & $14.5$ & $0.0$ & $0.0$ & $0.0$ & $25.5$ & $22.0$ & $16.5$ \\     & profile & $0.5$ & $1.5$ & $3.5$ & $9.5$ & $4.5$ & $0.5$ & $0.5$ & $1.5$ & $3.0$ & $10.0$ & $4.5$ & $0.5$ \\     & \textsc{tem} & $0.5$ & $3.0$ & $5.5$ & $4.5$ & $1.5$ & $0.0$ & $0.5$ & $2.5$ & $5.5$ & $5.0$ & $2.0$ & $0.0$ \\     & Severini (\textsc{tem}) & $0.5$ & $1.5$ & $3.5$ & $8.0$ & $3.5$ & $0.5$ & $0.5$ & $1.5$ & $3.5$ & $8.5$ & $4.0$ & $0.5$ \\     & Severini (cov.) & $0.5$ & $1.5$ & $3.5$ & $8.0$ & $3.5$ & $0.5$ & $0.5$ & $1.5$ & $3.5$ & $8.5$ & $4.0$ & $0.5$ \\ \addlinespace[3pt]     $F_2$ & Wald & $0.0$ & $0.0$ & $0.0$ & $19.5$ & $16.5$ & $11.5$ & $0.0$ & $0.0$ & $0.0$ & $19.5$ & $16.5$ & $12.0$ \\     & profile & $0.5$ & $1.5$ & $3.5$ & $7.5$ & $4.0$ & $1.0$ & $0.5$ & $2.0$ & $3.5$ & $7.5$ & $4.0$ & $1.0$ \\     & \textsc{tem} & $0.5$ & $3.0$ & $5.5$ & $4.0$ & $2.0$ & $0.5$ & $0.5$ & $3.0$ & $6.0$ & $4.0$ & $1.5$ & $0.5$ \\     & Severini (\textsc{tem}) & $0.5$ & $2.0$ & $3.5$ & $6.5$ & $3.5$ & $0.5$ & $0.5$ & $2.0$ & $4.0$ & $6.0$ & $3.0$ & $0.5$ \\     & Severini (cov.) & $0.5$ & $2.0$ & $3.5$ & $6.5$ & $3.5$ & $0.5$ & $0.5$ & $2.0$ & $4.0$ & $6.0$ & $3.0$ & $0.5$ \\ \addlinespace[3pt]     $F_3$ & Wald & $0.0$ & $0.0$ & $0.0$ & $21.0$ & $18.0$ & $13.0$ & $0.0$ & $0.0$ & $0.0$ & $21.0$ & $18.0$ & $13.5$ \\     & profile & $0.0$ & $1.0$ & $2.0$ & $9.0$ & $5.5$ & $1.5$ & $0.0$ & $1.0$ & $2.5$ & $9.0$ & $5.0$ & $1.0$ \\     & \textsc{tem} & $0.0$ & $2.0$ & $4.0$ & $5.0$ & $2.5$ & $0.5$ & $0.5$ & $2.0$ & $5.0$ & $5.0$ & $2.5$ & $0.5$ \\     & Severini (\textsc{tem}) & $0.0$ & $1.0$ & $2.5$ & $8.0$ & $4.5$ & $1.0$ & $0.0$ & $1.0$ & $3.0$ & $8.0$ & $4.5$ & $1.0$ \\     & Severini (cov.) & $0.0$ & $1.0$ & $2.5$ & $8.0$ & $4.5$ & $1.0$ & $0.0$ & $1.5$ & $3.0$ & $8.0$ & $4.5$ & $1.0$ \\ \addlinespace[3pt]     $F_4$ & Wald & $0.0$ & $0.0$ & $0.0$ & $30.0$ & $26.5$ & $20.5$ & $0.0$ & $0.0$ & $0.0$ & $32.0$ & $28.0$ & $22.5$ \\     & profile & $0.5$ & $1.5$ & $3.0$ & $11.0$ & $6.0$ & $1.0$ & $0.5$ & $1.0$ & $2.5$ & $11.5$ & $6.0$ & $1.0$ \\     & \textsc{tem} & $0.5$ & $3.0$ & $5.0$ & $4.5$ & $2.0$ & $0.0$ & $0.5$ & $2.5$ & $4.5$ & $5.0$ & $2.0$ & $0.0$ \\     & Severini (\textsc{tem}) & $0.0$ & $1.0$ & $2.5$ & $9.5$ & $4.5$ & $0.5$ & $0.0$ & $1.0$ & $2.0$ & $9.5$ & $5.0$ & $0.5$ \\     & Severini (cov.) & $0.0$ & $1.0$ & $2.5$ & $9.0$ & $4.5$ & $0.5$ & $0.0$ & $1.0$ & $2.5$ & $9.5$ & $4.5$ & $0.5$ \\ \addlinespace[3pt]     $F_5$ & Wald & $0.0$ & $0.0$ & $0.5$ & $17.0$ & $14.0$ & $9.5$ & $0.0$ & $0.0$ & $0.0$ & $17.5$ & $14.5$ & $10.0$ \\     & profile & $0.5$ & $2.0$ & $3.5$ & $7.5$ & $4.0$ & $1.0$ & $0.5$ & $2.0$ & $4.0$ & $7.0$ & $4.0$ & $1.0$ \\     & \textsc{tem} & $0.5$ & $3.0$ & $5.5$ & $4.5$ & $2.0$ & $0.5$ & $1.0$ & $3.5$ & $7.0$ & $4.5$ & $2.0$ & $0.5$ \\     & Severini (\textsc{tem}) & $0.5$ & $2.0$ & $4.0$ & $6.5$ & $3.5$ & $1.0$ & $0.5$ & $2.5$ & $5.0$ & $6.5$ & $3.5$ & $0.5$ \\     & Severini (cov.) & $0.5$ & $2.0$ & $4.0$ & $6.5$ & $3.5$ & $1.0$ & $0.5$ & $2.5$ & $5.0$ & $6.5$ & $3.5$ & $0.5$ \\ \addlinespace[3pt]     $F_6$ & Wald & $0.0$ & $0.0$ & $0.0$ & $27.0$ & $23.5$ & $18.0$ & $0.0$ & $0.0$ & $0.0$ & $29.0$ & $25.0$ & $19.5$ \\     & profile & $0.5$ & $2.0$ & $3.5$ & $12.5$ & $7.0$ & $2.0$ & $0.5$ & $1.5$ & $3.0$ & $13.5$ & $7.5$ & $2.0$ \\     & \textsc{tem} & $0.5$ & $3.0$ & $5.5$ & $7.0$ & $3.5$ & $0.5$ & $0.5$ & $2.5$ & $5.5$ & $7.5$ & $4.0$ & $0.5$ \\     & Severini (\textsc{tem}) & $0.5$ & $1.5$ & $3.5$ & $11.0$ & $6.5$ & $1.5$ & $0.5$ & $1.5$ & $3.5$ & $11.5$ & $6.5$ & $1.5$ \\     & Severini (cov.) & $0.5$ & $1.5$ & $3.5$ & $11.0$ & $6.5$ & $1.5$ & $0.5$ & $1.5$ & $3.5$ & $11.5$ & $6.5$ & $1.5$ \\ \addlinespace[3pt]     $F_7$ & Wald & $0.0$ & $0.0$ & $0.0$ & $21.5$ & $18.0$ & $13.0$ & $0.0$ & $0.0$ & $0.0$ & $22.5$ & $19.0$ & $14.0$ \\     & profile & $0.5$ & $1.5$ & $3.5$ & $9.5$ & $5.5$ & $1.5$ & $0.5$ & $1.5$ & $3.5$ & $10.0$ & $5.5$ & $1.5$ \\     & \textsc{tem} & $0.5$ & $3.0$ & $5.5$ & $5.5$ & $3.0$ & $0.5$ & $0.5$ & $3.0$ & $6.0$ & $5.5$ & $3.0$ & $0.5$ \\     & Severini (\textsc{tem}) & $0.5$ & $2.0$ & $4.0$ & $8.5$ & $5.0$ & $1.0$ & $0.5$ & $2.0$ & $4.0$ & $8.5$ & $5.0$ & $1.0$ \\     & Severini (cov.) & $0.5$ & $2.0$ & $4.0$ & $8.5$ & $5.0$ & $1.0$ & $0.5$ & $2.0$ & $4.0$ & $8.5$ & $4.5$ & $1.0$ \\ \addlinespace[3pt]     $F_8$ & Wald & $0.0$ & $0.0$ & $0.0$ & $23.0$ & $19.5$ & $14.5$ & $0.0$ & $0.0$ & $0.0$ & $23.5$ & $20.5$ & $15.5$ \\     & profile & $0.5$ & $1.5$ & $3.5$ & $9.0$ & $5.0$ & $1.0$ & $0.5$ & $1.5$ & $3.5$ & $9.0$ & $5.0$ & $1.0$ \\     & \textsc{tem} & $0.5$ & $3.0$ & $5.5$ & $4.5$ & $2.0$ & $0.5$ & $0.5$ & $3.0$ & $5.5$ & $4.5$ & $2.0$ & $0.5$ \\     & Severini (\textsc{tem}) & $0.0$ & $1.5$ & $3.5$ & $7.5$ & $4.0$ & $1.0$ & $0.0$ & $1.5$ & $3.5$ & $7.5$ & $4.0$ & $1.0$ \\     & Severini (cov.) & $0.5$ & $1.5$ & $3.5$ & $7.5$ & $4.0$ & $1.0$ & $0.0$ & $1.5$ & $3.5$ & $7.5$ & $4.0$ & $0.5$ \\ \addlinespace[3pt]     $F_9$ & Wald & $0.0$ & $0.0$ & $0.0$ & $27.5$ & $23.5$ & $18.0$ & $0.0$ & $0.0$ & $0.0$ & $28.0$ & $24.5$ & $19.0$ \\     & profile & $0.5$ & $1.5$ & $2.5$ & $9.5$ & $5.0$ & $0.5$ & $0.5$ & $1.0$ & $2.5$ & $9.5$ & $5.0$ & $0.5$ \\     & \textsc{tem} & $0.5$ & $2.5$ & $5.0$ & $4.0$ & $1.5$ & $0.0$ & $0.5$ & $2.5$ & $5.0$ & $3.5$ & $1.5$ & $0.0$ \\     & Severini (\textsc{tem}) & $0.0$ & $1.0$ & $2.5$ & $8.0$ & $4.0$ & $0.5$ & $0.0$ & $1.0$ & $2.5$ & $8.0$ & $3.5$ & $0.5$ \\     & Severini (cov.) & $0.0$ & $1.0$ & $2.5$ & $8.0$ & $3.5$ & $0.5$ & $0.0$ & $1.0$ & $2.5$ & $8.0$ & $3.5$ & $0.5$ \\     \bottomrule   \end{tabular}   \caption{One-sided empirical error rates (\%) for lower (first to third columns) and upper (fourth to sixth columns) confidence limits, peaks-over-threshold method with $k=60$.  
The distributions (from top to bottom) are Burr ($F_1$), Weibull ($F_2$), generalized gamma ($F_3$), normal ($F_4$), lognormal ($F_5$), Student $t$ ($F_6$), $\mathdis{GP}(\xi =0.1)$ ($F_7$), exponential ($F_8$) and $\mathdis{GP}(\xi=-0.1)$ ($F_9$).}   \label{tab_nomerr_l1_pot} \end{table}

\begin{table}[t!]   \centering\small   \begin{tabular}{*{2}{l}*{9}{r}}     \toprule     Parameter & Method \textbar\ $F$ & \multicolumn{1}{c}{$F_1$} & \multicolumn{1}{c}{$F_2$} & \multicolumn{1}{c}{$F_3$} & \multicolumn{1}{c}{$F_4$} & \multicolumn{1}{c}{$F_5$} & \multicolumn{1}{c}{$F_6$} & \multicolumn{1}{c}{$F_7$} & \multicolumn{1}{c}{$F_8$} & \multicolumn{1}{c}{$F_9$} \\     \midrule     Quantile & \textsc{mle} & $-1$ & $5$ & $2$ & $-3$ & $9$ & $-3$ & $1$ & $0$ & $-2$ \\     & \textsc{tem} & $1$ & $7$ & $5$ & $-1$ & $10$ & $-1$ & $3$ & $2$ & $0$ \\     & Severini (\textsc{tem}) & $-2$ & $3$ & $0$ & $-3$ & $4$ & $-4$ & $-1$ & $-1$ & $-2$ \\     & Severini (cov.) & $0$ & $5$ & $2$ & $-2$ & $7$ & $-3$ & $0$ & $0$ & $-1$ \\ \addlinespace[3pt]     $N$-obs. median & \textsc{mle} & $-1$ & $7$ & $4$ & $-3$ & $11$ & $-4$ & $1$ & $0$ & $-2$ \\     & \textsc{tem} & $1$ & $9$ & $7$ & $-1$ & $13$ & $-1$ & $3$ & $2$ & $1$ \\     & Severini (\textsc{tem}) & $-2$ & $4$ & $1$ & $-4$ & $6$ & $-5$ & $-1$ & $-1$ & $-2$ \\     & Severini (cov.) & $1$ & $6$ & $3$ & $-2$ & $9$ & $-3$ & $1$ & $0$ & $-1$ \\ \addlinespace[3pt]     $N$-obs. mean & \textsc{mle} & $-1$ & $10$ & $7$ & $-3$ & $19$ & $-4$ & $2$ & $0$ & $-2$ \\     & \textsc{tem} & $1$ & $13$ & $10$ & $-1$ & $20$ & $-1$ & $5$ & $3$ & $1$ \\     & Severini (\textsc{tem}) & $-2$ & $6$ & $3$ & $-4$ & $12$ & $-5$ & $0$ & $-1$ & $-2$ \\     & Severini (cov.) & $1$ & $10$ & $6$ & $-2$ & $17$ & $-4$ & $2$ & $1$ & $-1$ \\     \bottomrule   \end{tabular}   \caption{Truncated mean ($\alpha=0.1$) of relative bias (in \%), block maximum method with $m=45, k=40$. The largest standard error, obtained using a nonparametric bootstrap, is 0.51\%. The distributions (from top to bottom) are Burr ($F_1$), Weibull ($F_2$), generalized gamma ($F_3$), normal ($F_4$), lognormal ($F_5$), Student $t$ ($F_6$), $\mathdis{GEV}(\xi =0.1)$ ($F_7$), Gumbel ($F_8$) and $\mathdis{GEV}(\xi=-0.1)$ ($F_9$).}   \label{tab_RB_l2_bm} \end{table}
\begin{table}[t!]   \centering\small   \begin{tabular}{*{2}{l}*{9}{r}}     \toprule     Parameter & Method \textbar\ $F$ & \multicolumn{1}{c}{$F_1$} & \multicolumn{1}{c}{$F_2$} & \multicolumn{1}{c}{$F_3$} & \multicolumn{1}{c}{$F_4$} & \multicolumn{1}{c}{$F_5$} & \multicolumn{1}{c}{$F_6$} & \multicolumn{1}{c}{$F_7$} & \multicolumn{1}{c}{$F_8$} & \multicolumn{1}{c}{$F_9$} \\     \midrule     Quantile & \textsc{mle} & $0$ & $3$ & $1$ & $-2$ & $7$ & $-2$ & $1$ & $0$ & $-2$ \\     & \textsc{tem} & $2$ & $6$ & $5$ & $1$ & $10$ & $2$ & $4$ & $3$ & $1$ \\     & Severini (\textsc{tem}) & $-2$ & $-1$ & $-3$ & $-3$ & $0$ & $-4$ & $-3$ & $-2$ & $-3$ \\     & Severini (cov.) & $3$ & $4$ & $2$ & $0$ & $8$ & $1$ & $2$ & $2$ & $0$ \\ \addlinespace[3pt]     $N$-obs. median & \textsc{mle} & $1$ & $4$ & $2$ & $-2$ & $10$ & $-1$ & $2$ & $0$ & $-2$ \\     & \textsc{tem} & $3$ & $8$ & $6$ & $1$ & $13$ & $2$ & $5$ & $4$ & $2$ \\     & Severini (\textsc{tem}) & $-2$ & $-1$ & $-2$ & $-3$ & $1$ & $-4$ & $-3$ & $-2$ & $-3$ \\     & Severini (cov.) & $4$ & $6$ & $4$ & $0$ & $11$ & $1$ & $4$ & $3$ & $1$ \\ \addlinespace[3pt]     $N$-obs. mean & \textsc{mle} & $2$ & $9$ & $6$ & $-2$ & $21$ & $-1$ & $5$ & $2$ & $-1$ \\     & \textsc{tem} & $5$ & $13$ & $10$ & $1$ & $23$ & $3$ & $8$ & $6$ & $2$ \\     & Severini (\textsc{tem}) & $-1$ & $2$ & $0$ & $-3$ & $7$ & $-4$ & $-1$ & $-1$ & $-3$ \\     & Severini (cov.) & $9$ & $12$ & $9$ & $1$ & $25$ & $3$ & $8$ & $5$ & $2$ \\     \bottomrule   \end{tabular}   \caption{Truncated mean ($\alpha=0.1$) of relative bias (in \%), block maximum method with $m=90, k=20$. The largest standard error, obtained using a nonparametric bootstrap, is 0.81\%. The distributions (from top to bottom) are Burr ($F_1$), Weibull ($F_2$), generalized gamma ($F_3$), normal ($F_4$), lognormal ($F_5$), Student $t$ ($F_6$), $\mathdis{GEV}(\xi =0.1)$ ($F_7$), Gumbel ($F_8$) and $\mathdis{GEV}(\xi=-0.1)$ ($F_9$).}   \label{tab_RB_l3_bm} \end{table}
\begin{table}[t!]   \centering\small   \begin{tabular}{*{2}{l}*{9}{r}}     \toprule     Parameter & Method \textbar\ $F$ & \multicolumn{1}{c}{$F_1$} & \multicolumn{1}{c}{$F_2$} & \multicolumn{1}{c}{$F_3$} & \multicolumn{1}{c}{$F_4$} & \multicolumn{1}{c}{$F_5$} & \multicolumn{1}{c}{$F_6$} & \multicolumn{1}{c}{$F_7$} & \multicolumn{1}{c}{$F_8$} & \multicolumn{1}{c}{$F_9$} \\     \midrule     Quantile & \textsc{mle} & $-3$ & $-4$ & $-5$ & $-2$ & $-6$ & $-4$ & $-4$ & $-3$ & $-2$ \\     & \textsc{tem} & $2$ & $3$ & $2$ & $1$ & $3$ & $2$ & $3$ & $3$ & $1$ \\     & Severini (\textsc{tem}) & $-1$ & $-1$ & $-2$ & $-1$ & $-2$ & $-1$ & $-1$ & $0$ & $-1$ \\     & Severini (cov.) & $-1$ & $-1$ & $-2$ & $-1$ & $-2$ & $-1$ & $-1$ & $-1$ & $-1$ \\ \addlinespace[3pt]     $N$-obs. median & \textsc{mle} & $-3$ & $-4$ & $-5$ & $-3$ & $-6$ & $-4$ & $-5$ & $-3$ & $-2$ \\     & \textsc{tem} & $2$ & $4$ & $3$ & $1$ & $5$ & $2$ & $3$ & $3$ & $2$ \\     & Severini (\textsc{tem}) & $0$ & $0$ & $-1$ & $-1$ & $-1$ & $-1$ & $-1$ & $0$ & $0$ \\     & Severini (cov.) & $0$ & $0$ & $-1$ & $-1$ & $0$ & $-1$ & $0$ & $0$ & $0$ \\ \addlinespace[3pt]     $N$-obs. mean & \textsc{mle} & $-3$ & $-3$ & $-4$ & $-3$ & $-3$ & $-4$ & $-4$ & $-2$ & $-2$ \\     & \textsc{tem} & $4$ & $8$ & $7$ & $2$ & $14$ & $4$ & $7$ & $5$ & $3$ \\     & Severini (\textsc{tem}) & $1$ & $3$ & $2$ & $-1$ & $6$ & $0$ & $2$ & $1$ & $0$ \\     & Severini (cov.) & $1$ & $4$ & $3$ & $0$ & $8$ & $1$ & $3$ & $2$ & $0$ \\     \bottomrule   \end{tabular}   \caption{Truncated mean ($\alpha=0.1$) of relative bias (in \%), peaks-over-threshold method with $k=20$. The largest standard error, obtained using a nonparametric bootstrap, is 0.77\%. The distributions (from top to bottom) are Burr ($F_1$), Weibull ($F_2$), generalized gamma ($F_3$), normal ($F_4$), lognormal ($F_5$), Student $t$ ($F_6$), $\mathdis{GP}(\xi =0.1)$ ($F_7$), exponential ($F_8$) and $\mathdis{GP}(\xi=-0.1)$ ($F_9$).}   \label{tab_RB_l3_pot} \end{table}
\begin{table}[t!]   \centering\small   \begin{tabular}{*{2}{l}*{9}{r}}     \toprule     Parameter & Method \textbar\ $F$ & \multicolumn{1}{c}{$F_1$} & \multicolumn{1}{c}{$F_2$} & \multicolumn{1}{c}{$F_3$} & \multicolumn{1}{c}{$F_4$} & \multicolumn{1}{c}{$F_5$} & \multicolumn{1}{c}{$F_6$} & \multicolumn{1}{c}{$F_7$} & \multicolumn{1}{c}{$F_8$} & \multicolumn{1}{c}{$F_9$} \\     \midrule     Quantile & \textsc{mle} & $-3$ & $-2$ & $-4$ & $-4$ & $-1$ & $-5$ & $-3$ & $-3$ & $-3$ \\     & \textsc{tem} & $-1$ & $2$ & $0$ & $-2$ & $4$ & $-3$ & $0$ & $0$ & $-1$ \\     & Severini (\textsc{tem}) & $-2$ & $0$ & $-2$ & $-3$ & $2$ & $-4$ & $-2$ & $-2$ & $-2$ \\     & Severini (cov.) & $-2$ & $0$ & $-2$ & $-3$ & $2$ & $-4$ & $-2$ & $-2$ & $-2$ \\ \addlinespace[3pt]     $N$-obs. median & \textsc{mle} & $-3$ & $-1$ & $-4$ & $-4$ & $0$ & $-6$ & $-4$ & $-3$ & $-3$ \\     & \textsc{tem} & $-1$ & $3$ & $0$ & $-2$ & $5$ & $-3$ & $0$ & $0$ & $-1$ \\     & Severini (\textsc{tem}) & $-2$ & $1$ & $-2$ & $-4$ & $3$ & $-5$ & $-2$ & $-2$ & $-3$ \\     & Severini (cov.) & $-2$ & $1$ & $-2$ & $-4$ & $3$ & $-5$ & $-2$ & $-2$ & $-3$ \\ \addlinespace[3pt]     $N$-obs. mean & \textsc{mle} & $-4$ & $0$ & $-3$ & $-5$ & $4$ & $-7$ & $-3$ & $-3$ & $-3$ \\     & \textsc{tem} & $-1$ & $5$ & $2$ & $-3$ & $11$ & $-4$ & $1$ & $0$ & $-1$ \\     & Severini (\textsc{tem}) & $-2$ & $3$ & $0$ & $-4$ & $8$ & $-5$ & $-1$ & $-1$ & $-3$ \\     & Severini (cov.) & $-2$ & $3$ & $0$ & $-4$ & $8$ & $-5$ & $-1$ & $-1$ & $-2$ \\     \bottomrule   \end{tabular}   \caption{Truncated mean ($\alpha=0.1$) of relative bias (in \%), peaks-over-threshold method with $k=60$. The largest standard error, obtained using a nonparametric bootstrap, is 0.49\%. The distributions (from top to bottom) are Burr ($F_1$), Weibull ($F_2$), generalized gamma ($F_3$), normal ($F_4$), lognormal ($F_5$), Student $t$ ($F_6$), $\mathdis{GP}(\xi =0.1)$ ($F_7$), exponential ($F_8$) and $\mathdis{GP}(\xi=-0.1)$ ($F_9$).}   \label{tab_RB_l1_pot} \end{table}

\begin{table}[t!]   \centering\small   \begin{tabular}{*{2}{l}*{9}{r}}     \toprule      & Parameter & \multicolumn{3}{c}{Quantile} & \multicolumn{3}{c}{$N$-obs. median} & \multicolumn{3}{c}{$N$-obs. mean} \\     \cmidrule(lr){3-5} \cmidrule(lr){6-8} \cmidrule(lr){9-11}     $F$ & Method \textbar\ Conf. level (\%) & \multicolumn{1}{c}{90} & \multicolumn{1}{c}{95} & \multicolumn{1}{c}{99} & \multicolumn{1}{c}{90} & \multicolumn{1}{c}{95} & \multicolumn{1}{c}{99} & \multicolumn{1}{c}{90} & \multicolumn{1}{c}{95} & \multicolumn{1}{c}{99} \\     \midrule     $F_1$ & Wald & $68$ & $59$ & $44$ & $67$ & $58$ & $42$ & $61$ & $52$ & $35$ \\     & \textsc{tem} & $103$ & $103$ & $103$ & $104$ & $103$ & $103$ & $104$ & $104$ & $104$ \\     & Severini (\textsc{tem}) & $93$ & $93$ & $92$ & $93$ & $93$ & $92$ & $92$ & $91$ & $91$ \\     & Severini (cov.) & $105$ & $105$ & $105$ & $106$ & $106$ & $105$ & $111$ & $111$ & $112$ \\ \addlinespace[3pt]     $F_2$ & Wald & $66$ & $58$ & $45$ & $65$ & $58$ & $44$ & $60$ & $52$ & $39$ \\     & \textsc{tem} & $102$ & $102$ & $102$ & $102$ & $102$ & $102$ & $102$ & $102$ & $102$ \\     & Severini (\textsc{tem}) & $92$ & $92$ & $92$ & $92$ & $92$ & $92$ & $91$ & $91$ & $91$ \\     & Severini (cov.) & $97$ & $96$ & $96$ & $97$ & $97$ & $96$ & $97$ & $97$ & $97$ \\ \addlinespace[3pt]     $F_3$ & Wald & $66$ & $58$ & $44$ & $65$ & $57$ & $44$ & $59$ & $51$ & $37$ \\     & \textsc{tem} & $102$ & $102$ & $102$ & $102$ & $102$ & $102$ & $102$ & $102$ & $102$ \\     & Severini (\textsc{tem}) & $92$ & $92$ & $92$ & $92$ & $92$ & $92$ & $91$ & $91$ & $91$ \\     & Severini (cov.) & $96$ & $96$ & $96$ & $96$ & $96$ & $96$ & $96$ & $96$ & $96$ \\ \addlinespace[3pt]     $F_4$ & Wald & $64$ & $58$ & $47$ & $64$ & $58$ & $46$ & $62$ & $55$ & $43$ \\     & \textsc{tem} & $108$ & $108$ & $107$ & $108$ & $108$ & $107$ & $108$ & $108$ & $108$ \\     & Severini (\textsc{tem}) & $97$ & $96$ & $96$ & $96$ & $96$ & $96$ & $96$ & $96$ & $96$ \\     & Severini (cov.) & $104$ & $104$ & $104$ & $104$ & $104$ & $103$ & $105$ & $105$ & $104$ \\ \addlinespace[3pt]     $F_5$ & Wald & $67$ & $58$ & $42$ & $66$ & $57$ & $40$ & $55$ & $44$ & $28$ \\     & \textsc{tem} & $99$ & $99$ & $98$ & $99$ & $99$ & $99$ & $98$ & $98$ & $97$ \\     & Severini (\textsc{tem}) & $90$ & $89$ & $88$ & $89$ & $89$ & $88$ & $86$ & $85$ & $85$ \\     & Severini (cov.) & $94$ & $94$ & $93$ & $94$ & $94$ & $94$ & $93$ & $93$ & $93$ \\ \addlinespace[3pt]     $F_6$ & Wald & $68$ & $60$ & $46$ & $67$ & $59$ & $44$ & $63$ & $54$ & $38$ \\     & \textsc{tem} & $104$ & $104$ & $104$ & $105$ & $105$ & $105$ & $105$ & $105$ & $105$ \\     & Severini (\textsc{tem}) & $94$ & $94$ & $93$ & $94$ & $94$ & $93$ & $93$ & $93$ & $92$ \\     & Severini (cov.) & $100$ & $99$ & $99$ & $100$ & $99$ & $99$ & $100$ & $100$ & $99$ \\ \addlinespace[3pt]     $F_7$ & Wald & $67$ & $59$ & $44$ & $66$ & $58$ & $43$ & $60$ & $51$ & $35$ \\     & \textsc{tem} & $102$ & $102$ & $102$ & $103$ & $102$ & $102$ & $103$ & $103$ & $102$ \\     & Severini (\textsc{tem}) & $92$ & $92$ & $91$ & $92$ & $92$ & $91$ & $91$ & $90$ & $90$ \\     & Severini (cov.) & $97$ & $96$ & $96$ & $96$ & $96$ & $96$ & $96$ & $96$ & $95$ \\ \addlinespace[3pt]     $F_8$ & Wald & $66$ & $59$ & $46$ & $65$ & $58$ & $46$ & $62$ & $54$ & $42$ \\     & \textsc{tem} & $105$ & $105$ & $104$ & $105$ & $105$ & $105$ & $105$ & $105$ & $105$ \\     & Severini (\textsc{tem}) & $94$ & $94$ & $94$ & $94$ & $94$ & $94$ & $93$ & $93$ & $94$ \\     & Severini (cov.) & $99$ & $99$ & $98$ & $99$ & $99$ & $98$ & $99$ & $99$ & $99$ \\ \addlinespace[3pt]     $F_9$ & Wald & $64$ & $59$ & $49$ & $64$ & $59$ & $49$ & $62$ & $57$ & $46$ \\     & \textsc{tem} & $107$ & $107$ & $106$ & $107$ & $107$ & $106$ & $108$ & $107$ & $106$ \\     & Severini (\textsc{tem}) & $96$ & $96$ & $96$ & $96$ & $96$ & $96$ & $96$ & $96$ & $95$ \\     & Severini (cov.) & $102$ & $102$ & $101$ & $102$ & $102$ & $101$ & $102$ & $102$ & $101$ \\     \bottomrule   \end{tabular}   \caption{Truncated mean ($\alpha=0.1$) of the ratio of the confidence interval width relative to the width of profile confidence interval (in \%), block maximum method with $m=45, k=40$. The largest standard error, obtained using a nonparametric bootstrap, is 0.13\%. The distributions (from top to bottom) are Burr ($F_1$), Weibull ($F_2$), generalized gamma ($F_3$), normal ($F_4$), lognormal ($F_5$), Student $t$ ($F_6$), $\mathdis{GEV}(\xi =0.1)$ ($F_7$), Gumbel ($F_8$) and $\mathdis{GEV}(\xi=-0.1)$ ($F_9$).}   \label{tab_CIW_l2_bm} \end{table}
\begin{table}[t!]   \centering\small   \begin{tabular}{*{2}{l}*{9}{r}}     \toprule      & Parameter & \multicolumn{3}{c}{Quantile} & \multicolumn{3}{c}{$N$-obs. median} & \multicolumn{3}{c}{$N$-obs. mean} \\     \cmidrule(lr){3-5} \cmidrule(lr){6-8} \cmidrule(lr){9-11}     $F$ & Method \textbar\ Conf. level (\%) & \multicolumn{1}{c}{90} & \multicolumn{1}{c}{95} & \multicolumn{1}{c}{99} & \multicolumn{1}{c}{90} & \multicolumn{1}{c}{95} & \multicolumn{1}{c}{99} & \multicolumn{1}{c}{90} & \multicolumn{1}{c}{95} & \multicolumn{1}{c}{99} \\     \midrule     $F_1$ & Wald & $46$ & $37$ & $22$ & $45$ & $36$ & $22$ & $37$ & $28$ & $15$ \\     & \textsc{tem} & $105$ & $105$ & $105$ & $106$ & $106$ & $106$ & $106$ & $106$ & $106$ \\     & Severini (\textsc{tem}) & $85$ & $84$ & $83$ & $85$ & $84$ & $83$ & $81$ & $80$ & $81$ \\     & Severini (cov.) & $117$ & $118$ & $118$ & $118$ & $118$ & $117$ & $149$ & $151$ & $149$ \\ \addlinespace[3pt]     $F_2$ & Wald & $47$ & $39$ & $26$ & $47$ & $38$ & $26$ & $40$ & $32$ & $22$ \\     & \textsc{tem} & $104$ & $105$ & $104$ & $105$ & $105$ & $104$ & $105$ & $105$ & $104$ \\     & Severini (\textsc{tem}) & $85$ & $85$ & $85$ & $85$ & $85$ & $86$ & $82$ & $83$ & $85$ \\     & Severini (cov.) & $102$ & $101$ & $100$ & $102$ & $102$ & $101$ & $103$ & $102$ & $101$ \\ \addlinespace[3pt]     $F_3$ & Wald & $47$ & $39$ & $25$ & $46$ & $38$ & $25$ & $39$ & $31$ & $20$ \\     & \textsc{tem} & $104$ & $105$ & $104$ & $105$ & $105$ & $104$ & $104$ & $104$ & $104$ \\     & Severini (\textsc{tem}) & $85$ & $85$ & $85$ & $85$ & $85$ & $85$ & $82$ & $82$ & $84$ \\     & Severini (cov.) & $101$ & $100$ & $99$ & $101$ & $101$ & $100$ & $101$ & $101$ & $100$ \\ \addlinespace[3pt]     $F_4$ & Wald & $47$ & $40$ & $29$ & $47$ & $40$ & $29$ & $43$ & $36$ & $25$ \\     & \textsc{tem} & $114$ & $114$ & $112$ & $115$ & $114$ & $111$ & $115$ & $114$ & $111$ \\     & Severini (\textsc{tem}) & $92$ & $92$ & $91$ & $92$ & $91$ & $91$ & $90$ & $89$ & $89$ \\     & Severini (cov.) & $116$ & $116$ & $113$ & $116$ & $115$ & $112$ & $122$ & $121$ & $118$ \\ \addlinespace[3pt]     $F_5$ & Wald & $46$ & $36$ & $21$ & $44$ & $34$ & $20$ & $31$ & $22$ & $12$ \\     & \textsc{tem} & $98$ & $98$ & $98$ & $98$ & $98$ & $98$ & $96$ & $96$ & $96$ \\     & Severini (\textsc{tem}) & $80$ & $78$ & $77$ & $79$ & $78$ & $78$ & $72$ & $72$ & $75$ \\     & Severini (cov.) & $98$ & $97$ & $97$ & $98$ & $98$ & $98$ & $101$ & $100$ & $101$ \\ \addlinespace[3pt]     $F_6$ & Wald & $47$ & $38$ & $24$ & $46$ & $37$ & $23$ & $39$ & $30$ & $17$ \\     & \textsc{tem} & $107$ & $107$ & $107$ & $108$ & $108$ & $108$ & $108$ & $108$ & $108$ \\     & Severini (\textsc{tem}) & $87$ & $86$ & $84$ & $86$ & $85$ & $85$ & $83$ & $82$ & $83$ \\     & Severini (cov.) & $108$ & $107$ & $107$ & $107$ & $107$ & $106$ & $113$ & $113$ & $112$ \\ \addlinespace[3pt]     $F_7$ & Wald & $47$ & $37$ & $23$ & $45$ & $36$ & $22$ & $37$ & $28$ & $16$ \\     & \textsc{tem} & $104$ & $104$ & $104$ & $104$ & $104$ & $104$ & $104$ & $104$ & $104$ \\     & Severini (\textsc{tem}) & $84$ & $83$ & $82$ & $84$ & $83$ & $83$ & $80$ & $79$ & $80$ \\     & Severini (cov.) & $101$ & $101$ & $100$ & $101$ & $101$ & $100$ & $103$ & $102$ & $101$ \\ \addlinespace[3pt]     $F_8$ & Wald & $48$ & $40$ & $28$ & $47$ & $39$ & $28$ & $42$ & $35$ & $24$ \\     & \textsc{tem} & $109$ & $109$ & $108$ & $109$ & $109$ & $108$ & $110$ & $109$ & $107$ \\     & Severini (\textsc{tem}) & $89$ & $88$ & $88$ & $88$ & $88$ & $88$ & $86$ & $86$ & $87$ \\     & Severini (cov.) & $106$ & $105$ & $104$ & $106$ & $105$ & $103$ & $108$ & $107$ & $105$ \\ \addlinespace[3pt]     $F_9$ & Wald & $49$ & $42$ & $32$ & $48$ & $42$ & $32$ & $45$ & $39$ & $30$ \\     & \textsc{tem} & $114$ & $113$ & $110$ & $114$ & $112$ & $109$ & $114$ & $112$ & $108$ \\     & Severini (\textsc{tem}) & $92$ & $92$ & $91$ & $92$ & $91$ & $91$ & $90$ & $90$ & $89$ \\     & Severini (cov.) & $112$ & $110$ & $108$ & $111$ & $109$ & $107$ & $113$ & $112$ & $108$ \\     \bottomrule   \end{tabular}   \caption{Truncated mean ($\alpha=0.1$) of the ratio of the confidence interval width relative to the width of profile confidence interval (in \%), block maximum method with $m=90, k=20$. The largest standard error, obtained using a nonparametric bootstrap, is 0.34\%. The distributions (from top to bottom) are Burr ($F_1$), Weibull ($F_2$), generalized gamma ($F_3$), normal ($F_4$), lognormal ($F_5$), Student $t$ ($F_6$), $\mathdis{GEV}(\xi =0.1)$ ($F_7$), Gumbel ($F_8$) and $\mathdis{GEV}(\xi=-0.1)$ ($F_9$).}   \label{tab_CIW_l3_bm} \end{table}
\begin{table}[t!]   \centering\small   \begin{tabular}{*{2}{l}*{9}{r}}     \toprule      & Parameter & \multicolumn{3}{c}{Quantile} & \multicolumn{3}{c}{$N$-obs. median} & \multicolumn{3}{c}{$N$-obs. mean} \\     \cmidrule(lr){3-5} \cmidrule(lr){6-8} \cmidrule(lr){9-11}     $F$ & Method \textbar\ Conf. level (\%) & \multicolumn{1}{c}{90} & \multicolumn{1}{c}{95} & \multicolumn{1}{c}{99} & \multicolumn{1}{c}{90} & \multicolumn{1}{c}{95} & \multicolumn{1}{c}{99} & \multicolumn{1}{c}{90} & \multicolumn{1}{c}{95} & \multicolumn{1}{c}{99} \\     \midrule     $F_1$ & Wald & $34$ & $24$ & $11$ & $32$ & $22$ & $11$ & $22$ & $14$ & $6$ \\     & \textsc{tem} & $203$ & $202$ & $179$ & $207$ & $202$ & $172$ & $253$ & $238$ & $184$ \\     & Severini (\textsc{tem}) & $127$ & $127$ & $127$ & $128$ & $128$ & $127$ & $135$ & $133$ & $126$ \\     & Severini (cov.) & $131$ & $131$ & $131$ & $133$ & $133$ & $131$ & $142$ & $140$ & $130$ \\ \addlinespace[3pt]     $F_2$ & Wald & $36$ & $26$ & $14$ & $35$ & $25$ & $14$ & $26$ & $18$ & $10$ \\     & \textsc{tem} & $192$ & $186$ & $158$ & $193$ & $184$ & $152$ & $221$ & $205$ & $168$ \\     & Severini (\textsc{tem}) & $126$ & $126$ & $125$ & $126$ & $126$ & $124$ & $129$ & $127$ & $121$ \\     & Severini (cov.) & $130$ & $130$ & $128$ & $131$ & $131$ & $128$ & $135$ & $133$ & $125$ \\ \addlinespace[3pt]     $F_3$ & Wald & $36$ & $26$ & $14$ & $34$ & $25$ & $14$ & $24$ & $17$ & $9$ \\     & \textsc{tem} & $193$ & $189$ & $162$ & $195$ & $187$ & $155$ & $224$ & $206$ & $168$ \\     & Severini (\textsc{tem}) & $126$ & $126$ & $125$ & $126$ & $126$ & $124$ & $130$ & $128$ & $123$ \\     & Severini (cov.) & $131$ & $131$ & $129$ & $132$ & $132$ & $129$ & $137$ & $134$ & $126$ \\ \addlinespace[3pt]     $F_4$ & Wald & $37$ & $29$ & $18$ & $36$ & $28$ & $18$ & $30$ & $22$ & $14$ \\     & \textsc{tem} & $182$ & $171$ & $144$ & $182$ & $168$ & $140$ & $201$ & $187$ & $157$ \\     & Severini (\textsc{tem}) & $124$ & $123$ & $120$ & $124$ & $122$ & $118$ & $122$ & $119$ & $112$ \\     & Severini (cov.) & $128$ & $126$ & $121$ & $129$ & $127$ & $121$ & $128$ & $125$ & $116$ \\ \addlinespace[3pt]     $F_5$ & Wald & $34$ & $23$ & $10$ & $32$ & $21$ & $9$ & $18$ & $10$ & $4$ \\     & \textsc{tem} & $211$ & $213$ & $189$ & $217$ & $214$ & $179$ & $293$ & $266$ & $184$ \\     & Severini (\textsc{tem}) & $130$ & $130$ & $129$ & $131$ & $131$ & $128$ & $144$ & $141$ & $129$ \\     & Severini (cov.) & $134$ & $135$ & $132$ & $136$ & $136$ & $132$ & $152$ & $148$ & $132$ \\ \addlinespace[3pt]     $F_6$ & Wald & $35$ & $25$ & $12$ & $33$ & $23$ & $11$ & $24$ & $15$ & $7$ \\     & \textsc{tem} & $203$ & $202$ & $179$ & $207$ & $202$ & $172$ & $248$ & $235$ & $185$ \\     & Severini (\textsc{tem}) & $126$ & $127$ & $127$ & $127$ & $128$ & $127$ & $133$ & $132$ & $126$ \\     & Severini (cov.) & $130$ & $131$ & $130$ & $132$ & $133$ & $131$ & $140$ & $138$ & $129$ \\ \addlinespace[3pt]     $F_7$ & Wald & $35$ & $24$ & $12$ & $33$ & $23$ & $11$ & $22$ & $14$ & $6$ \\     & \textsc{tem} & $203$ & $201$ & $176$ & $207$ & $201$ & $168$ & $252$ & $235$ & $180$ \\     & Severini (\textsc{tem}) & $127$ & $127$ & $127$ & $128$ & $128$ & $127$ & $135$ & $133$ & $126$ \\     & Severini (cov.) & $131$ & $132$ & $131$ & $133$ & $133$ & $131$ & $142$ & $140$ & $129$ \\ \addlinespace[3pt]     $F_8$ & Wald & $37$ & $28$ & $16$ & $35$ & $27$ & $16$ & $28$ & $20$ & $12$ \\     & \textsc{tem} & $186$ & $178$ & $151$ & $187$ & $175$ & $146$ & $209$ & $195$ & $162$ \\     & Severini (\textsc{tem}) & $124$ & $124$ & $122$ & $125$ & $124$ & $121$ & $125$ & $122$ & $115$ \\     & Severini (cov.) & $129$ & $129$ & $125$ & $130$ & $129$ & $125$ & $131$ & $129$ & $120$ \\ \addlinespace[3pt]     $F_9$ & Wald & $39$ & $31$ & $21$ & $37$ & $30$ & $21$ & $33$ & $26$ & $18$ \\     & \textsc{tem} & $170$ & $158$ & $136$ & $168$ & $155$ & $133$ & $185$ & $171$ & $148$ \\     & Severini (\textsc{tem}) & $122$ & $121$ & $117$ & $122$ & $120$ & $115$ & $120$ & $116$ & $110$ \\     & Severini (cov.) & $126$ & $123$ & $117$ & $126$ & $123$ & $117$ & $125$ & $120$ & $112$ \\     \bottomrule   \end{tabular}   \caption{Truncated mean ($\alpha=0.1$) of the ratio of the confidence interval width relative to the width of profile confidence interval (in \%), peaks-over-threshold method with $k=20$. The largest standard error, obtained using a nonparametric bootstrap, is 1.17\%. The distributions (from top to bottom) are Burr ($F_1$), Weibull ($F_2$), generalized gamma ($F_3$), normal ($F_4$), lognormal ($F_5$), Student $t$ ($F_6$), $\mathdis{GP}(\xi =0.1)$ ($F_7$), exponential ($F_8$) and $\mathdis{GP}(\xi=-0.1)$ ($F_9$).}   \label{tab_CIW_l3_pot} \end{table}
\begin{table}[t!]   \centering\small   \begin{tabular}{*{2}{l}*{9}{r}}     \toprule      & Parameter & \multicolumn{3}{c}{Quantile} & \multicolumn{3}{c}{$N$-obs. median} & \multicolumn{3}{c}{$N$-obs. mean} \\     \cmidrule(lr){3-5} \cmidrule(lr){6-8} \cmidrule(lr){9-11}     $F$ & Method \textbar\ Conf. level (\%) & \multicolumn{1}{c}{90} & \multicolumn{1}{c}{95} & \multicolumn{1}{c}{99} & \multicolumn{1}{c}{90} & \multicolumn{1}{c}{95} & \multicolumn{1}{c}{99} & \multicolumn{1}{c}{90} & \multicolumn{1}{c}{95} & \multicolumn{1}{c}{99} \\     \midrule     $F_1$ & Wald & $64$ & $53$ & $36$ & $62$ & $52$ & $34$ & $55$ & $44$ & $26$ \\     & \textsc{tem} & $127$ & $128$ & $130$ & $130$ & $131$ & $131$ & $137$ & $138$ & $138$ \\     & Severini (\textsc{tem}) & $108$ & $108$ & $108$ & $109$ & $109$ & $109$ & $111$ & $111$ & $111$ \\     & Severini (cov.) & $108$ & $108$ & $108$ & $109$ & $109$ & $109$ & $111$ & $111$ & $111$ \\ \addlinespace[3pt]     $F_2$ & Wald & $62$ & $53$ & $38$ & $61$ & $52$ & $37$ & $55$ & $45$ & $30$ \\     & \textsc{tem} & $126$ & $127$ & $127$ & $128$ & $128$ & $128$ & $133$ & $133$ & $131$ \\     & Severini (\textsc{tem}) & $108$ & $108$ & $108$ & $108$ & $108$ & $108$ & $110$ & $110$ & $109$ \\     & Severini (cov.) & $108$ & $108$ & $108$ & $108$ & $109$ & $109$ & $110$ & $110$ & $110$ \\ \addlinespace[3pt]     $F_3$ & Wald & $62$ & $53$ & $37$ & $61$ & $52$ & $36$ & $55$ & $44$ & $29$ \\     & \textsc{tem} & $126$ & $127$ & $128$ & $128$ & $129$ & $129$ & $134$ & $135$ & $132$ \\     & Severini (\textsc{tem}) & $108$ & $108$ & $108$ & $108$ & $108$ & $109$ & $110$ & $110$ & $110$ \\     & Severini (cov.) & $108$ & $108$ & $109$ & $108$ & $109$ & $109$ & $110$ & $111$ & $110$ \\ \addlinespace[3pt]     $F_4$ & Wald & $60$ & $53$ & $40$ & $59$ & $52$ & $39$ & $57$ & $49$ & $36$ \\     & \textsc{tem} & $126$ & $127$ & $126$ & $128$ & $128$ & $127$ & $131$ & $131$ & $129$ \\     & Severini (\textsc{tem}) & $108$ & $108$ & $107$ & $108$ & $108$ & $107$ & $109$ & $109$ & $108$ \\     & Severini (cov.) & $108$ & $108$ & $108$ & $108$ & $108$ & $108$ & $109$ & $109$ & $108$ \\ \addlinespace[3pt]     $F_5$ & Wald & $64$ & $53$ & $34$ & $63$ & $51$ & $32$ & $50$ & $37$ & $20$ \\     & \textsc{tem} & $127$ & $129$ & $132$ & $130$ & $132$ & $133$ & $144$ & $146$ & $143$ \\     & Severini (\textsc{tem}) & $108$ & $109$ & $109$ & $109$ & $110$ & $110$ & $113$ & $114$ & $114$ \\     & Severini (cov.) & $109$ & $109$ & $110$ & $110$ & $110$ & $110$ & $114$ & $114$ & $114$ \\ \addlinespace[3pt]     $F_6$ & Wald & $64$ & $54$ & $37$ & $63$ & $53$ & $36$ & $57$ & $46$ & $29$ \\     & \textsc{tem} & $127$ & $128$ & $130$ & $130$ & $131$ & $131$ & $137$ & $137$ & $137$ \\     & Severini (\textsc{tem}) & $108$ & $108$ & $108$ & $109$ & $109$ & $109$ & $111$ & $110$ & $110$ \\     & Severini (cov.) & $108$ & $108$ & $108$ & $109$ & $109$ & $109$ & $111$ & $111$ & $111$ \\ \addlinespace[3pt]     $F_7$ & Wald & $63$ & $53$ & $36$ & $62$ & $52$ & $35$ & $55$ & $43$ & $26$ \\     & \textsc{tem} & $128$ & $128$ & $130$ & $130$ & $131$ & $131$ & $138$ & $138$ & $137$ \\     & Severini (\textsc{tem}) & $108$ & $108$ & $108$ & $109$ & $109$ & $109$ & $111$ & $111$ & $111$ \\     & Severini (cov.) & $108$ & $108$ & $109$ & $109$ & $109$ & $109$ & $111$ & $111$ & $111$ \\ \addlinespace[3pt]     $F_8$ & Wald & $62$ & $54$ & $39$ & $61$ & $53$ & $38$ & $57$ & $48$ & $34$ \\     & \textsc{tem} & $126$ & $126$ & $126$ & $127$ & $128$ & $127$ & $131$ & $131$ & $129$ \\     & Severini (\textsc{tem}) & $107$ & $108$ & $108$ & $108$ & $108$ & $108$ & $109$ & $109$ & $109$ \\     & Severini (cov.) & $108$ & $108$ & $108$ & $108$ & $108$ & $108$ & $109$ & $109$ & $109$ \\ \addlinespace[3pt]     $F_9$ & Wald & $60$ & $54$ & $42$ & $60$ & $53$ & $42$ & $58$ & $51$ & $39$ \\     & \textsc{tem} & $125$ & $125$ & $123$ & $126$ & $126$ & $123$ & $128$ & $127$ & $124$ \\     & Severini (\textsc{tem}) & $108$ & $107$ & $107$ & $108$ & $107$ & $106$ & $108$ & $108$ & $106$ \\     & Severini (cov.) & $108$ & $108$ & $107$ & $108$ & $108$ & $107$ & $109$ & $108$ & $107$ \\     \bottomrule   \end{tabular}   \caption{Truncated mean ($\alpha=0.1$) of the ratio of the confidence interval width relative to the width of profile confidence interval (in \%), peaks-over-threshold method with $k=60$. The largest standard error, obtained using a nonparametric bootstrap, is 0.12\%. The distributions (from top to bottom) are Burr ($F_1$), Weibull ($F_2$), generalized gamma ($F_3$), normal ($F_4$), lognormal ($F_5$), Student $t$ ($F_6$), $\mathdis{GP}(\xi =0.1)$ ($F_7$), exponential ($F_8$) and $\mathdis{GP}(\xi=-0.1)$ ($F_9$).}   \label{tab_CIW_l1_pot} \end{table}
}
\clearpage 
\section{\texorpdfstring{Fisher information matrix for the $r$-largest observations}{Fisher information matrix for the r-largest observations}} \label{sec:rlarginfomat}
The information matrices for the $r$-largest order statistics of a generalized extreme value distribution are of the form $I_{ii} =  m_{ii}^{(a)} - (r-1) m_{ii}^{(b)}$, where the respective entries of the $3 \times 3$ matrix for $\xi > -0.5, \xi \neq 0$, are 
\begin{align*}
m^{a}_{11} &= \frac{{\left(\xi^2 + 2\xi + r\right)} \Gamma\left(r + 2 \, \xi\right)}{\tau^{2} \Gamma\left(r\right)}, \\
m^{a}_{12} &=-\frac{{\left(\xi^2 + 2\xi + r\right)} \Gamma\left(r + 2 \, \xi\right) - \Gamma\left(r + \xi + 1\right)}{\tau^{2} \xi \Gamma\left(r\right)}, \\
m^{a}_{13} &= \frac{r \xi \Gamma\left(r + \xi\right) + {\left(\xi^2 + 2\xi + r\right)} \Gamma\left(r + 2 \, \xi\right) - {\left\{\xi \psi^{(0)}\left(r + \xi + 1\right) + \xi + 1\right\}} \Gamma\left(r + \xi + 1\right)}{\tau \xi^{2} \Gamma\left(r\right)},\\
m^{a}_{22} &= 
\frac{{\left(\xi^2 + 2\xi + r\right)} \Gamma\left(r + 2 \, \xi\right) - 2 \, \Gamma\left(r + \xi + 1\right) + \Gamma\left(r + 1\right)}{\tau^{2} \xi^{2} \Gamma\left(r\right)}, \\
m^{a}_{23} & = -\frac{{\left(\xi^2 + 2\xi + r\right)} \Gamma\left(r + 2 \, \xi\right) - {\left\{{\left(r + \xi\right)} \xi \psi^{(0)}\left(r + \xi + 1\right) + \xi^2 + 2\xi + 2 \, r\right\}} \Gamma\left(r + \xi\right)}{\tau \xi^{3} \Gamma\left(r\right)}
\\ & \qquad + \frac{{\left\{\xi \psi^{(0)}\left(r + 1\right) + 1\right\}} \Gamma\left(r + 1\right)}{\tau \xi^{3} \Gamma\left(r\right)},\\
m^{a}_{33} & = \frac{{\left(\xi^2 + 2\xi + r\right)} \Gamma\left(r + 2 \, \xi\right) - 2 \, {\left\{{\left(r + \xi\right)} \xi \psi^{(0)}\left(r + \xi + 1\right) + \xi^{2} + r + \xi\right\}} \Gamma\left(r + \xi\right)}{\xi^{4} \Gamma\left(r\right)}\\&\qquad  -
\frac{{\left({\left[{\left\{\psi^{(0)}\left(r + 1\right)^{2} + \psi^{(1)}\left(r + 1\right)\right\}} \xi + 2 \, \psi^{(0)}\left(r\right)\right]} r \xi + \xi^2 + 2\xi + r\right)} \Gamma\left(r\right)}{\xi^{4} \Gamma\left(r\right)},
\\
m^{(b)}_{11} &= \frac{\xi^{2} \Gamma\left(r + 2 \, \xi\right)}{\tau^{2} {\left(2 \, \xi + 1\right)} \Gamma\left(r\right)}, \\
m^{(b)}_{12} &=  -\frac{\xi \Gamma\left(r + 2 \, \xi\right)}{\tau^{2} {\left(2 \, \xi + 1\right)} \Gamma\left(r\right)},
\\
 m^{(b)}_{13} &=\frac{(\xi + 1)\Gamma\left(r + 2 \, \xi\right)- (2 \, \xi + 1)\Gamma\left(r + \xi\right) }{\tau (2 \, \xi + 1)({\xi + 1})\Gamma\left(r\right)}, \\
m^{(b)}_{22} &= \frac{\Gamma\left(r + 2 \, \xi\right)}{\tau^{2} {\left(2 \, \xi + 1\right)} \Gamma\left(r\right)},
\\
m^{(b)}_{23} &= \frac{({2 \, \xi + 1})\Gamma\left(r + \xi\right) - ({\xi + 1})\Gamma\left(r + 2 \, \xi\right)}{\tau \xi ({\xi + 1})({2 \, \xi + 1})\Gamma\left(r\right)},
\\
m^{(b)}_{33} &= \frac{\Gamma\left(r + 2 \, \xi\right)(\xi + 1) - 2 \, \Gamma\left(r + \xi\right)(2\xi + 1) +\Gamma(r) (\xi + 1)(2\xi + 1)}{\xi^{2}(\xi + 1) (2\xi + 1) \Gamma(r)},
\end{align*}
where $\psi^{(0)}(x)=\Gamma'(x)/\Gamma(x)$ denotes the digamma function and $\psi^{(1)}(x)$ its first derivative.
\end{document}